\documentclass{aa}  

\usepackage{graphicx}
\usepackage[varg]{txfonts}
\usepackage{natbib}
\usepackage{booktabs}
\usepackage{longtable} 
\usepackage[maxfloats=256]{morefloats}
\usepackage{hyperref}
\usepackage[dvipsnames]{xcolor}
\definecolor{Sa_ra}{rgb}{0.1, 0.2, 0.7}
\definecolor{i_sa}{rgb}{0.1,0.7,0.1}

\begin{document} 

   \title{A search for ionised gas outflows in an H$\alpha$ imaging atlas of nearby LINERs
\thanks{Based on observations made with the Nordic Optical Telescope, operated by the Nordic Optical Telescope Scientific Association at the Observatorio del Roque de los Muchachos, La Palma, Spain, of the Instituto de Astrofisica de Canarias.}}

   \author{L. Hermosa Mu{\~n}oz
        \and
        I. M{\'a}rquez
        \and
        S. Cazzoli
        \and
        J. Masegosa
        \and
        B. Ag{\'i}s-Gonz{\'a}lez
        }

   \institute{
        Instituto de Astrof\'isica de Andaluc\'ia - CSIC, Glorieta de la Astronom\'ia s/n, 18008 Granada, Spain \\
              \email{lhermosa@iaa.es}
             }

   \date{Received M DD, YYYY; accepted M DD, YYYY}

 
  \abstract{
  Outflows play a major role in the evolution of galaxies. However, we do not have yet a complete picture of their properties (extension, geometry, orientation and clumpiness). For low-luminosity Active Galactic Nuclei (AGNs), in particular, low-ionisation nuclear emission line regions (LINERs), the rate of outflows and their properties are largely unknown.}
  {The main goal of this work is to create the largest, up-to-date atlas of ionised gas outflow candidates in a sample of 70 nearby LINERs. We aim to use narrow-band, imaging data to analyse the morphological properties of the ionised gas nuclear emission of these galaxies and to identify signatures of extended emission with distinctive outflow-like morphologies.} 
  {We obtained new imaging data from Alhambra Faint Object Spectrograph and Camera (ALFOSC)/Nordic Optical Telescope (NOT) for a total of 32 LINERs, and complemented it with Hubble Space Telescope archival data (HST) for 6 objects. We extracted the H$\alpha$ emission of the galaxies and used it to morphologically classify the circumnuclear emission. We combined our results with those from the literature for additional 32 targets. We additionally obtained soft X-ray data from Chandra archive to compare this emission with the ionised gas.}
  {The distribution of the ionised gas in these LINER indicates that $\sim$32\% show bubble-like emission, $\sim$28\% show a \lq Core-halo\rq, unresolved emission, and $\sim$21\% of the sample have a disky-like distribution. Dust lanes prevent any detailed classification for $\sim$11\% of the sample, that we call as \lq Dusty\rq. The soft X-ray emission is in most cases ($\sim$60\%) co-spatial with the ionised gas. If we account for the kinematical information which is available for a total of 60 galaxies, we end up with a total of 48\% of the LINERs with detected outflows/inflows in the emission lines (50\% considering only kinematical information based on Integral Field Spectroscopic data).}
  {Our results suggest that the incidence of outflows in LINERs may vary from 41\% up to 56\%, based on both the H$\alpha$ morphology and the kinematical information from the literature. The ionised gas seems to be correlated with the soft X-ray emission, so that they may have a common origin. We discuss the use of H$\alpha$ imaging for the pre-selection of candidates likely hosting ionised gas outflows.}


   \maketitle
%

 \section{Introduction}

   \noindent Outflows are believed to play an important role in the evolution of galaxies \citep[e.g.][]{Kormendy2013,Cresci2018,Veilleux2020}. Those driven by Active Galactic Nuclei (AGNs) interact with the gas in the host galaxy generating both negative and positive feedback processes that affect the evolution of the host with notable effects, such as the gas-recycling or the suppression of the star formation \citep[e.g.][]{Fabian2012,Cresci2018}.
   Outflows are characterised by their multi-wavelength phases (cold, warm and ionised gas), whose analysis is needed to fully understand their importance on the evolution of their host galaxies \citep[e.g.][]{Cazzoli2014, Cicone2014,RamosAlmeida2017,Morganti2017,Veilleux2020,Fluetsch2021}. 

   \noindent These galactic outflows are commonly seen in all AGN types \citep[e.g.][]{Veilleux2005,Morganti2017,Veilleux2020}, including a handful of Low-Ionisation Nuclear Emission-line Regions (LINERs) \citep{Heckman1980}. Given that the presence of outflows has been suggested to be ubiquitous within the AGN population \citep[e.g.][]{Veilleux2005,Concas2019}, it is also expected to be the case for low-luminosity AGNs, as LINERs. Although that, this field is largely unexplored except for few works \citep{Cazzoli2018,HM2020}. This is probably a result of a bias towards the search of outflows in powerful AGNs, where these outflows are more easily identified, as the outflow rate scales with the AGN luminosity \citep{Fluetsch2019}. Nevertheless, much less is known for the largest population of AGNs in the Local Universe, which is dominated by LINERs \citep{Ho1997}. Mainly hosted in early-type galaxies \citep{Ho2008}, their AGN nature is not yet clear for all the objects \citep{Marquez2017}, given the existence of other mechanisms, such as shocks \citep{Heckman1980,Dopita1995, Molina2018}, that could explain the observed spectra of these systems.
   
   \noindent One of the best ways to identify the presence of outflows is using kinematical information via 2D spectroscopy \citep[e.g.][Cazzoli et al. in prep]{Davies2014,Harrison2016,Mingozzi2019,Davies2020,Cazzoli2020,Raimundo2021}. These spectroscopic measurements allow us to obtain resolved information of the gaseous component of a galaxy at low \citep[e.g.][]{Raimundo2021} and high redshift \citep[e.g.][]{Harrison2016}. In more luminous AGNs, the kinematical component of the emission lines associated to the outflow can dominate and be the more predominant feature of the spectra, above other rotational movements of the internal gas \citep{Davies2020}. However, in the absence of spectroscopic data, outflows may be detected, or at least hinted by using imaging techniques, even when the activity of the host galaxy is low \citep{Masegosa2011}. 
   In the works by \cite{Pogge2000} and \cite{Masegosa2011} they studied the extended ionised H$\alpha$ emission of several LINER-like galaxies. In the latter work, this emission was classified in different morphological types depending on the distribution of the ionised gas. One of the possible classification is outflow-like, referred to extended filamentary or bubble-like gas extending out of the galaxy nuclei. For some galaxies, this ionised gas emission with morphological signatures of outflowing gas, was later confirmed as an outflow by means of the use of kinematic data obtained from 2D spectroscopy \citep[see e.g. NGC\,4676A in][]{Wild2014}. \\
   
   \noindent In this paper, our objectives are: (i) to create an atlas of the ionised gas morphology of a sample of nearby (z$<$0.025) LINERs that almost doubles that previously existing; (ii) search for morphological evidences of possible outflowing gas that could be associated to feedback processes, in order to determine how common outflows/inflows may be for the LINER AGN family; (iii) compare the H$\alpha$ morphology to that seen in X-rays (when possible), to investigate possible correlations between both emissions, as seen in different AGN types in previous works \citep{Bianchi2006,Masegosa2011,Bianchi2019}; and (iv) look for kinematically-identified outflows in the literature within our targets, in order to test the goodness and eventual relation of the morphological signatures in the actual detection of outflows.
   
   \noindent This paper is organised as follows: 
   Section~\ref{Sect_sample_data} describes the selection of the sample and the data gathering. In Section~\ref{Sect_datareduction} we present the data reduction. In Section~\ref{Section_results} we show the main results of the H$\alpha$ and X-ray morphologies. We discuss the kinematical signatures of outflows identified in the literature for our targets, its relation with the gas morphology, as well as the implications of our work in Section~\ref{Section_discussion}. Finally, Section~\ref{summary_conclusions} summarises the main conclusions. Individual comments on each galaxy are in Appendix~\ref{Appendix_A}; all the figures can be found in Appendix~\ref{Appendix_B}.

\section{Sample and data}
\label{Sect_sample_data}
   The total number of objects analysed in this work is 70 LINERs (see Table~\ref{Table_galaxieslisted}), that were selected as explained below.
   The parent sample of our data set comes from \cite{GM2009}, where the X-rays properties of 82 LINERs were analysed, and is complemented with 11 LINERs from \cite{Cazzoli2018} and 2 from \cite{Pogge2000}. We included in the sample one additional LINER (NGC\,5957) not initially in either of the initial samples, but for which we have integral field spectroscopic data (Hermosa Mu{\~n}oz et al. in prep). To select the targets, we applied the following criteria: (i) all galaxies should have distances not larger than 100 Mpc in order to ensure that the galaxies fitted in the field-of-view (FoV) allowing to detect parsec-scale structures; (ii) we do not consider systems on a high level of interaction. The first criterion excluded 14 galaxies and the second excluded 5 objects. This lead to 77 objects, from which 36 were already studied either in \cite{Masegosa2011} (hereafter M11) or in \cite{Pogge2000} (32 from the first, 4 from the second). Among the 32 targets in M11 we found NGC\,4676A and B, that are on a high interacting level, so we further excluded them. We added NGC\,3379 and NGC\,4278 to the remaining 41 targets, as they were analysed in M11 based only in the H$\beta$, but not in the H$\alpha$ emission. Thus, we end up with 43 LINERs; from them, we were able to gather data for 38 targets (see Table~\ref{Table_obslogHST}, Table~\ref{Table_obslogNOT} and Sect.~\ref{SubSec_datagather}). 
   
\begin{figure}
   \includegraphics[width=\columnwidth]{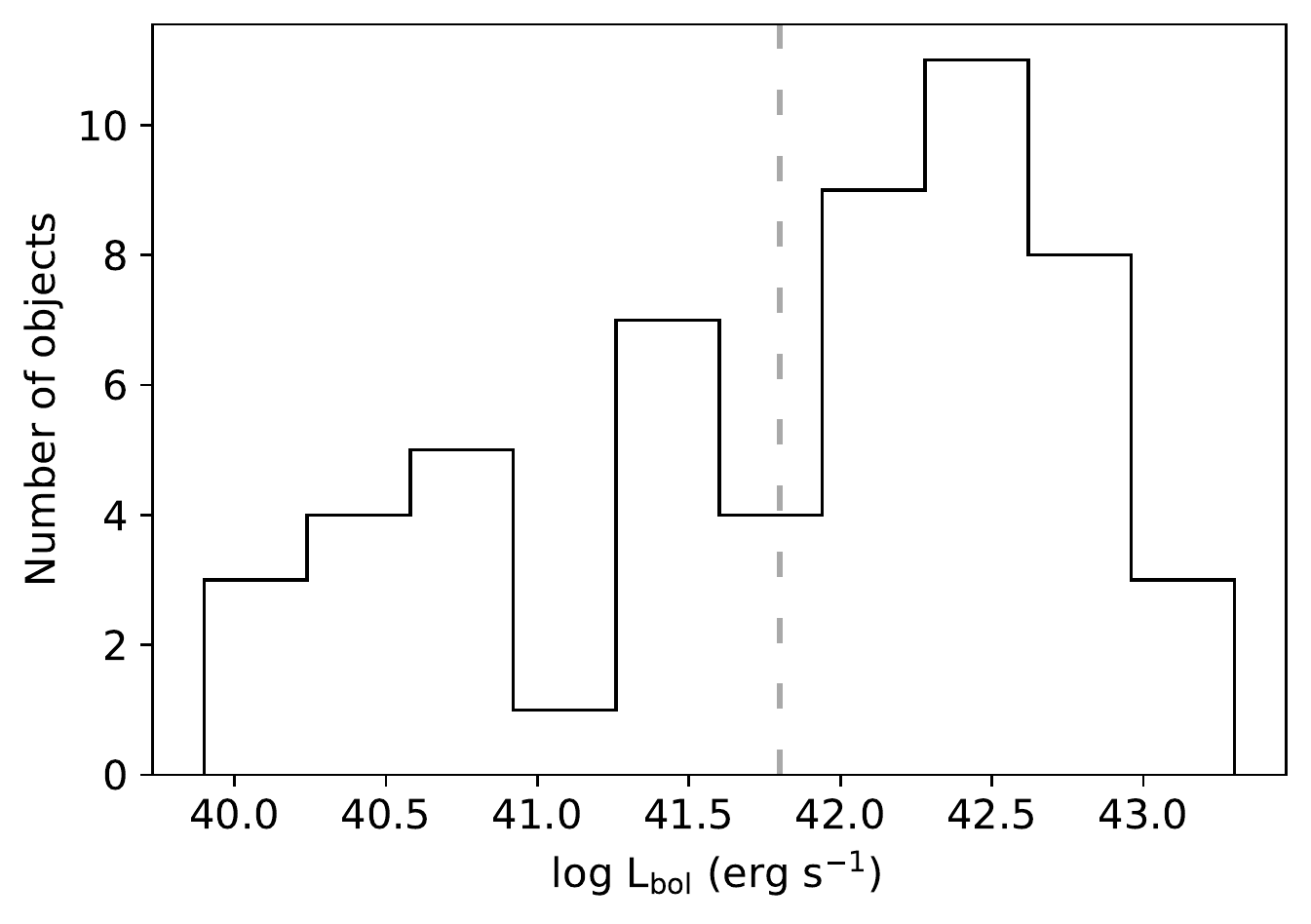}
   \caption{Histogram of the bolometric luminosities (in log units) of the selected sample of LINERs. The luminosities were estimated from \cite{GM2009b} (see Table~\ref{Table_galaxieslisted}). The dashed grey line indicates the average value (10$^{41.8}$ erg\,s$^{-1}$).}
   \label{Fig_histLum}
   \end{figure}

   \noindent To achieve the largest statistics of H$\alpha$ images possible, our final sample encompass the 38 new targets with those from M11 and \cite{Pogge2000} with previous observations (32 targets, excluding NGC\,3379, NGC\,4278, NGC\,4676A and B). Hence, summing up, the complete sample we analysed is 70 LINERs. We estimated the bolometric luminosity of the sources with the X-ray luminosity (2-10\,keV) following \cite{GM2009b} (see Fig.~\ref{Fig_histLum}). The typical bolometric luminosity of our sources is 10$^{41.8}$ erg\,s$^{-1}$ (see Table~\ref{Table_galaxieslisted}).

   \subsection{Data gathering}
   \label{SubSec_datagather}
   
   \noindent We searched for archival data in any available narrow filter from the HST for the 43 objects that corresponded to the H$\alpha$ line at the redshift of the targets, finding 6 objects with at least one image available (see Table~\ref{Table_obslogHST}). The selected filters are indicated for each object in Table~\ref{Table_obslogHST}. We aimed to observe the remaining 37 objects with the Alhambra Faint Object Spectrograph and Camera (ALFOSC), at the Nordic Optical Telescope (NOT), during several observing campaigns (see Table~\ref{Table_obslogNOT}). 
   
   \noindent The ALFOSC instrument is located at the NOT telescope (2.56 m), in \lq Roque de los Muchachos\rq\ observatory, in La Palma, Spain. We used its camera in imaging mode, with a total FoV of 6.4\arcmin$\times$6.4\arcmin\ and a pixel scale of 0.214\arcsec pix$^{-1}$.
   
   \noindent The data were gathered in a total of 14 nights (PIs: S. Cazzoli and L. Hermosa Mu{\~n}oz; see Table~\ref{Table_obslogNOT}). We obtained 3 individual exposures for both narrow and broad band filters per galaxy, with typical exposure times of 1200\,s and 300\,s, respectively. For some objects, the broad filter images were observed with smaller exposure time to avoid saturation of the galaxy centre. The total time on source per galaxy was 1.25 hours. The average seeing and airmass were 0.92\arcsec and 1.24, respectively (the corresponding information for the different nights is listed in Table~\ref{Table_obslogNOT}).
   
   \noindent We used the narrow filters from NOT that allowed us to observe the ionised gas of the LINERs at their given redshift, that corresponded to the H$\alpha$ line. These are (as indicated in the NOT web page\footnote{\url{http://www.not.iac.es/instruments/filters/filters.php}}): \#49 (661\_5 - FWHM 50\AA), \#50 (665\_5 - FWHM 55\AA), \#68 (664\_4 - FWHM 40\AA), \#77 (662\_4 - FWHM 39\AA), \#78 ([N\,II]658\_4 - FWHM 36\AA) and \#123 ([S\,II] 672\_5 - FWHM 50\AA). As for the broad filter, we used the standard R band filter from the Sloan Digital Sky Survey (as indicated in the NOT web page: r'\_SDSS 625\_140 - FWHM 1400\AA). The selected narrow filters are wide enough to also include one of the [N\,II]$\lambda\lambda$6548,6584\AA\,emission lines. However, we have chosen the filters such that the only other emission line that could also fit within the filter is the [N\,II]$\lambda$6548\AA\,line, which is weaker than the [N\,II]$\lambda$6584\AA\,line \citep[relation 1:3, see][]{Osterbrock2006}. Contamination of the H$\alpha$ line due star-formation processes could exist. However, we do not expect this to be important, as all our sources are early-type galaxies (see morphological types in Table~\ref{Table_galaxieslisted}), where the star formation is expected to be less important than in later host types.
   
    \noindent The presence of several ghosts contaminated the images of NGC\,0474 and NGC\,0524, observed with the NOT filter \#77. This prevented a reliable analysis and therefore, these two galaxies were further excluded from the sample. NGC\,5957 was observed while high clouds appeared during at least one individual exposure, dimming the object flux. Finally, due to bad weather conditions and time constraints to finish the last observing run, 3 objects (NGC\,0833, NGC\,0835 \& NGC\,2655) were not observed. Thus, the campaign is completed up to a $\sim$90\% of the sample (37 objects).
   
   \subsection{X-ray data}
   \label{SubSec_xrays}

   \noindent We searched for X-ray data in the Chandra Data Archive for all the new galaxies in the sample (38 objects) for which we obtained optical data, to compare their soft emission (0.3 - 2 KeV) with the ionised gas. We retrieved X-ray images taken with the Advanced CCD Imaging Spectrometer (ACIS) for a total of 28 objects. We estimated the 3$\sigma$ contours of the emission and superimposed them to the H$\alpha$ images. 
   
   \noindent For the data coming from M11, there is a similar analysis with both hard and soft X-ray Chandra images. They gathered data for a total of 26 objects, whose results we discuss in Sect.~\ref{SubSec_DiscXray}. The images from their objects can be found in their Fig.~6.
   
   \subsection{Kinematical information}
   \label{SubSec_kinematicInfo}
   
   \noindent By searching in the literature, we retrieved kinematical information either for all the LINERs from Integral Field Spectroscopy (IFS) (30 galaxies) or long-slit spectroscopic data (30 galaxies). When a target was analysed with IFS data, we prioritised those 3D results over those coming from long-slit spectroscopy. The previous works analysing the different targets (see Sect.~\ref{SubSec_outflows} and Appendix~\ref{Appendix_A} for more details on the individual sources) discuss the possible presence of outflows/inflows or non-rotational motions in the galaxies for 28 objects. We compare the morphologically-identified outflows with those kinematically-identified (or undetected) in Sect.~\ref{SubSec_outflows}.
 
\longtab{1}{
\begin{longtable}{cllccccrcrr}
  \caption{\label{Table_galaxieslisted} General properties for all the 70 LINERs in this paper. The line separates the new observed targets from those selected from \cite{Masegosa2011} and \cite{Pogge2000}. (3) RA and (4) DEC: coordinates; (5) Morphology, (6) z and (7) scale distance: from the Local Group from NED; (8) \textit{i}: inclination angle from \citet{Ho1997}; (9) V magnitude from HyperLeda; and (10) bolometric luminosity estimated from \citep{GM2009b}. $^{(*)}$ indicates galaxies observed with high clouds; $^{\dagger}$ indicates targets already analysed in \citet{Masegosa2011}.}\\
    \hline 
    \hline
    \# & ID & Other name & RA & DEC & Morphology & z & Scale & \textit{i} & V mag & L$_{\rm bol}$ (10$^{X}$) \\
    & & & (hh mm ss) & (dd mm ss) &  &  & (pc\,arcsec$^{-1}$) & (deg) &  & (erg\,s$^{-1}$) \\ 
    & (1) & (2) & (3) & (4) & (5) & (6) & (7) & (8) & (9) & (10) \\
   \hline
    \endfirsthead
    \caption{Continue.}    \\
    \hline
    \# & ID & Other name & RA & DEC & Morphology & z & Scale & \textit{i} & V mag & L$_{\rm bol}$ (10$^{X}$) \\
    & & & (hh mm ss) & (dd mm ss) &  &  & (pc\,arcsec$^{-1}$) & (deg) &  & (erg\,s$^{-1}$) \\
    & (1) & (2) & (3) & (4) & (5) & (6) & (7) & (8) & (9) & (10)\\
    \hline 
    \endhead 
    \hline
    \endfoot
    1& NGC 0266 & UGC\,508 & 00 49 47.80 & +32 16 39.79 & SB(rs)ab & 0.0155 & 351 & 12 & 11.8 & -- \\
    2& NGC 0410 & UGC\,735 & 01 10 58.90 & +33 09 06.81 & E+ & 0.0177 & 395 & -- & 11.5 & 42.25 \\
    3& NGC 0841 & IRAS\,02082+3715 & 02 11 17.36 & +37 29 49.80 & (R')SAB(s)ab & 0.0151 & 339 & 57 & 13.1 & -- \\
    4& NGC 2685 & IRAS\,08517+5855 & 08 55 34.71 & +58 44 03.83 & (R)SB0+pec & 0.0030 & 70 & 60 & 11.4 & 42.35 \\
    5& NGC 3185 & HCG\,44c & 10 17 38.56 & +21 41 17.70 & (R)SB(r)a & 0.0041 & 79 & 48 & 12.2 & 42.68 \\
    6& NGC 3379$^{\dagger}$ & M\,105 & 10 47 49.59 & +12 34 53.85 & E1 & 0.0030 & 55 & -- & 9.3 & 41.44 \\
    7& NGC 3414 & IRAS\,10485+2814 & 10 51 16.21 & +27 58 30.36 & S0 pec & 0.0049 & 100 & 44 & 11.1 & 41.39 \\
    8& NGC 3507 & UGC\,6123 & 11 03 25.36 & +18 08 07.62 & SB(s)b & 0.0033 & 62 & 32 & -- & $<$40.51 \\
    9& NGC 3608 & UGC\,6299 & 11 16 58.95 & +18 08 55.26 & E2 & 0.0041 & 80 & -- & 10.6 & 41.51 \\
    10& NGC 3628 & IRAS\,11176+1351 & 11 20 16.97 & +13 35 22.86 & Sb & 0.0028 & 51 & -- & 9.5 & 41.47 \\
    11& NGC 3642 & IRAS\,11194+5920 & 11 22 17.89 & +59 04 28.25 & SA(r)bc & 0.0053 & 121 & 34 & 12.2 & -- \\
    12& NGC 3884 &  UGC\,6746 & 11 46 12.18 & +20 23 29.93 & SA(r)0/a & 0.0233 & 496 & 51 & 12.6 & -- \\
    13& NGC 3898 & IRAS\,11465+5621 & 11 49 15.37 & +56 05 03.69 & SA(s)ab & 0.0039 & 91 & 55 & 10.7 & $<$42.08 \\
    14& NGC 3945 & IRAS\,11506+6056 & 11 53 13.73 & +60 40 32.00 & SB(rs)0 & 0.0043 & 100 & 50 & 10.8 & 40.66 \\
    15& NGC 4125 & IRAS\,12055+6527 & 12 08 06.02 & +65 10 26.90 & E6 pec & 0.0060 & 107 & -- & 9.7 & $<$42.04 \\
    16& NGC 4143 & UGC\,7142 & 12 09 36.06 & +42 32 03.00 & SAB(s)0 & 0.0032 & 70 & 52 & 12.1 & -- \\
    17& NGC 4203 & IRAS\,12125+3328 & 12 15 05.05 & +33 11 50.38 & SAB0- & 0.0036 & 77 & 21 & 11.6 & -- \\
    18& NGC 4261 & 3C\,270 & 12 19 23.22 & +05 49 30.78 & E2-3 & 0.0074 & 148 & -- & 11.1 & 42.60 \\
    19& NGC 4278$^{\dagger}$ & IRAS\,12175+2933 & 12 20 06.82 & +29 16 50.72 & E1-2 & 0.0021 & 42 &  -- & 10.2 & 42.53 \\
    20& NGC 4321 & M\,100 & 12 22 54.83 & +15 49 18.54 & SAB(s)bc & 0.0052 & 106 & 32 & 9.5 & 42.03 \\
    21& NGC 4450 & IRAS\,12259+1721 & 12 28 29.63 & +17 05 05.81 & SA(s)ab & 0.0065 & 134 & 43 & 10.9 & -- \\
    22& NGC 4457 & IRAS\,12264+0350 & 12 28 59.01 & +03 34 14.09 & (R)SAB(s)0/a & 0.0029 & 53 & 32 & 10.6 & 42.12 \\
    23& NGC 4459 & IRAS\,12264+1415 & 12 29 00.01 & +13 58 42.14 & SA(r)0+ & 0.0040 & 78 & 41 & 10.3 & 39.90 \\
    24& NGC 4494 & IRAS\,12288+2603 & 12 31 24.10 & +25 46 30.91 & E1-2 & 0.0045 & 93 & -- & 9.8 & 40.31 \\
    25& NGC 4589 & IRAS\,12353+7428 & 12 37 24.99 & +74 11 30.92 & E2 & 0.0066 & 155 & -- & 10.7 & 42.23 \\
    26& NGC 4596 & IRAS\,12373+1027 & 12 39 55.95 & +10 10 34.10 & SB(r)0+ & 0.0063 & 128 & 43 & 10.5 & 40.00 \\
    27& NGC 4698 & IRAS\,12458+0845 & 12 48 22.91 & +08 29 14.58 & SA(s)ab & 0.0034 & 64 & 53 & 10.7 & 42.05 \\
    28& NGC 4750 & IRAS\,12483+7308 & 12 50 07.27 & +72 52 28.72 & (R)SA(rs)ab & 0.0054 & 129 & 24 & 12.1 & -- \\
    29& NGC 4772 &  UGC\,8021 & 12 53 29.16 & +02 10 06.16 & SA(s)a & 0.0035 & 65 & 62 & 11.3 & -- \\
    30& NGC 5077 & UGCA\,347 & 13 19 31.67 & $-$12 39 25.07 & E3-4 & 0.0094 & 188 & -- & 11.7 & -- \\
    31& NGC 5363 & IRAS\,13356+0529 & 13 56 07.21 & +05 15 17.18 & I0? & 0.0038 & 76 & 51 & 10.2 & 43.09 \\
    32& NGC 5746 & IRAS\,14424+0209 & 14 44 55.92 & +01 57 18.01 & SAB(rs)b & 0.0058 & 120 & -- & 10.6 & 41.75 \\
    33& NGC 5813 &  UGC\,9655 & 15 01 11.23 & +01 42 07.13 & E1-2 & 0.0065 & 138 & -- & 10.5 & 42.08 \\
    34& NGC 5838 & IRAS\,15029+0217 & 15 05 26.26 & +02 05 57.59 & SA0- & 0.0045 & 94 & 72 & 10.8 & 42.51 \\
    35& NGC 5957$^{(*)}$ & IRAS\,15330+1212 & 15 35 23.21 & +12 02 51.36 & (R')SAB(r)b & 0.0061 & 134 & -- & 12.3 & -- \\
    36& NGC 6482 & UGC\,11009 & 17 51 48.81 & +23 04 18.99 & E & 0.0131 & 296 & -- & 11.3 & 42.64 \\
    37& NGC 7331 & IRAS\,22347+3409 & 22 37 04.01 & +34 24 55.87 & SA(s)b & 0.0038 & 80 & 72 & 9.4 & 41.76 \\
    38& NGC 7743 & IRAS\,23417+0939 & 23 44 21.14 & +09 56 02.69 & (R)SB(s)0+ & 0.0057 & 138 & 32 & 12.4 & $<$42.86 \\
    \hline
    39& IC 1459 & IRAS\,22544-3643 & 22 57 10.61 & -36 27 44.00 & E3-4 & 0.0060 & 131 & -- & 10.5 & -- \\
    40& NGC 0315 & UGC\,00597 & 00 57 48.88 & +30 21 08.81 & E+ & 0.0165 & 370 & 52 & 11.6 & 43.30 \\
    41& NGC 0404 & IRAS\,01066+3527 & 01 09 27.02 & +35 43 05.27 & SA0(s) & -0.0002 & - & -- & 10.6 & -- \\
    42& NGC 1052 & IRAS\,02386-0828 & 02 41 04.80 & -08 15 20.75 & E4 & 0.005 & 108 & -- & 11.0 & 42.77 \\
    43& NGC 2639 & IRAS\,08400+5023 & 08 43 38.08 & +50 12 20.00 & (R)SA(r)a & 0.0111 & 242 & 54 & 11.8 & $<$41.59 \\
    44& NGC 2681 & IRAS\,08499+5130 & 08 53 32.73 & +51 18 49.30 & (R)SAB(rs)0/a & 0.0023 & 24 & -- & 10.9 & 42.58 \\
    45& NGC 2787 & IRAS\,09148+6924 & 09 19 18.56 & +69 12 12.00 & SB0+(r) & 0.0023 & 60 & 51 & 11.3 & $<$40.34 \\
    46& NGC 2841 & IRAS\,09185+5111 & 09 22 02.63 & +50 58 35.47 & SA(r)b & 0.0021 & 49 & 66 & 10.2 & 40.75 \\
    47& NGC 3226 & ARP\,094 & 10 23 27.01 & +19 53 54.68 & E2 pec & 0.0044 & 86 & -- & 12.9 & 42.33 \\
    48& NGC 3245 & IRAS\,10244+2845 & 10 27 18.39 & +28 30 26.56 & SA0$^0$(r) & 0.0044 & 90 & 58 & 10.8 & 42.29 \\
    49& NGC 3607 & UGC\,06297 & 11 16 54.66 & +18 03 06.50 & SA0$^0$(s) & 0.0031 & 58 & 62 & 10.0 & 42.07 \\
    50& NGC 3623 & M\,65 & 11 18 55.96 & +13 05 32.00 & SAB(rs)a & 0.0027 & 48 & 77 & 9.3 & $<$40.91 \\
    51& NGC 3627 & M\,66 & 11 20 15.03 & +12 59 29.58 & SAB(s)b & 0.0024 & 42 & 65 & 10.3 & 42.72 \\
    52& NGC 3718 & ARP\,214 & 11 32 34.85 & +53 04 04.52 & SB(s)a pec & 0.0033 & 76 & 62 & 10.7 & -- \\
    53& NGC 3998 & UGC\,06946 & 11 57 56.13 & +55 27 12.92 & SA0$^0$(r) & 0.0035 & 82 & 34 & 11.3 & 42.85 \\
    54& NGC 4036 & IRAS\,11588+6210 & 12 01 26.75 & +61 53 44.81 & S0$^{-}$ & 0.0046 & 108 & 69 & 10.8 & 42.43 \\
    55& NGC 4111 & UGC\,07103 & 12 07 03.13 & +43 03 56.59 & SA0$^{+}$(r) & 0.0026 & 59 & 85 & 10.8 & $<$41.89 \\
    56& NGC 4192 & M\,98 & 12 13 48.29 & +14 54 01.20 & SAB(s)ab & -0.0005 & -- & 78 & 10.8 & -- \\
    57& NGC 4314 & IRAS\,12200+3010 & 12 22 31.82 & +29 53 45.19 & SB(rs)a & 0.0032 & 67 & 27 & 10.6 & $<$40.63 \\
    58& NGC 4374 & M\,84 & 12 25 03.74 & +12 53 13.14 & E1 & 0.0034 & 65 & -- & 9.8 & 42.84 \\
    50& NGC 4438 & IRAS\,12252+1317 & 12 27 45.59 & +13 00 31.78 & SA(s)0/a pec & 0.0002 & -- & 71 & 10.9 & $<$42.36 \\
    60& NGC 4486 & M\,87 & 12 30 49.42 & +12 23 28.04 & E & 0.0043 & 84 & -- & 9.0 & 42.35 \\
    61& NGC 4552 & M\,89 & 12 35 39.81 & +12 33 22.83 & E0-1 & 0.0011 & 17 & -- & 10.1 & 40.78 \\
    62& NGC 4579 & M\,58 & 12 37 43.52 & +11 49 05.50 & SAB(rs)b & 0.0051 & 101 & 38 & 10.3 & 42.70 \\
    63& NGC 4594 & M\,104 & 12 39 59.43 & -11 37 22.99 & SA(s)a & 0.0034 & 59 & 68 & 8.6 & 41.50 \\
    64& NGC 4636 & UGC\,07878 & 12 42 49.83 & +02 41 15.99 & E0-1 & 0.0031 & 57 & -- & 9.9 & $<$40.56 \\
    65& NGC 4696 & ABELL\,3526:BCG & 12 48 49.27 & -41 18 40.04 & cD1 pec & 0.0099 & 193 & -- & 10.3 & 41.51 \\
    66& NGC 4736 & M\,94 & 12 50 53.06 & +41 07 13.65 & (R)SA(r)ab & 0.0010 & 25 & 36 & 9.5 & 40.13 \\
    67& NGC 5005 & IRAS\,13086+3719 & 13 10 56.23 & +37 03 33.14 & SAB(rs)bc & 0.0032 & 70 & 63 & 10.7 & $<$43.16 \\
    68& NGC 5055 & M\,63 & 13 15 49.33 & +42 01 45.40 & SA(rs)bc & 0.0017 & 40 & 56 & 8.6 & 41.09 \\
    69& NGC 5846 & UGC\,09706 & 15 06 29.28 & +01 36 20.25 & E0-1 & 0.0057 & 121 & -- & 10.2 & $<$42.34 \\
    70& NGC 5866 & M\,102 & 15 06 29.49 & +55 45 47.57 & SA0+ & 0.00001 & 13 & 68 & 11.3 & 41.60 \\
    \hline
\end{longtable}
}  
   
\begin{table*}
        \caption{Observing log of HST data. Columns indicate: (1) galaxy name; (2) observing instrument; (3) file name of the image as indicated in the archive; (4) proposal ID of the observations; (5) Principal Investigator of the proposal; (6) date of the observation; (7) exposure time; (8) filter used, also from the HST archive.}
        \label{Table_obslogHST}
    \centering
        \begin{tabular}{l c c c c c c c}
        \hline \hline
        ID & Instrument & Filename & Proposal & P.I. & Obs. date & Exp. time & Filter\\ 
         & & (.fits) & ID &  & (yy-mm-dd) & (s) & \\
         (1) & (2) & (3) & (4) & (5) & (6) & (7) & (8) \\ \hline
        NGC 3642 & ACS/WFC & \textit{hst\_9788\_41\_acs\_wfc\_f658n} & $9788$ & L. Ho & 2003-12-16 & 700 & F658N \\
         & ACS/WFC & \textit{hst\_9788\_41\_acs\_wfc\_f814w} & $9788$ & L. Ho & 2003-12-16 & 120 & F814W \\
        NGC 4125 & WFPC2 & \textit{hst\_11966\_15\_wfpc2\_f658n\_wf} & 11966 & M. Regan & 2009-01-30 & 2100 & F658N \\
         & WFPC2 & \textit{hst\_06587\_23\_wfpc2\_f555w\_wf} & 6587 & D.Richstone & 1997-03-15 & 1400 & F555W \\
        NGC 4203 & ACS/WFC & \textit{hst\_9788\_72\_acs\_wfc\_f658n} & $9788$ & L. Ho & 2003-07-18 & 700 & F658N \\
         & ACS/WFC & \textit{hst\_9788\_72\_acs\_wfc\_f814w} & $9788$ & L. Ho & 2003-07-18 & 120 & F814W\\  
        NGC 4450 & WFPC2 & \textit{hst\_11966\_18\_wfpc2\_f658n\_wf} & 11966 & M. Regan & 2009-01-31 & $1800$ & F658N \\
         & WFPC2 & \textit{hst\_05375\_04\_wfpc2\_f555w\_wf} & 5375 & V.Rubin & 1994-05-09 & $520$ & F555W \\
        NGC 4750 & ACS/WFC & \textit{hst\_9788\_b4\_acs\_wfc\_f658n} & $9788$ & L. Ho & 2004-05-20 & $700$ & F658N \\
         & ACS/WFC & \textit{hst\_9788\_b4\_acs\_wfc\_f814w} & $9788$ & L. Ho & 2004-05-20 & $120$ & F814W \\
        NGC 7331 & WFC3 & \textit{hst\_14202\_01\_wfc3\_uvis\_f657n} & 14202 & D.Milisavljevic & 2015-08-22 & 2400 & F657N  \\
         & WFC3 & \textit{hst\_14202\_01\_wfc3\_uvis\_f814w} & 14202 & D.Milisavljevic & 2015-08-22 & 1350 & F814W \\
        \hline
        \end{tabular}
\end{table*}
 
\begin{table*}
        \caption{Observing log of ALFOSC/NOT data. Columns indicate: (1) observing campaign; (2) date of the observation; (3) number of galaxies observed ; (4) exposure time for the narrow (broad) filter; (5) range of the airmass; (6) seeing range during the observations.}
        \label{Table_obslogNOT}
    \centering
        \begin{tabular}{l c c c c c c}
        \hline \hline
        Observing & Observing date & Observed galaxies & Exposure time & Airmass & Seeing \\ 
        campaign & (yy-mm-dd) &  & NF-BF (s) &  &  \\
        (1) & (2) & (3) & (4) & (5) & (6) \\ \hline
        GT-Jan2019 & 2019-01-03 & NGC\,0841/NGC\,2685/NGC\,3185 & 1200 (300) & 1.01-1.42 & 0.7$\arcsec$-1.5$\arcsec$  \\
         & & NGC\,3884/NGC\,4278 &  &  &  \\
         & 2019-01-04 & NGC\,0266/NGC\,0410/NGC\,3414/NGC\,4143 & 1200 (300) & 1.03-1.7 & 0.6$\arcsec$-1.8$\arcsec$  \\
         & 2019-01-05 & NGC\,3507/NGC\,4698/NGC\,4772  & 1200 (300) & 1.06-1.5 & 0.7$\arcsec$-1.3$\arcsec$ \\
        \hline
        GT-May2019 & 2019-05-31 & NGC\,3628 & 1200 (300) & 1.10-1.30 & 0.8$\arcsec$-0.9$\arcsec$  \\
         & 2019-06-01 & NGC\,5363 & 1200 (100) & 1.09-1.16 & 0.9$\arcsec$-1.6$\arcsec$  \\
        \hline
        GT-Jan2020 & 2020-01-25 & NGC\,3379 & 1200 (300) & 1.04-1.19 & 0.7$\arcsec$-0.9$\arcsec$  \\
         & 2020-01-26 & NGC\,3608/NGC\,3945/NGC\,4596 & 1200 (300) & 1.07-1.29 & 0.6$\arcsec$-0.9$\arcsec$  \\
        \hline
        GT-Jun2021 & 2021-06-11 & NGC\,4494 & 1200 (100) & 1.18-1.30 & 0.8$\arcsec$-0.9$\arcsec$ \\ 
         & 2021-06-12 & NGC\,6482 & 1200 (300) & 1.04-1.10 & 0.8$\arcsec$-1.0$\arcsec$ \\ 
        \hline
        OT-Mar2020 & 2020-03-29 & NGC\,3898/NGC\,4321/NGC\,4457/NGC\,4459 & 1200 (300) & 1.03-1.21 & 0.6$\arcsec$-1.1$\arcsec$  \\ 
         & & NGC\,5746/NGC\,5813/NGC\,5957 &  &  &  \\ 
        \hline
        OT-Jun2020 & 2020-06-08 & NGC\,5077 & 1200 (300) & 1.33-1.43 & 0.7$\arcsec$-0.8$\arcsec$  \\  
         & 2020-06-09 & NGC\,4589 &  1200 (300) & 1.54-1.70 & 0.7$\arcsec$-1.0$\arcsec$  \\  
        \hline
        Other & 2020-11-09 & NGC\,7743 & 1200 (100) & 1.12-1.39 & 0.7$\arcsec$-0.8$\arcsec$  \\    
         & 2021-02-15 & NGC\,4261 \& NGC\,5838 & 1200 (100) & 1.10-1.42 & 0.8$\arcsec$-1.1$\arcsec$  \\    
        \hline
        \end{tabular}
\end{table*}

\section{Data reduction process}
\label{Sect_datareduction}

The data gathered from the HST archive were already fully reduced and the narrow-band images of the galaxies were already combined in those cases where there was more than one available. However, given that the HST observations are from the archive, some of the selected narrow filter (NF) images were not observed the same night as the broad filter (BF) images (see Table~\ref{Table_obslogHST}). We tried to select, when possible, the BF with the closest wavelength range to the NF range. Thus, the only procedure applied to HST data was the realignment of the NF and BF images when they were obtained in different observing campaigns (see Table~\ref{Table_obslogHST}). Specifically, we used the position angles on the header of the different frames to set the images to a common North-East axis reference. Then the outermost, elliptical isophotes were fitted to ellipses, providing us with the centres and position angles that were used to apply the corresponding shift and rotation for alignment. Finally, we trimmed the images to where both had information.\\

\begin{figure*}
   \includegraphics[width=\textwidth]{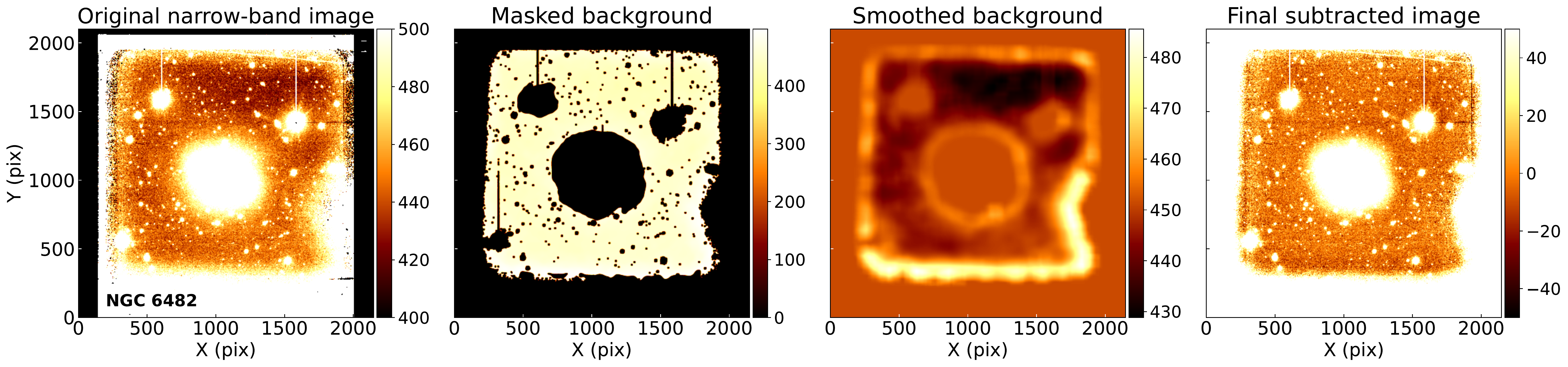}
   \caption{Example of the background subtraction process for NGC\,6482. \textit{Left:} Original narrow-band image of the galaxy. \textit{Middle-left:} Mask applied to the original image over imposed to the estimated background. \textit{Middle-right:} Final smoothed background. \textit{Right:} Final narrow-band image with the corresponding background subtracted. The scale on the left and right panels is around the background level to appreciate the differences before and after the subtraction. See Sect.~\ref{Sect_datareduction} for more details.}
   \label{Fig_BKG}
   \end{figure*}

\noindent The data from ALFOSC/NOT were reduced following standard procedures (bias and flat-fielding) with both \textsc{IRAF}\footnote{IRAF is the Image Reduction and Analysis Facility distributed by the National Optical Astronomy Observatories (NOAO) for the reduction and analysis of astronomical data. \url{http://iraf.noao.edu/}} and \textsc{python} routines. We performed a dedicated background subtraction to the ALFOSC/NOT images required for a reliable sky determination. An example of the procedure is shown in Fig.~\ref{Fig_BKG}. We used \textsc{photutils} routines under a \textsc{python} environment to mask external stars, galaxies and artefacts from the filter, and estimate a 2D background. We created masks for objects detected at least at 3$\sigma$ over the background in the BF images (as usually the flux is more extended in this filter) and applied them to the NF ones. The background in the masked pixels could not be estimated, so we used the median value of the background in the rest of the image. Then, we made a smoothing of the background image (100$\times$100 pixels), so that no structures were included due to the masks, and we subtracted the final smoothed background image per each galaxy and filter. \\

\noindent In order to obtain exclusively the emission of the galaxies in the H$\alpha$ line, we subtracted the continuum of the narrow-band images using broad-band images of the same targets. For ALFOSC/NOT, those BF images were observed the same night as the NF ones. The procedure follows that from M11. The NF images include not only the H$\alpha$ emission but also the underlying continuum. The line emission is usually less extended than the continuum, which is translated into the fact that, at large scales in the galaxy, both NF and BF images should return comparable emissions. Thus, we scaled the BF image to the NF image and then subtract them to obtain the H$\alpha$ image. \\ 

\noindent The peculiar morphologies of the galactic continuum in the studied galaxies are highlighted by the use of sharp-divided images \citep[e.g.][]{Marquez1996, Marquez1999, Marquez2003}. Briefly, this technique consists on dividing the BF image by its smoothed version, which is the original image convolved with a median filter of a 20-pixel box. Although simple, this method is very useful specially for detecting asymmetries on the image, such as bars, rings, spiral arms or dust regions. 
The sharp-divided images of the galaxies are shown in the right panels of Appendix~\ref{Appendix_B}. \\ 

\noindent The Chandra X-ray images were reduced following the Chandra threads aimed for the ACIS instrument, using the dedicated software \textsc{CIAO} v4.13 and the calibration package \textsc{CALDB} 4.9.5. After downloading the data, we applied the most recent calibrations running the script \lq chandra\_repro\rq\ and restricted to the energy range between 0.3-2 KeV. We extracted background light curves from the event files using the \textsc{CIAO} tool \textsc{DMEXTRACT} in order to eliminate high background events. Based on those background light curves, we generated Good Time Intervals for our images excluding time intervals deviated $1\sigma$ from the mean rate of counts in the background using the \textsc{CIAO} tool \textsc{DEFLARE}. 

\section{Results}
\label{Section_results}

   \begin{figure*}
   \includegraphics[trim=0.5cm 9cm 0.5cm 0cm,width=\textwidth]{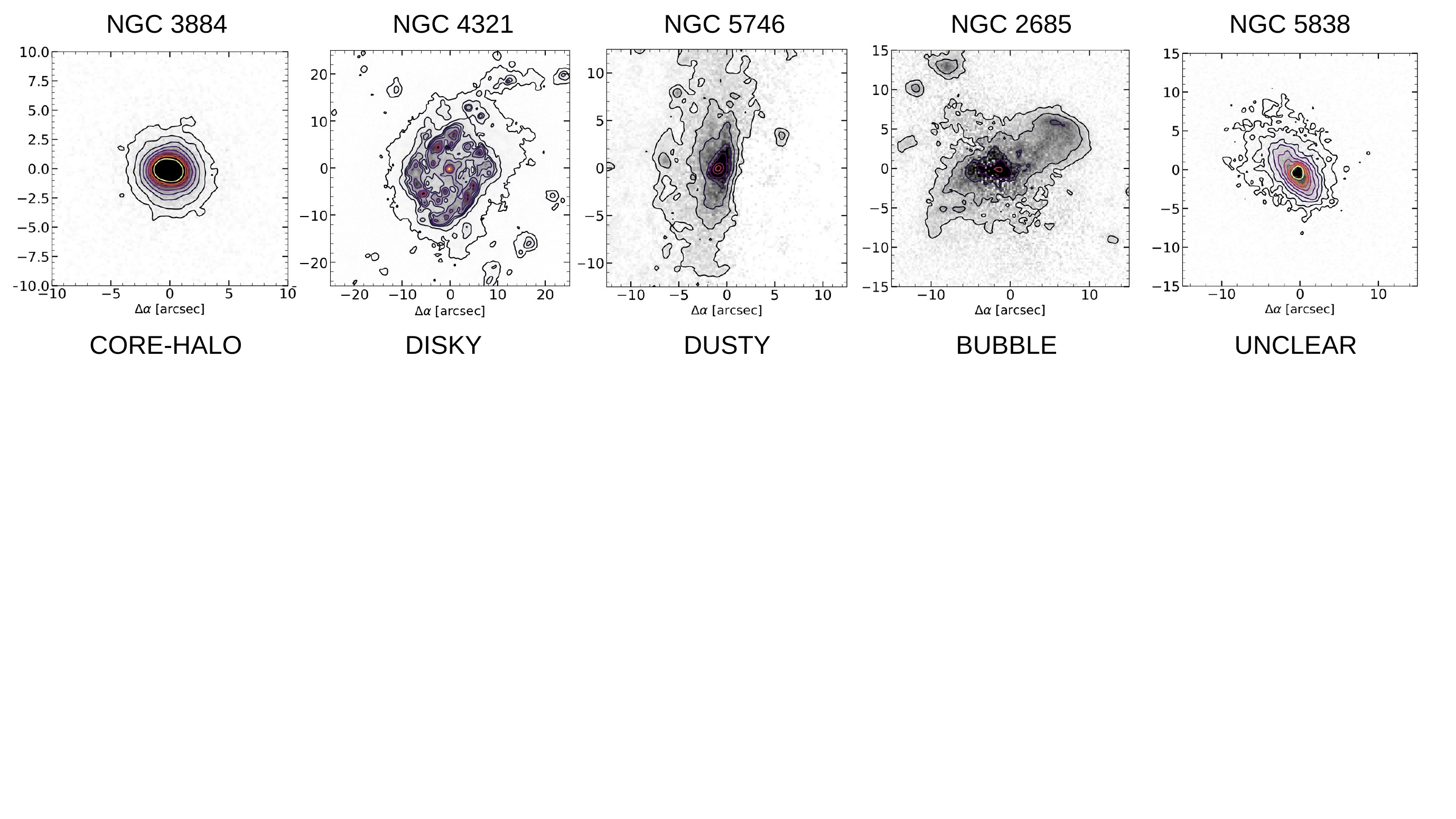}
   \caption{Examples of the five proposed morphological classifications of the nuclear H$\alpha$ emission (see Sect.~\ref{Section_results}). The complete images are in Appendix~\ref{Appendix_B}.}  
   \label{Fig_examples}
   \end{figure*}

We have classified the circumnuclear H$\alpha$ morphologies of the galaxies according to 4 different categories: \lq Core-halo\rq, \lq Disky\rq, \lq Bubble\rq\ and \lq Dusty\rq. An example of each morphological type is shown in Fig.~\ref{Fig_examples}. The properties of the ionised emission included in these categories are fully explained in M11. The basic definitions are: (i) \lq Core-halo\rq: unresolved nuclear emission in the galaxy centre; (ii) \lq Disky\rq: ionised emission along spiral arms, star forming rings or diffuse emission in the disc; (iii) \lq Bubble\rq: equivalent to the \lq Outflow\rq\,morphology in M11, refers to biconical, filamentary or bubble-like structures emerging from the nucleus; and (iv) \lq Dusty\rq: mostly edge-on galaxies with dust lanes obscuring the nuclear emission. We classified all the targets using visual inspection of the H$\alpha$ images, considering all the emission detected above 3$\sigma$ from the background. Our classification is given in Table~\ref{Table_classification}. We assigned a morphological class when 3 or more authors agreed on the classification. This happened for 87\% of the targets from the new data. The remaining 13\% (5 out of 38) had ambiguous morphologies that could be included in various classes. To enable a more clear discussion, we defined for them the additional category \lq Unclear\rq. In this Section we account only for the new observed data, as the results for the other targets are presented in M11 and \cite{Pogge2000}. More details on the individual galaxies can be found in Appendix~\ref{Appendix_A}; all H$\alpha$ and sharp-divided images are in Appendix~\ref{Appendix_B}, and the soft X-ray images (in contours) are in Fig.~\ref{Figure_Xrays}.

\subsection{Ionised-gas morphological classification}
\label{SubSec_Morphology}

\noindent We have identified a total of 12 LINERs that show \lq Core-halo\rq\ emission. Among them, NGC\,4261, NGC\,4589 and NGC\,6482 (see Figs.~\ref{Figure_AB5},~\ref{Figure_AB7} and~\ref{Figure_AB9}, respectively) show extended emission probably elongated along their galaxy disks; NGC\,3884, NGC\,4278, NGC\,4772 and NGC\,5957 (see Figs.~\ref{Figure_AB3},~\ref{Figure_AB5},~\ref{Figure_AB8} and~\ref{Figure_AB9}, respectively) have an unresolved, not elongated nuclear emission; in the remaining galaxies, NGC\,0410, NGC\,4450, NGC\,4494, NGC\,4698 and NGC\,7743 (see Figs.~\ref{Figure_AB1},~\ref{Figure_AB6},~\ref{Figure_AB7} and~\ref{Figure_AB10}), the H$\alpha$ emission is extended and detected at 3$\sigma$ level at distances larger than $>$10\arcsec from their nuclei.

\noindent The \lq Disky\rq\ morphology is found in 9 galaxies. For NGC\,0841, NGC\,3642, NGC\,4321 and NGC\,4457 (see Figs.~\ref{Figure_AB1},~\ref{Figure_AB3},~\ref{Figure_AB5} and~\ref{Figure_AB6}) the emission is found along a star-formation ring or arm, where several clumps are easily identified. In 3 galaxies, namely NGC\,3185, NGC\,3507 and NGC\,5077 (see Figs.~\ref{Figure_AB2} and~\ref{Figure_AB8}), the emission shows a twisted and elongated shape that may be ascribed to a bar or to large-scale spiral arms or discs. Finally, for 2 galaxies, NGC\,3608 and NGC\,3898 (see Figs.~\ref{Figure_AB3} and~\ref{Figure_AB4}, respectively), the emission is extended along the disc, with no clumps or evidence of star formation.

\begin{figure*}
   \includegraphics[trim=0.5cm 8cm 0.5cm 0cm,width=\textwidth]{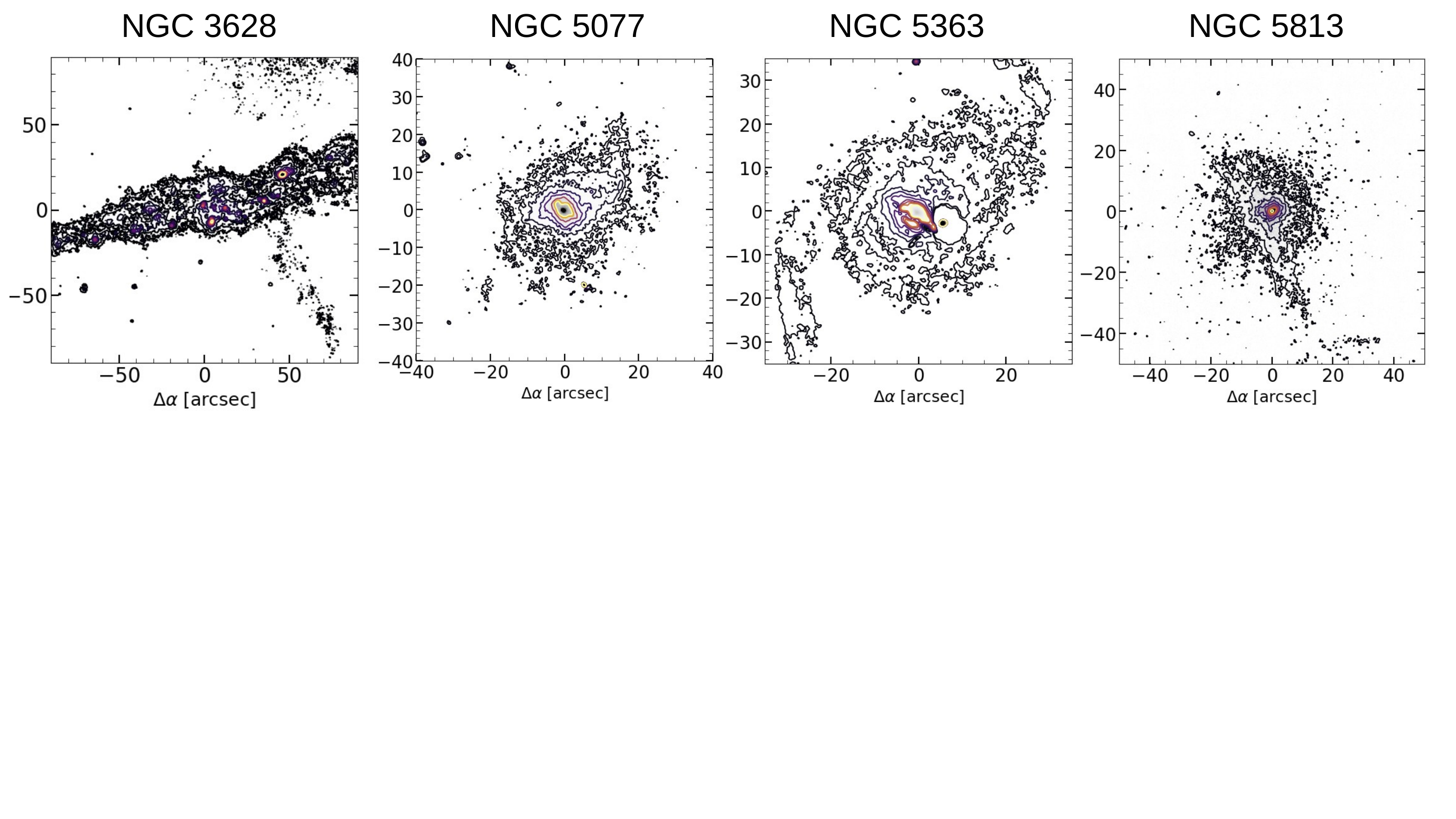}
   \caption{Large-scale H$\alpha$ images of NGC\,3628, NGC\,5077, NGC\,5363 and NGC\,5813. The largest contours indicate the 1$\sigma$ level over the background signal, except for NGC\,5363, that represents 3$\sigma$. The corresponding image on smaller scale ($\sim$10\arcsec$\times$10\arcsec) with the 3$\sigma$-detection are shown in Appendix~\ref{Appendix_B}.}
   \label{Figure_largescale}
   \end{figure*}

\noindent There are only 3 objects classified as \lq Dusty\rq, as they show clear dust lanes that prevent from analysing their nuclear emission (NGC\,3628, NGC\,4125 and NGC\,5746). However, note that this classification is based on the innermost regions ($\sim$20\arcsec) of the galaxies. NGC\,3628 (Fig.~\ref{Figure_AB3}) is known to have extended emission further out from the nuclei, a large scale H$\alpha$ emission clearly visible in the southern parts of the galaxy \citep{Fabbiano1990}. In our image, this plume is barely detected at 1$\sigma$ level (see Fig.~\ref{Figure_largescale}), whose analysis is out of the aim of this paper. NGC\,4125 (Fig.~\ref{Figure_AB4}) has some extended emission, but clearly around dust lanes, thus we were conservative in its classification. For NGC\,5746 (Fig.~\ref{Figure_AB8}), the emission is barely visible through some regions of the galactic disc, thus not reliable for a proper morphological classification.

\noindent We have identified 9 LINERs that may have an outflowing emission with complex ionised gas morphologies. NGC\,0266, NGC\,3414, NGC\,4596 and NGC\,4750 (see Figs.~\ref{Figure_AB1},~\ref{Figure_AB2} and~\ref{Figure_AB7}) show a faint extended emission parting asymmetrically from the photometric centre of the galaxies. NGC\,3945 image unveils a filamentary structure (Fig.~\ref{Figure_AB4}). For NGC\,2685, NGC\,3379 and NGC\,5813 (see Figs~\ref{Figure_AB1},~\ref{Figure_AB2} and~\ref{Figure_AB9}), the H$\alpha$ emission has a biconical or bubble-like shape oriented almost perpendicular to the major axis of the galaxies (inclination of NGC\,2685 is 60$\degr$, not measured for the other two; see Table~\ref{Table_galaxieslisted}). NGC\,4459 (see Fig.~\ref{Figure_AB6}) can also be included in this latter group, as the nuclear emission in the inner 3\arcsec$\times$3\arcsec resembles the base of a outflowing bubble. All the objects would benefit of detailed spectroscopic information with spatial resolution to determine the extent and origin of their emission. 

\noindent Finally, there are five galaxies (NGC\,4143, NGC\,4203, NGC\,5363, NGC\,5838 and NGC\,7331) with \lq Unclear\rq\ morphologies. NGC\,4143, NGC\,4203 and NGC\,5838 are double classified in this work as \lq Bubble\rq\,and \lq Disky\rq\ (see Figs.~\ref{Figure_AB4},~\ref{Figure_AB5} and~\ref{Figure_AB9}). In these cases, the emission seems to be extended along the galaxy disc, but with a non-symmetrical distribution. These asymmetry may be produced due to outflows instead of being associated to the disc, which lead to the unclear classification. For the case of NGC\,5363 (Fig.~\ref{Figure_AB8}), a nuclear dust lane of $\sim$1\,kpc is obscuring the nucleus, although at larger scales ($\geq$20\arcsec) there is a extended, filamentary emission of ionised gas (see Fig.~\ref{Figure_largescale}). The morphological differences at both scales lead to a double classification as \lq Dusty\rq, and \lq Bubble\rq. NGC\,7331 (Fig.~\ref{Figure_AB10}) is classified as both \lq Dusty\rq\ and \lq Bubble\rq, as we detect some extended emission out of the nucleus, but we cannot ensure this H$\alpha$ profile to be symmetrical, as a thick dust lane is visible within the galaxy disc. \\

\noindent There are two galaxies, NGC\,3379 and NGC\,4278 (Figs.~\ref{Figure_AB2} and~\ref{Figure_AB5}), that were observed in our sample despite they were also analysed in M11. The only available data in the HST archival were in the narrow filter corresponding to the [O\,III] line emission (filter F547M). In our work, we have obtained data for these galaxies in the H$\alpha$ wavelength range, to have an uniform comparison of all the nuclear emissions in the galaxies from the sample. For M11, NGC\,3379 (NGC\,4278) was classified as having a \lq Disky\rq\ (\lq Core-halo\rq) emission. We agree on the classification of NGC\,4278 but, on the contrary, we classified NGC\,3379 as having an \lq Bubble\rq, as the inner H$\alpha$ morphology shows an elongated shape that cannot be ascribed to either of the other possible classifications. This elongated emission is coincident with a dust lane crossing the nuclear region, visible in H$\beta$ \citep{Masegosa2011}. It could also be an effect produced by a bar (see individual comments on Appendix~\ref{Appendix_A}), as the sharp-divided image (see Fig.~\ref{Figure_AB2}) show an \lq X\rq-like shape which is usually ascribed to barred systems \citep{Laurikainen2017}.

   \subsection{Soft X-ray images properties}
   \label{SubSec_ResultXrays}
   
\begin{figure*}
   \includegraphics[width=0.4\columnwidth]{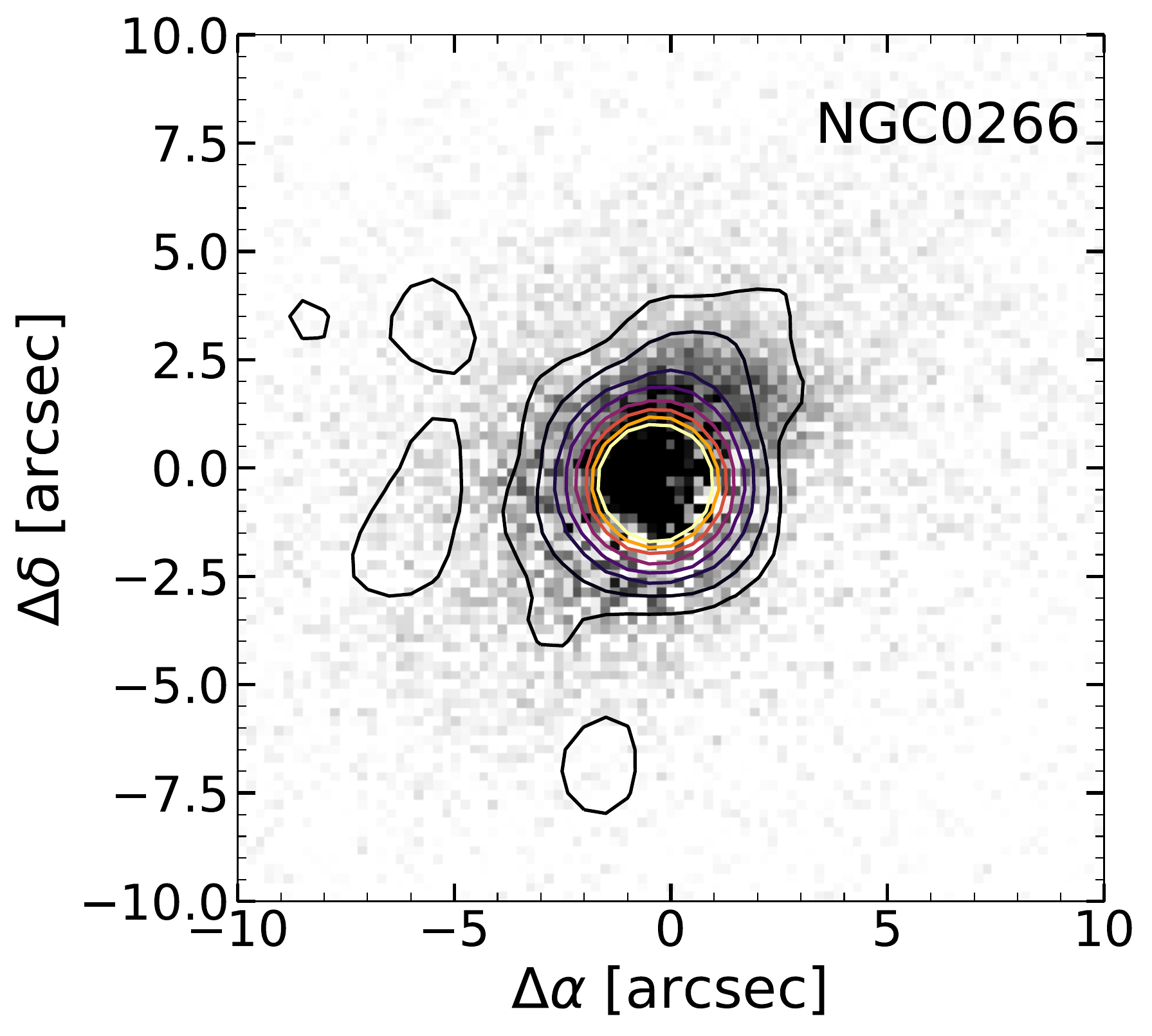}
   \includegraphics[width=0.37\columnwidth]{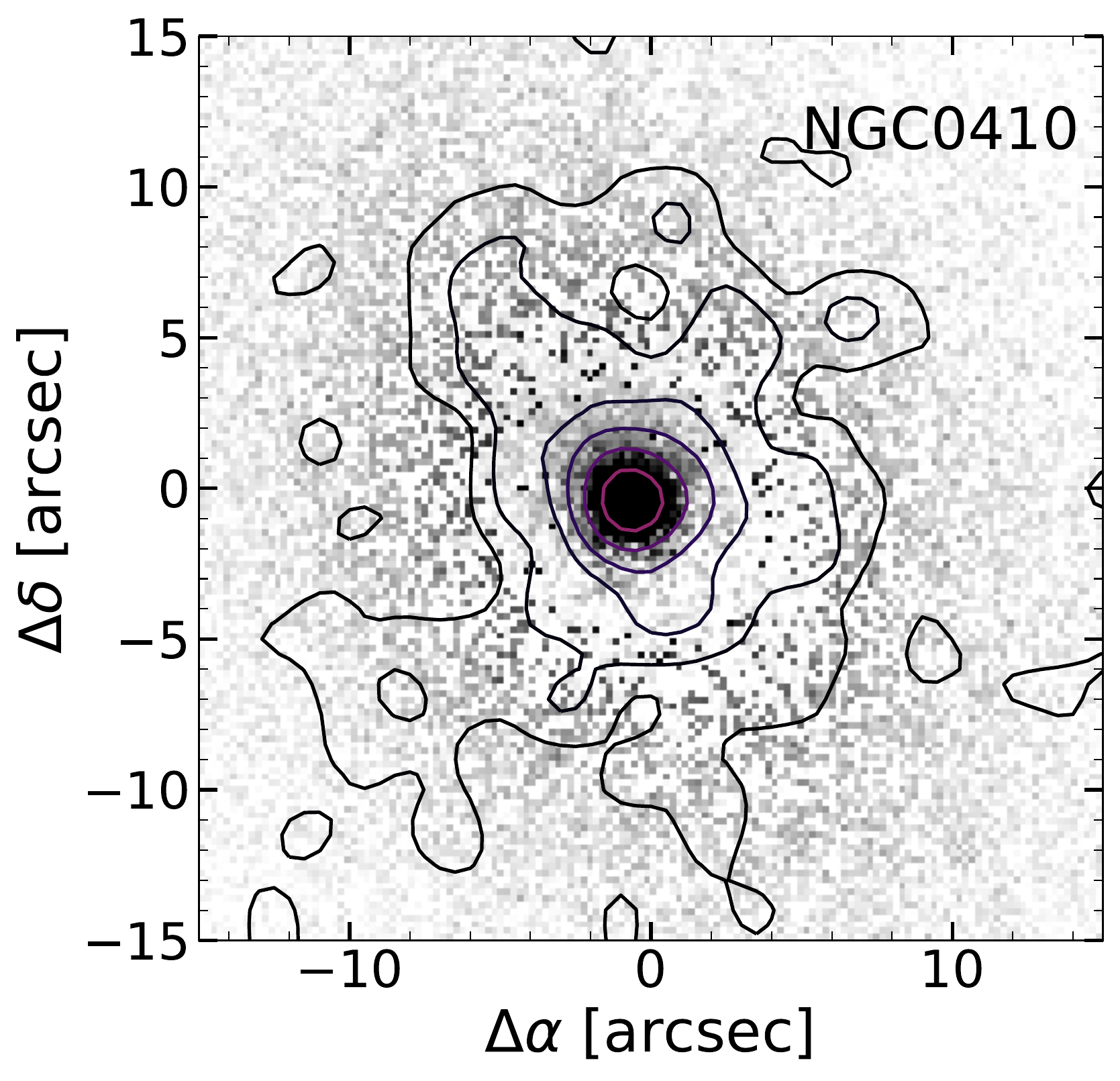}
   \includegraphics[width=0.37\columnwidth]{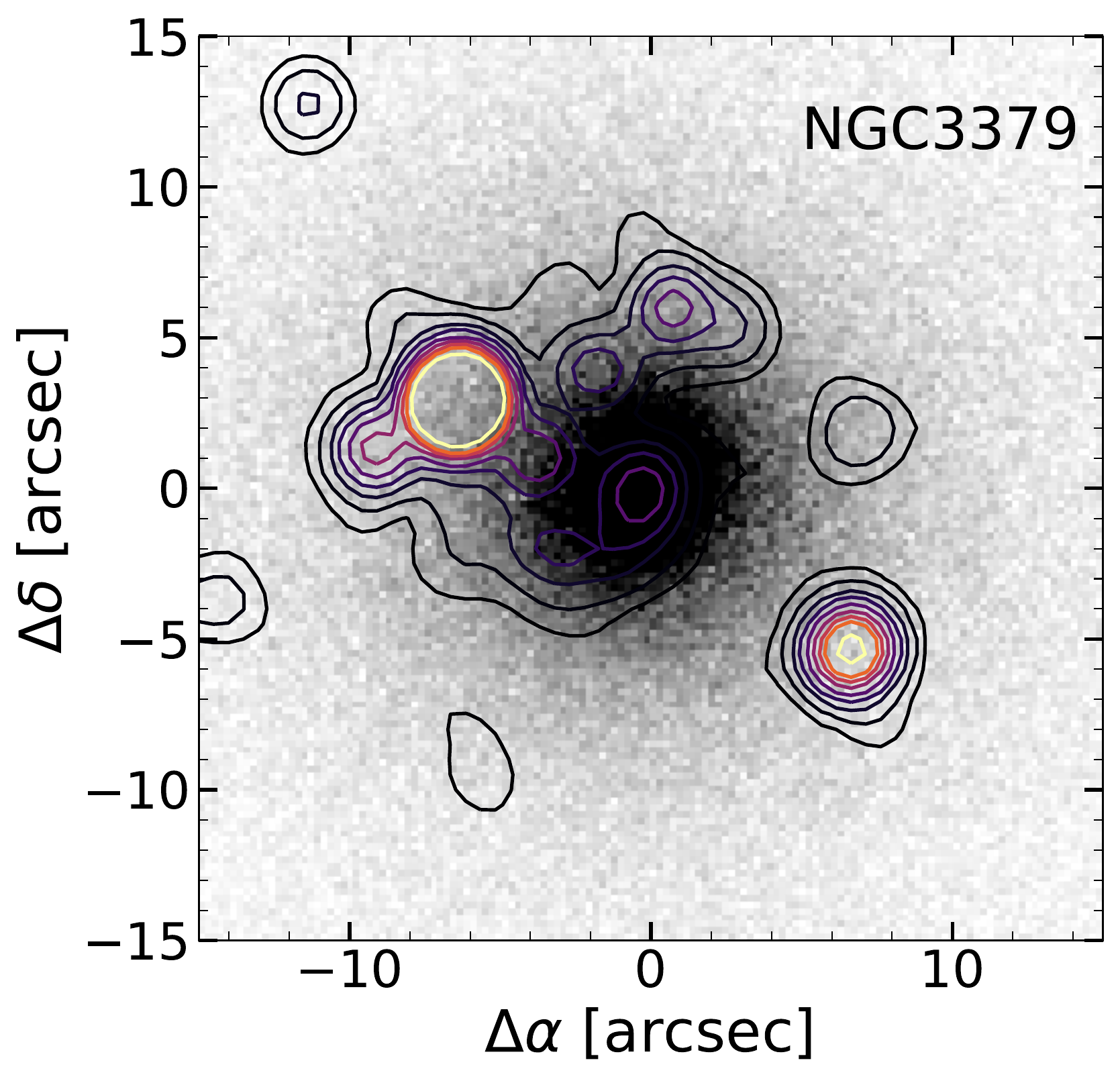}
   \includegraphics[width=0.38\columnwidth]{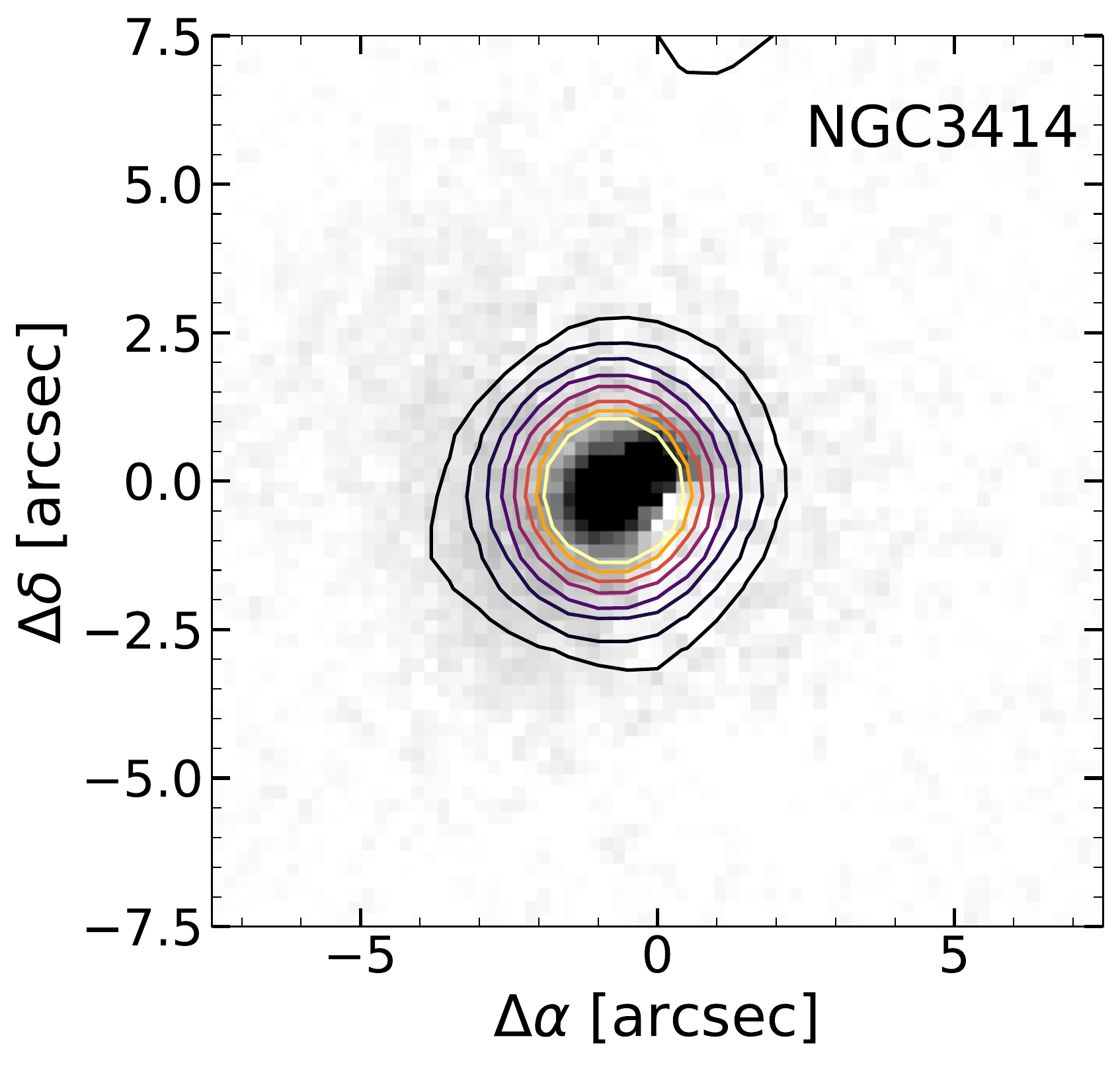}
   \includegraphics[width=0.4\columnwidth]{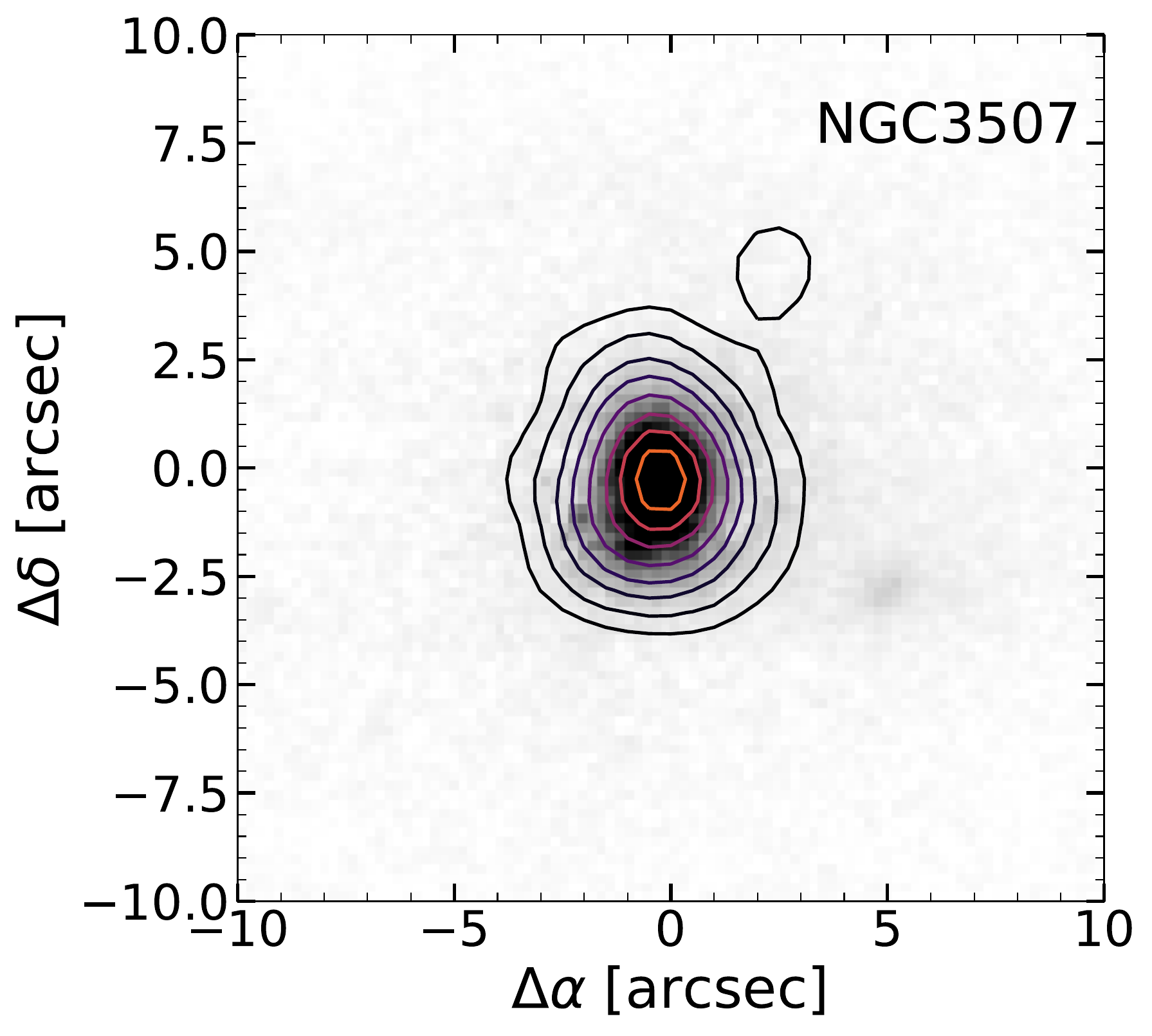}
   \includegraphics[width=0.4\columnwidth]{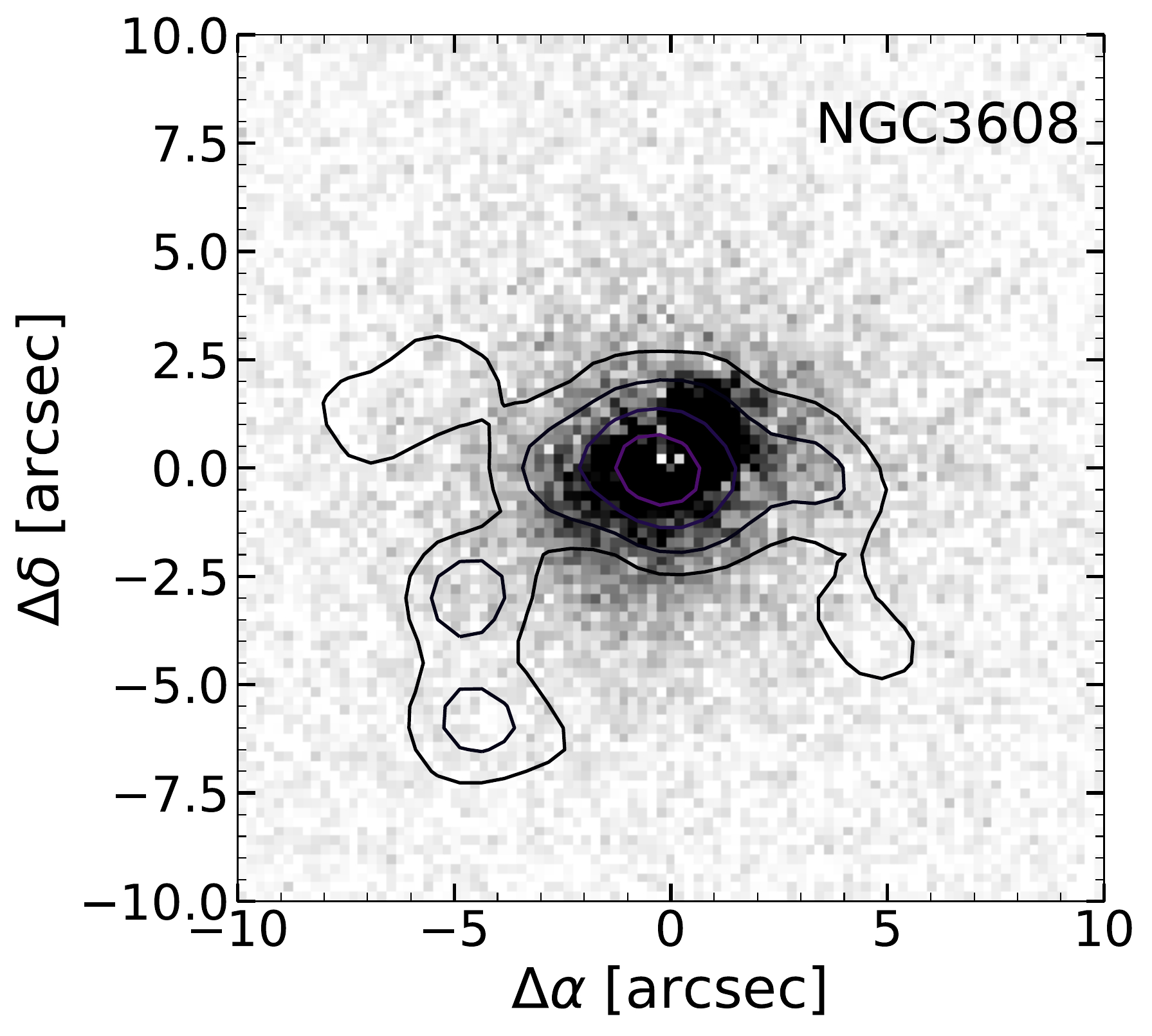}
   \includegraphics[width=0.37\columnwidth]{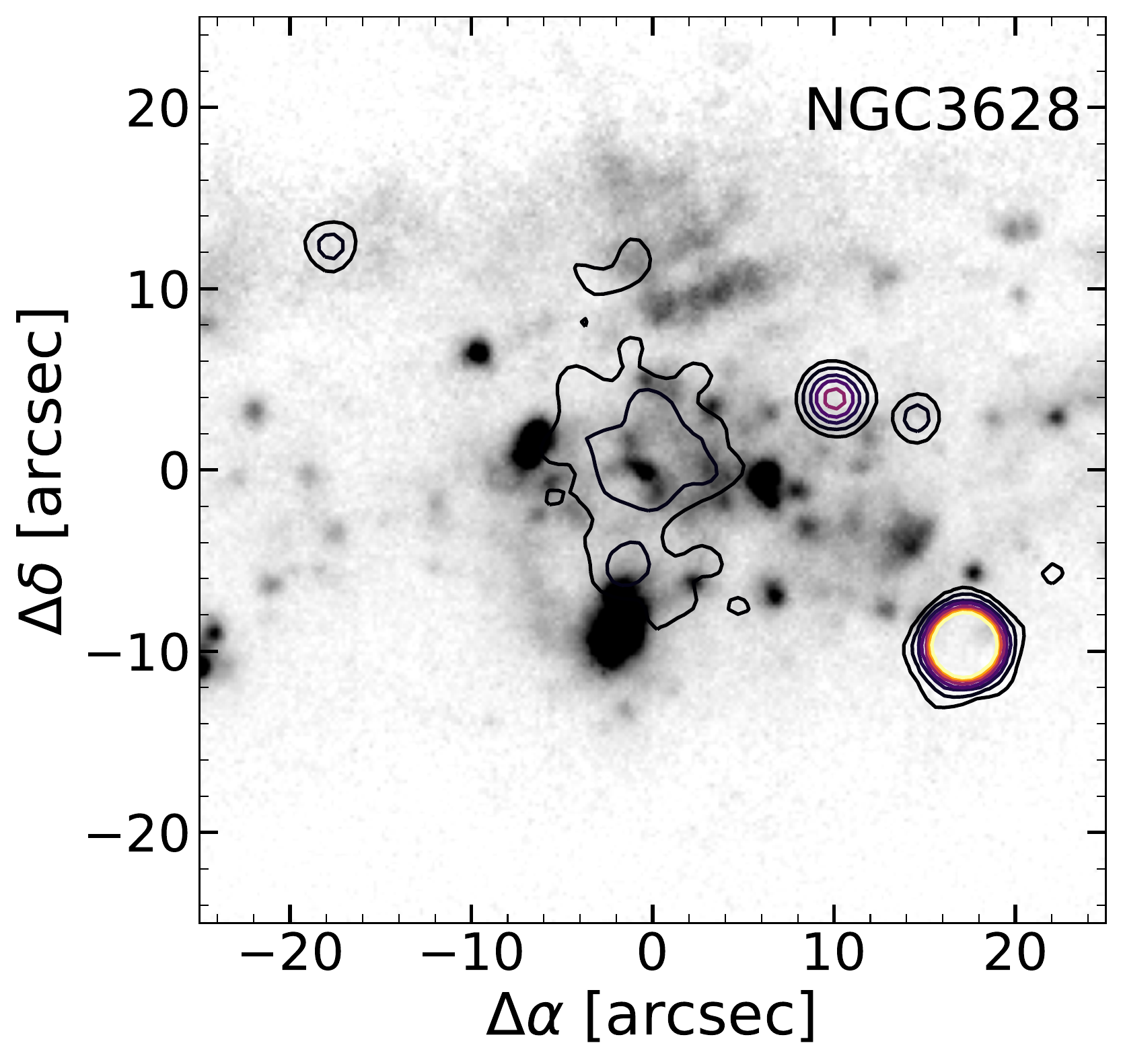}
   \includegraphics[width=0.4\columnwidth]{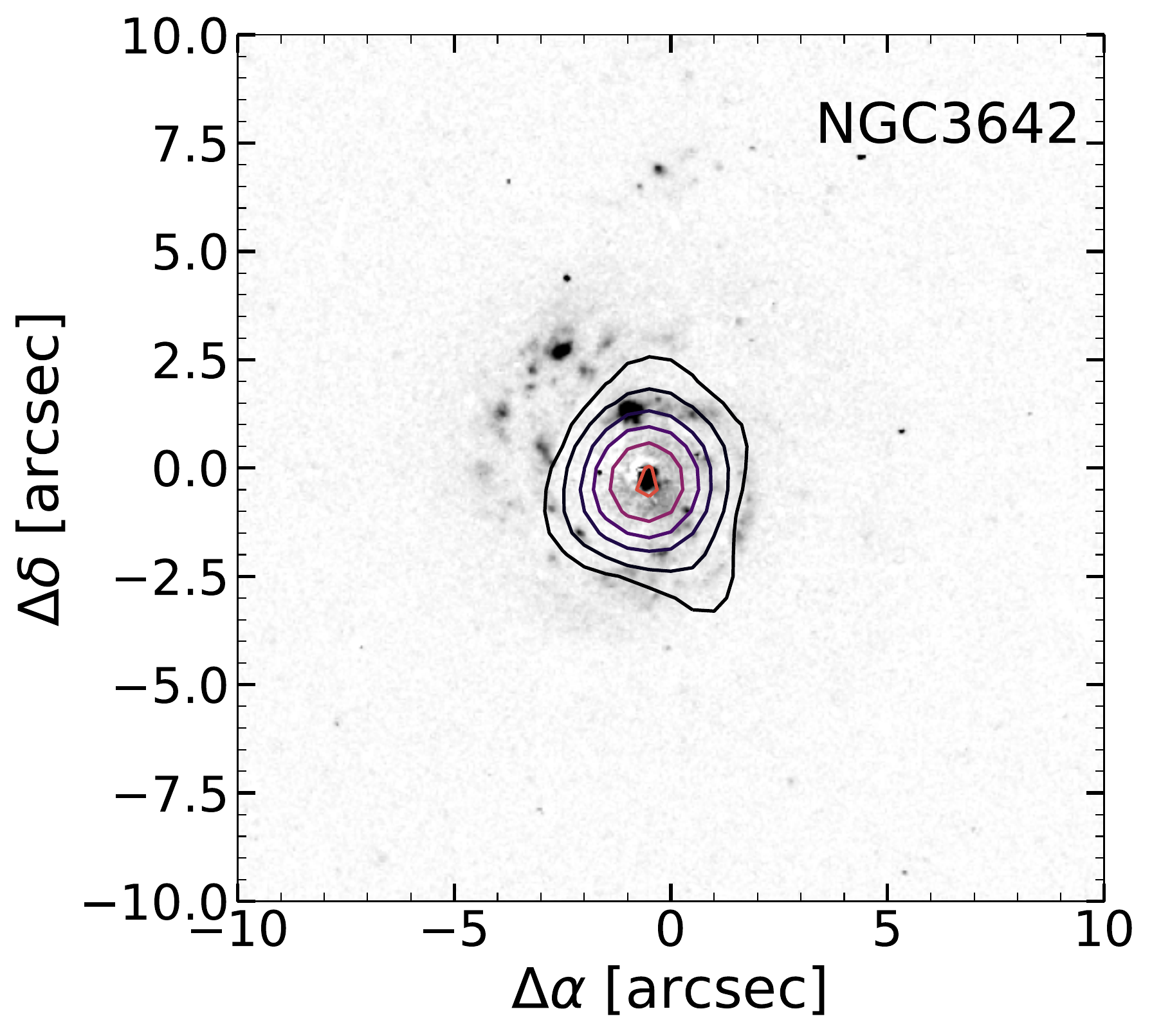}
   \includegraphics[width=0.4\columnwidth]{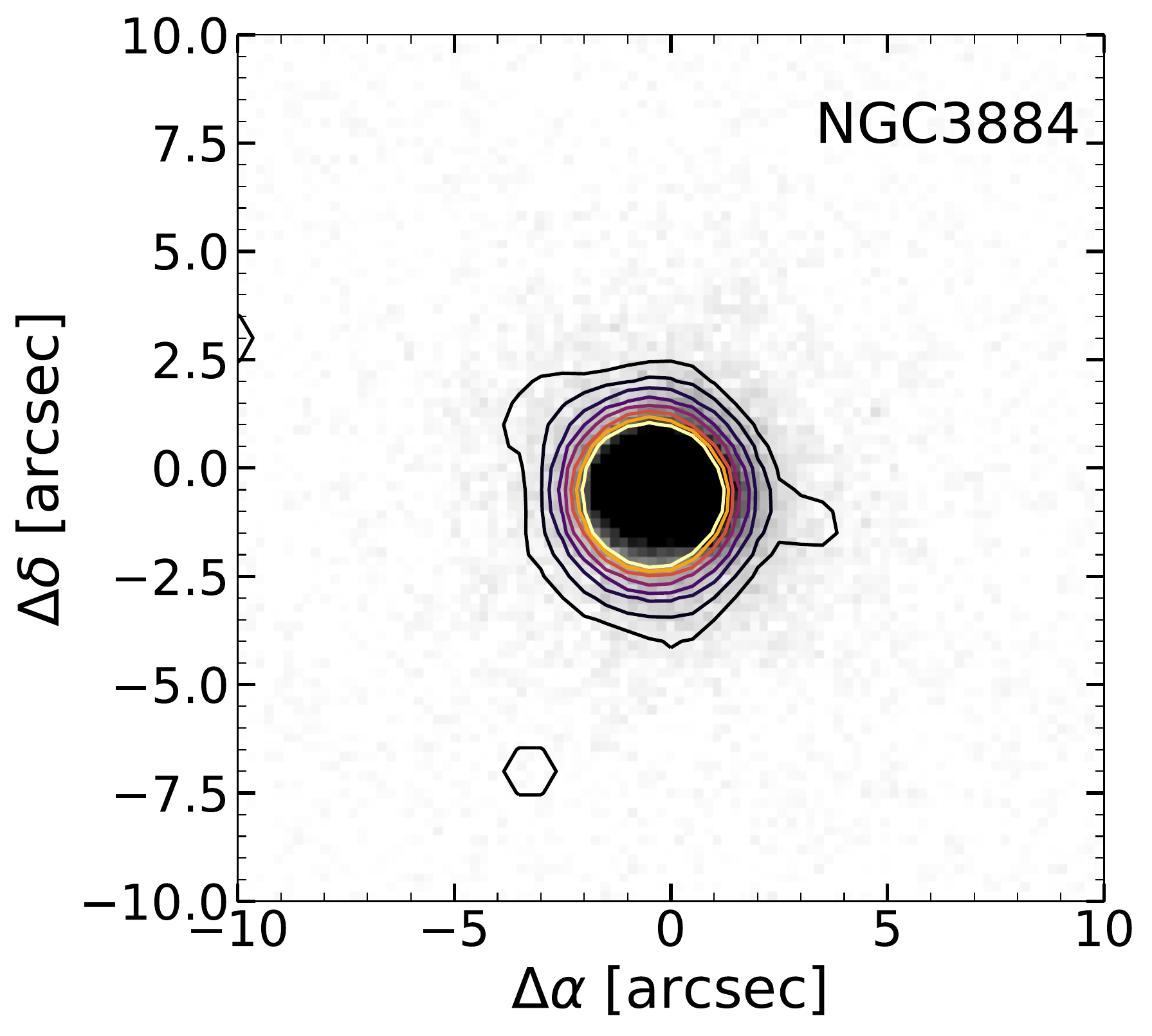}
   \includegraphics[width=0.37\columnwidth]{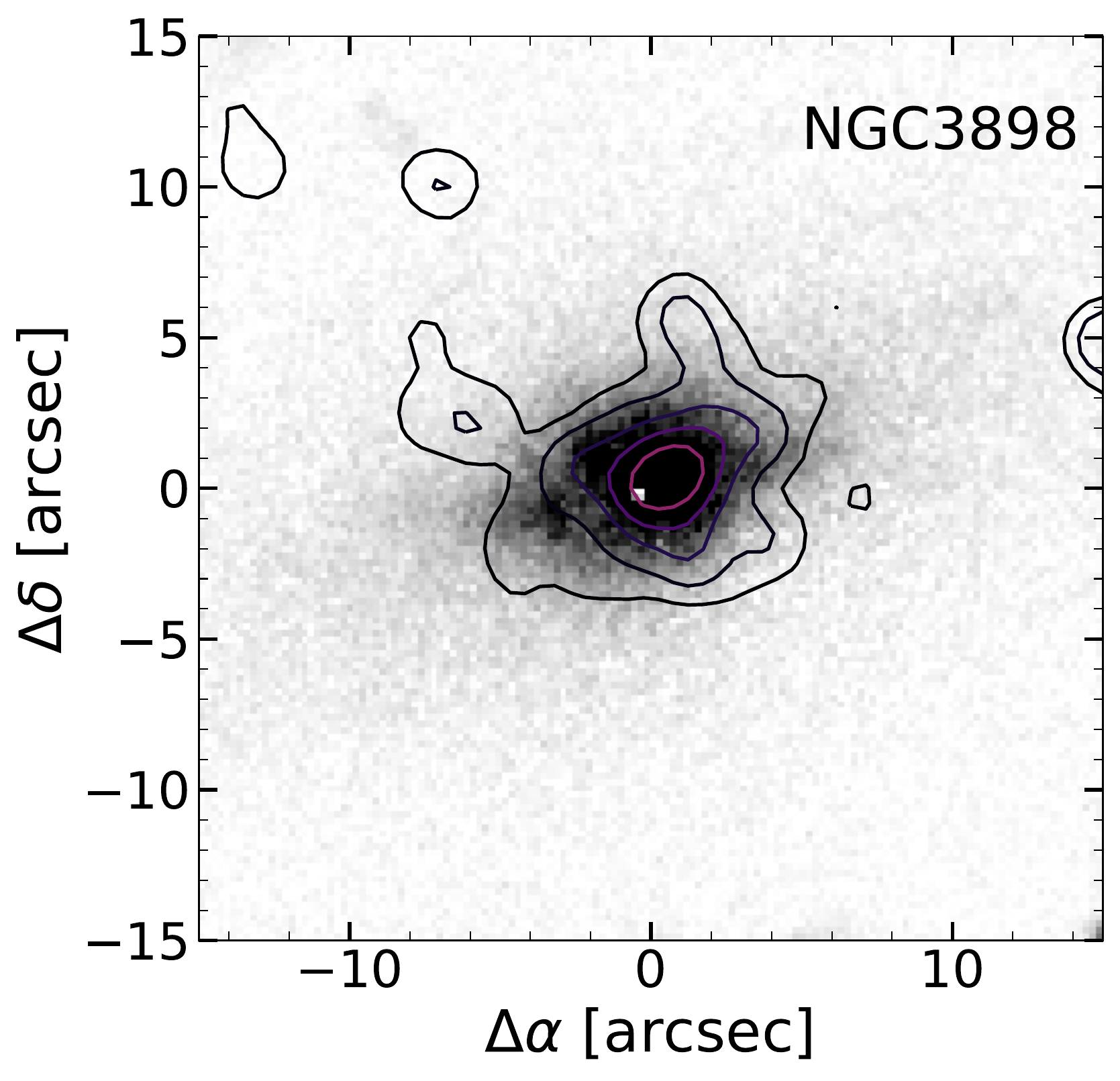}
   \includegraphics[width=0.39\columnwidth]{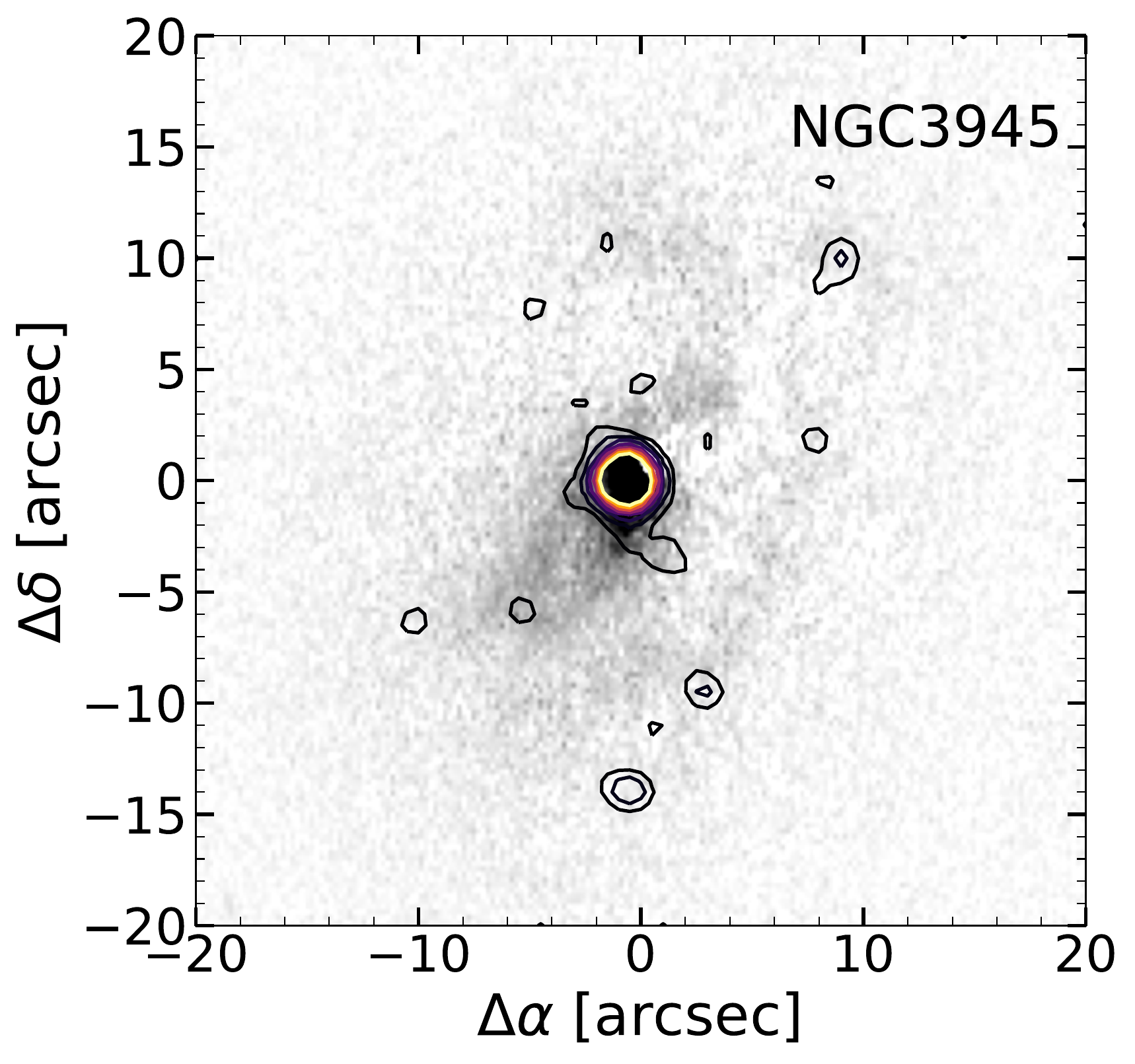}
   \includegraphics[width=0.4\columnwidth]{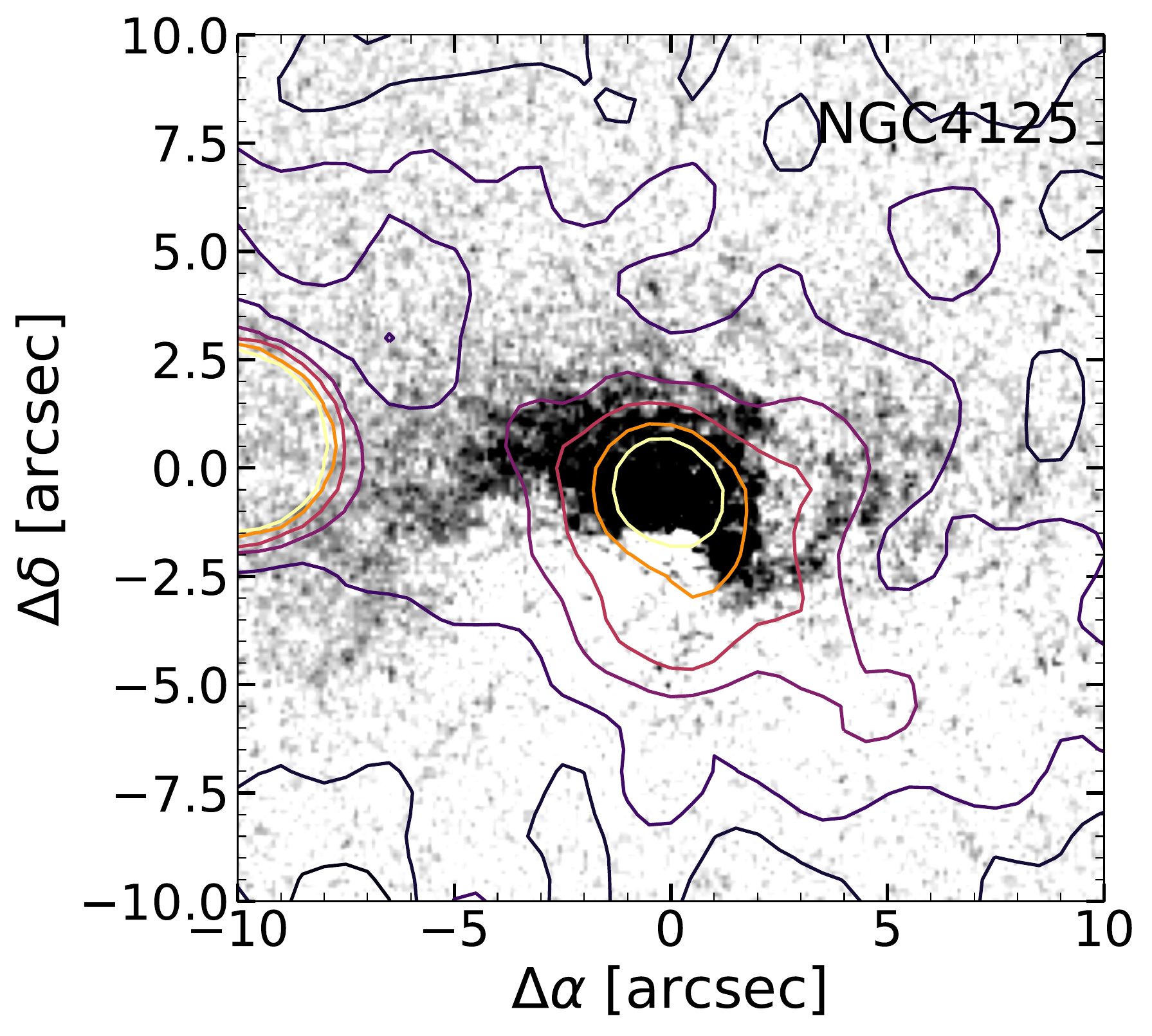}
   \includegraphics[width=0.4\columnwidth]{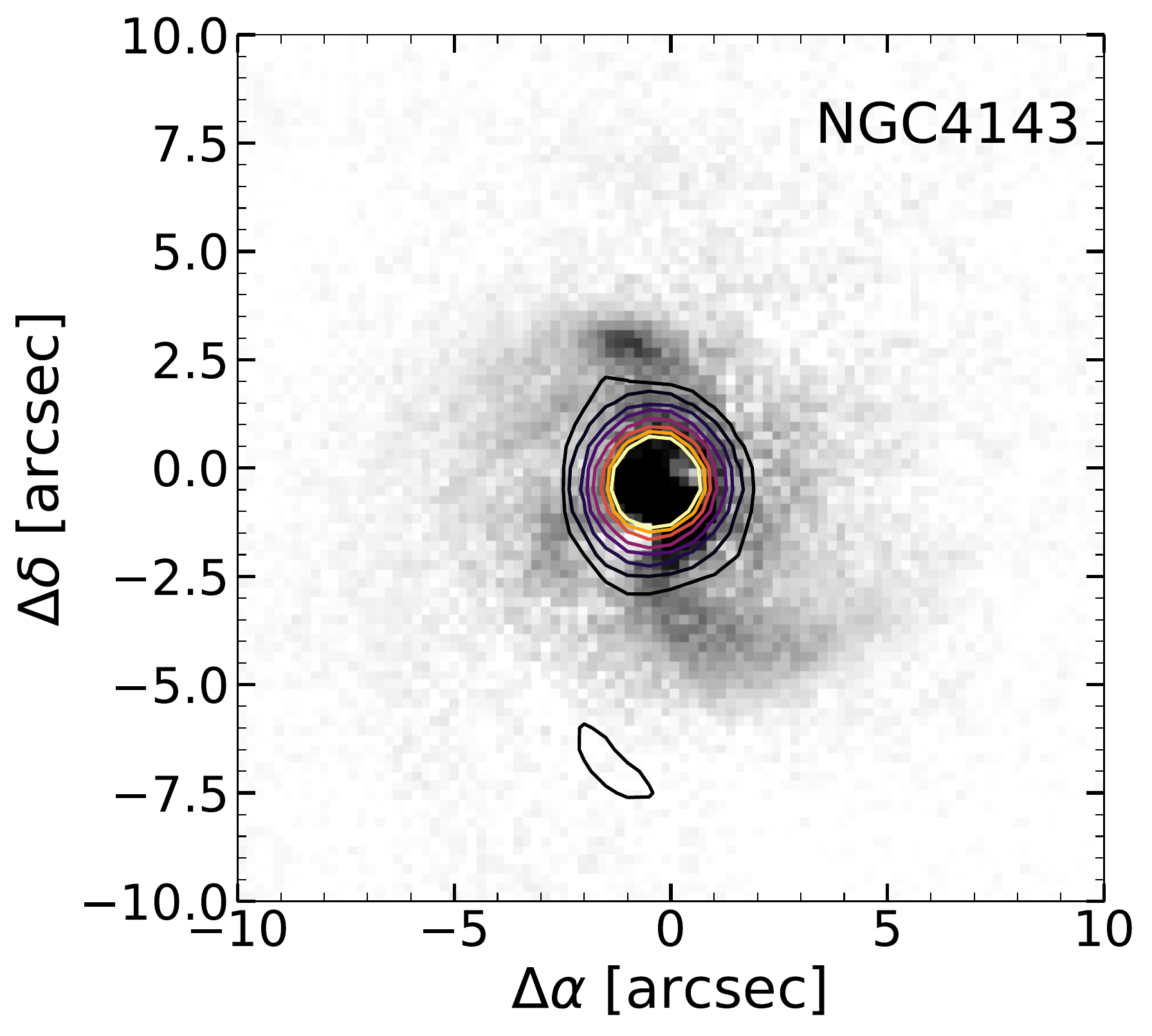}
   \includegraphics[width=0.38\columnwidth]{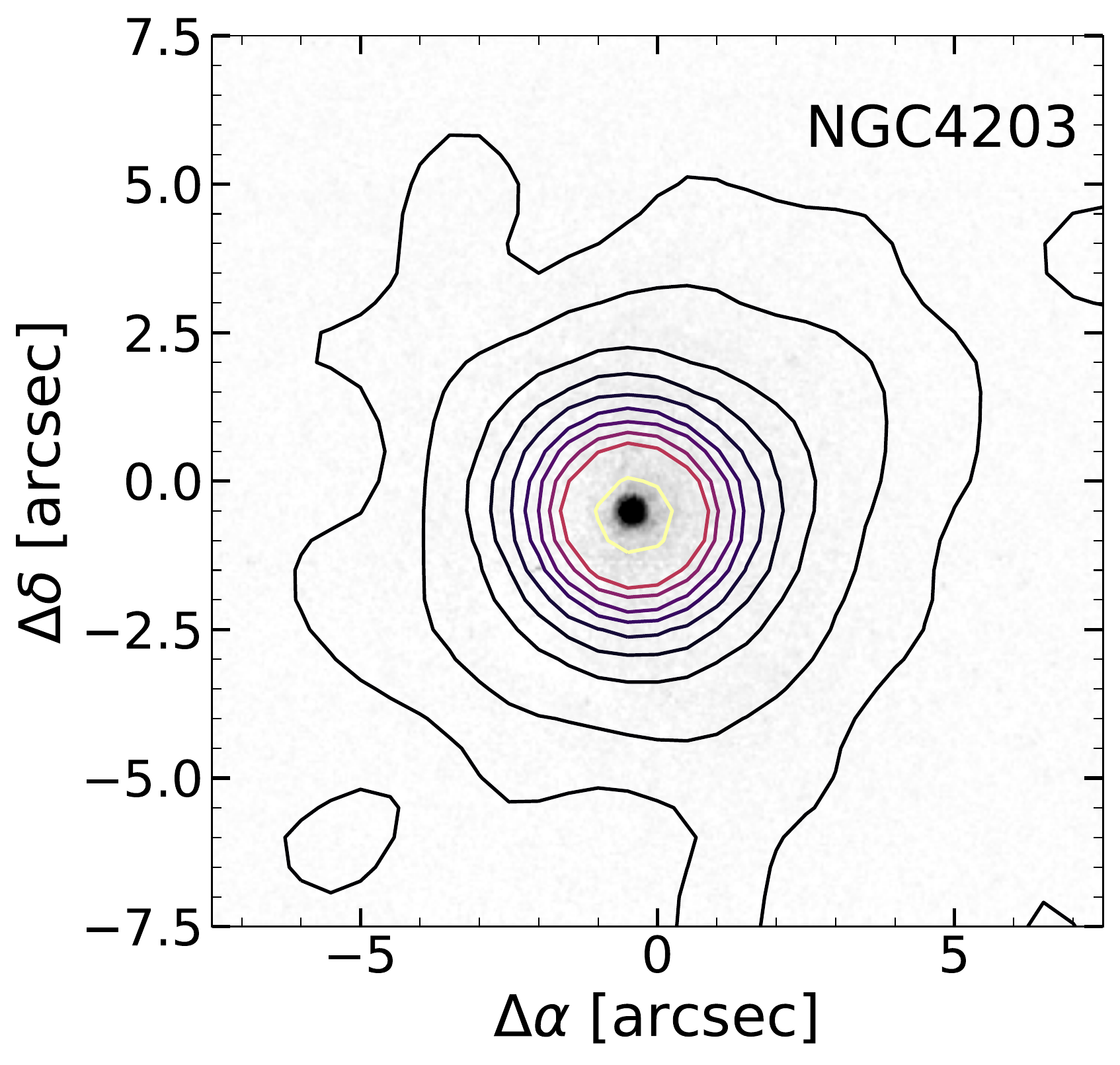}
   \includegraphics[width=0.37\columnwidth]{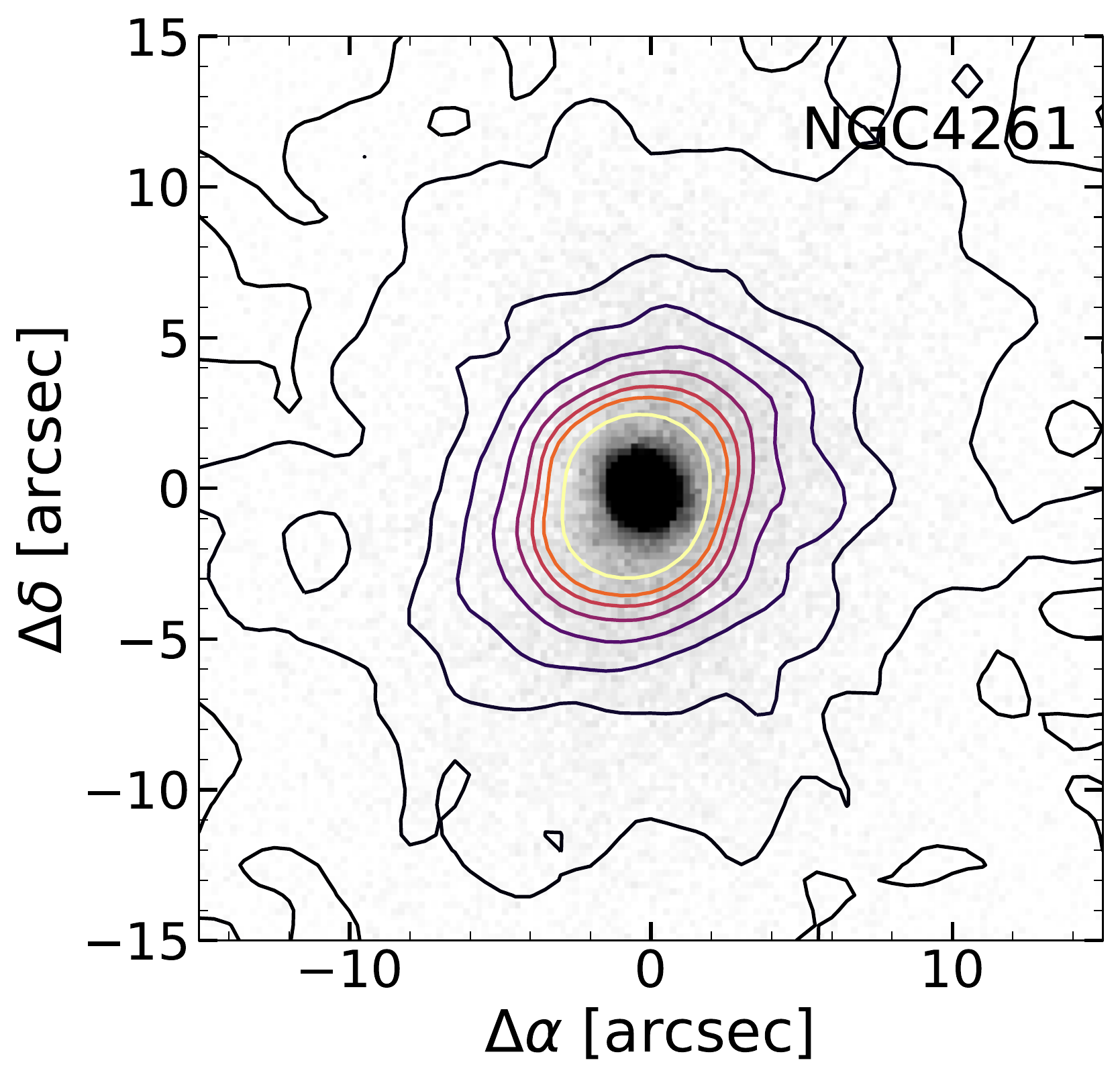}
   \includegraphics[width=0.4\columnwidth]{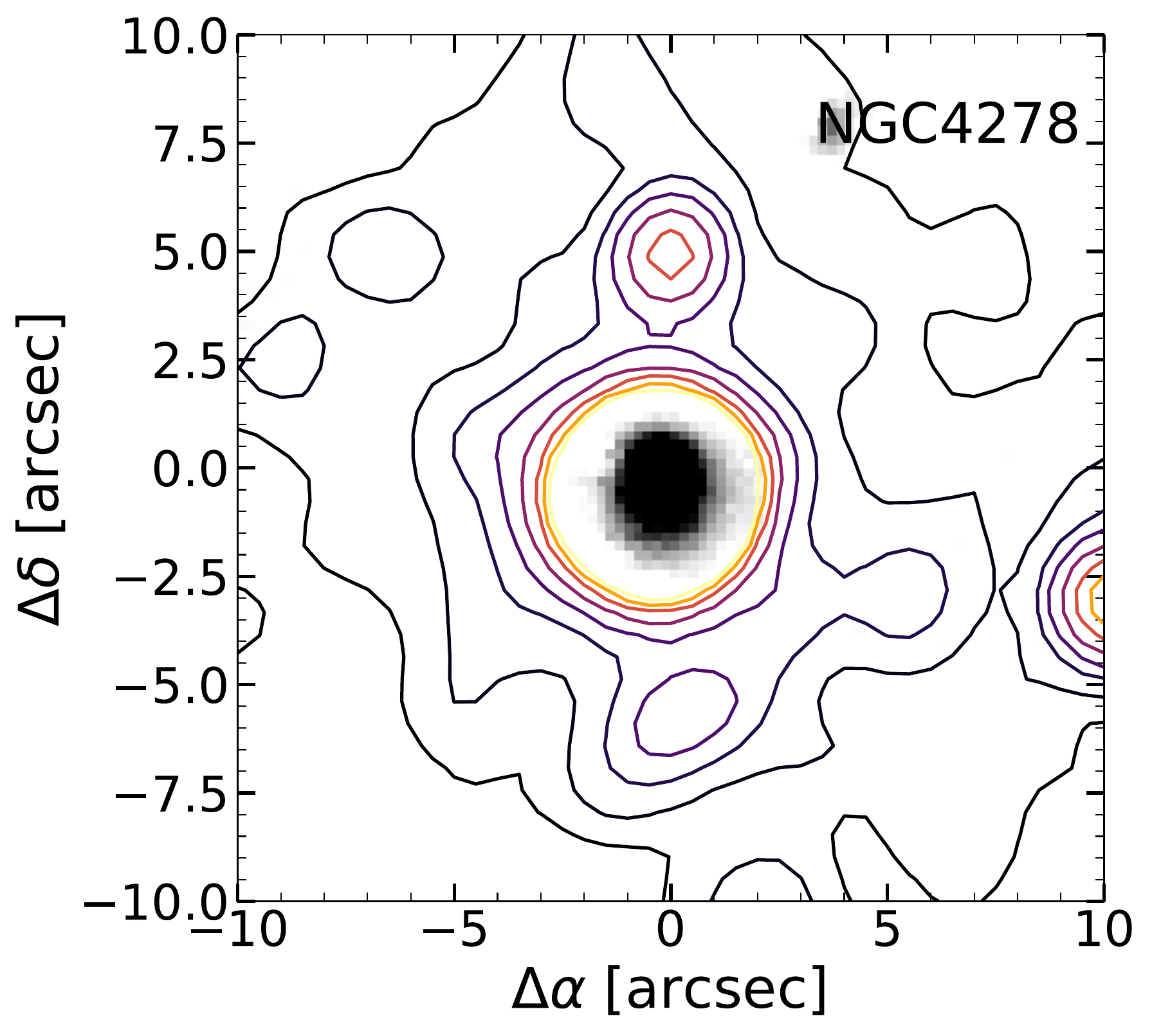}
   \includegraphics[width=0.39\columnwidth]{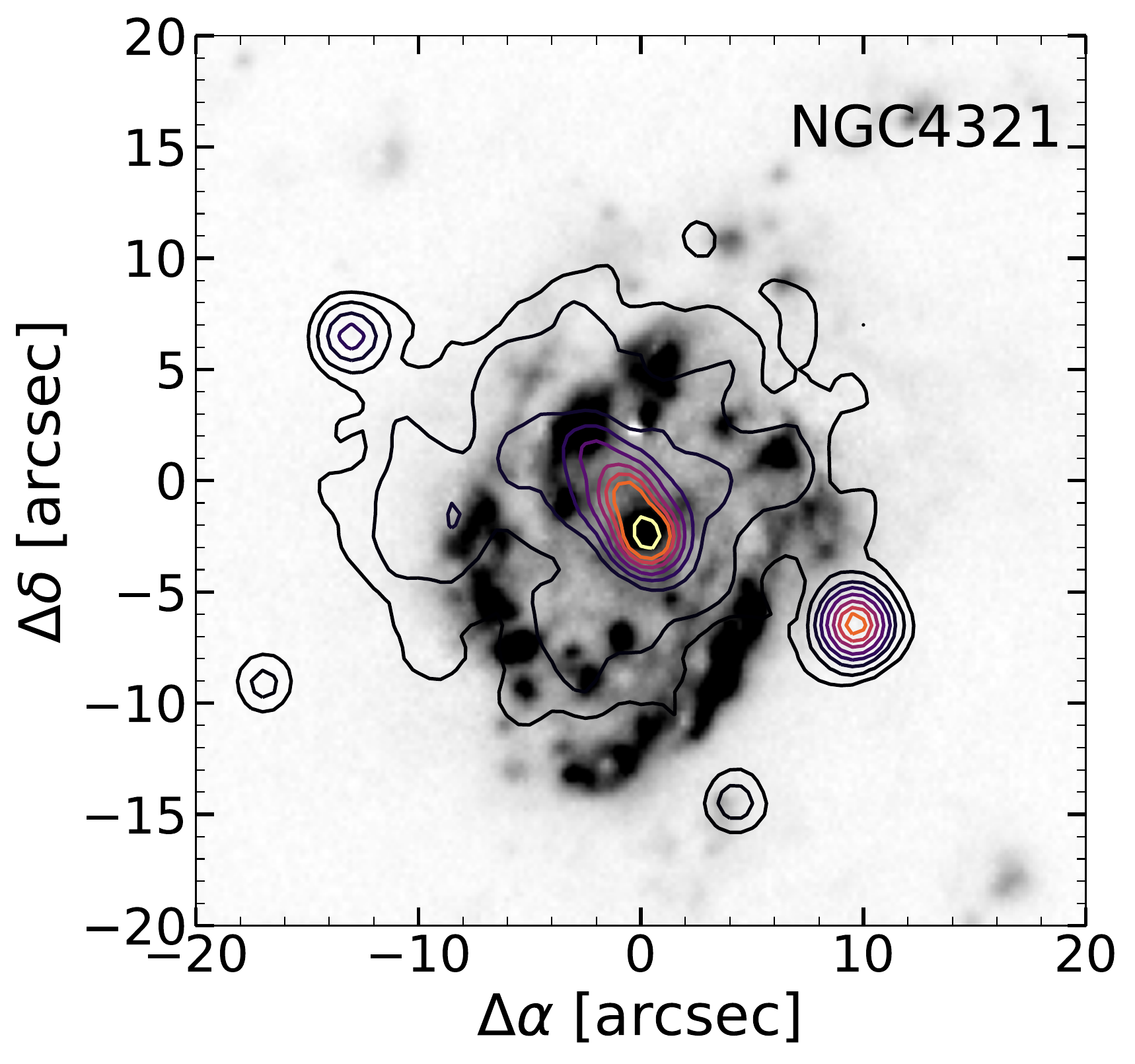}
   \includegraphics[width=0.38\columnwidth]{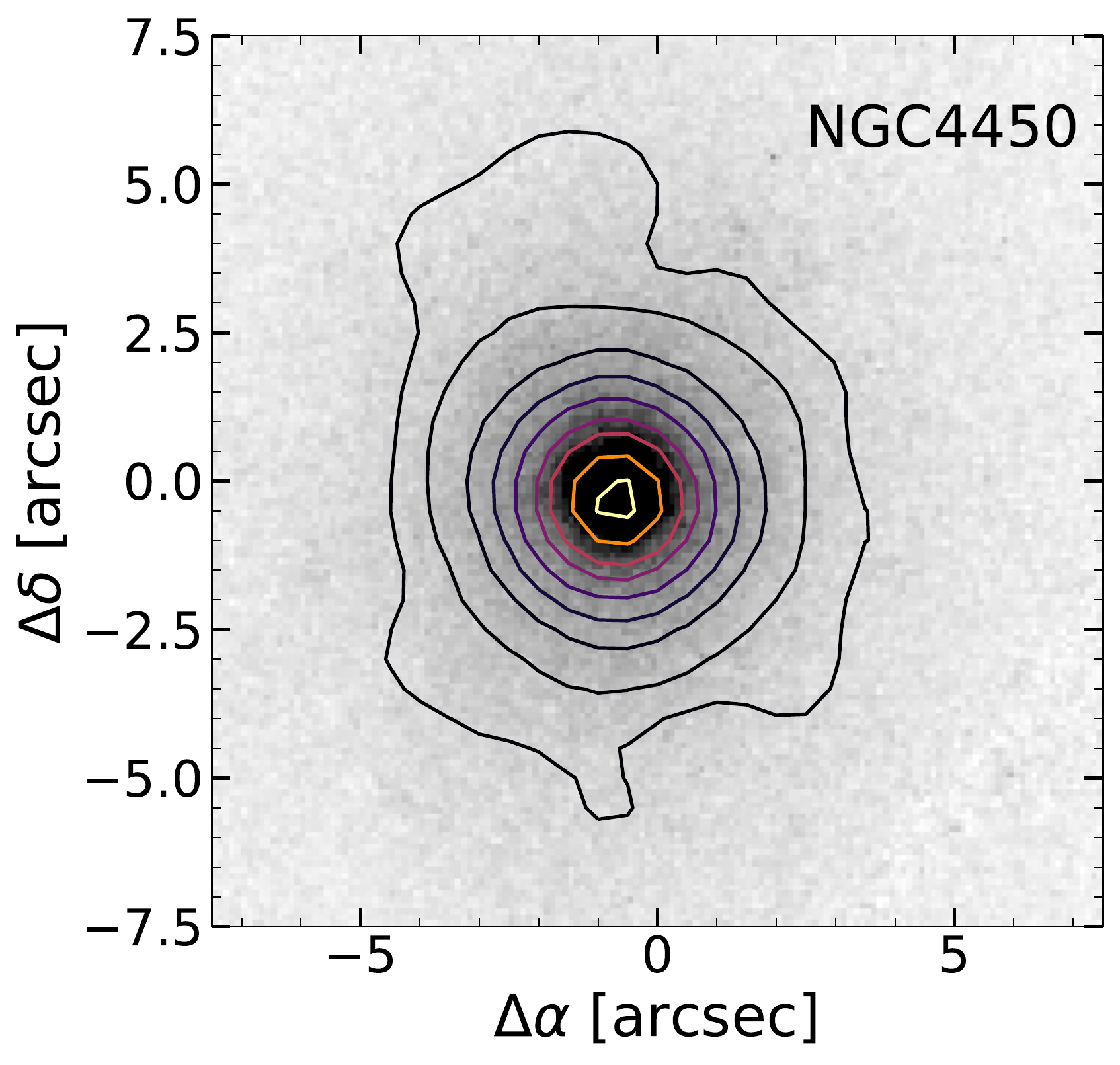}
   \includegraphics[width=0.37\columnwidth]{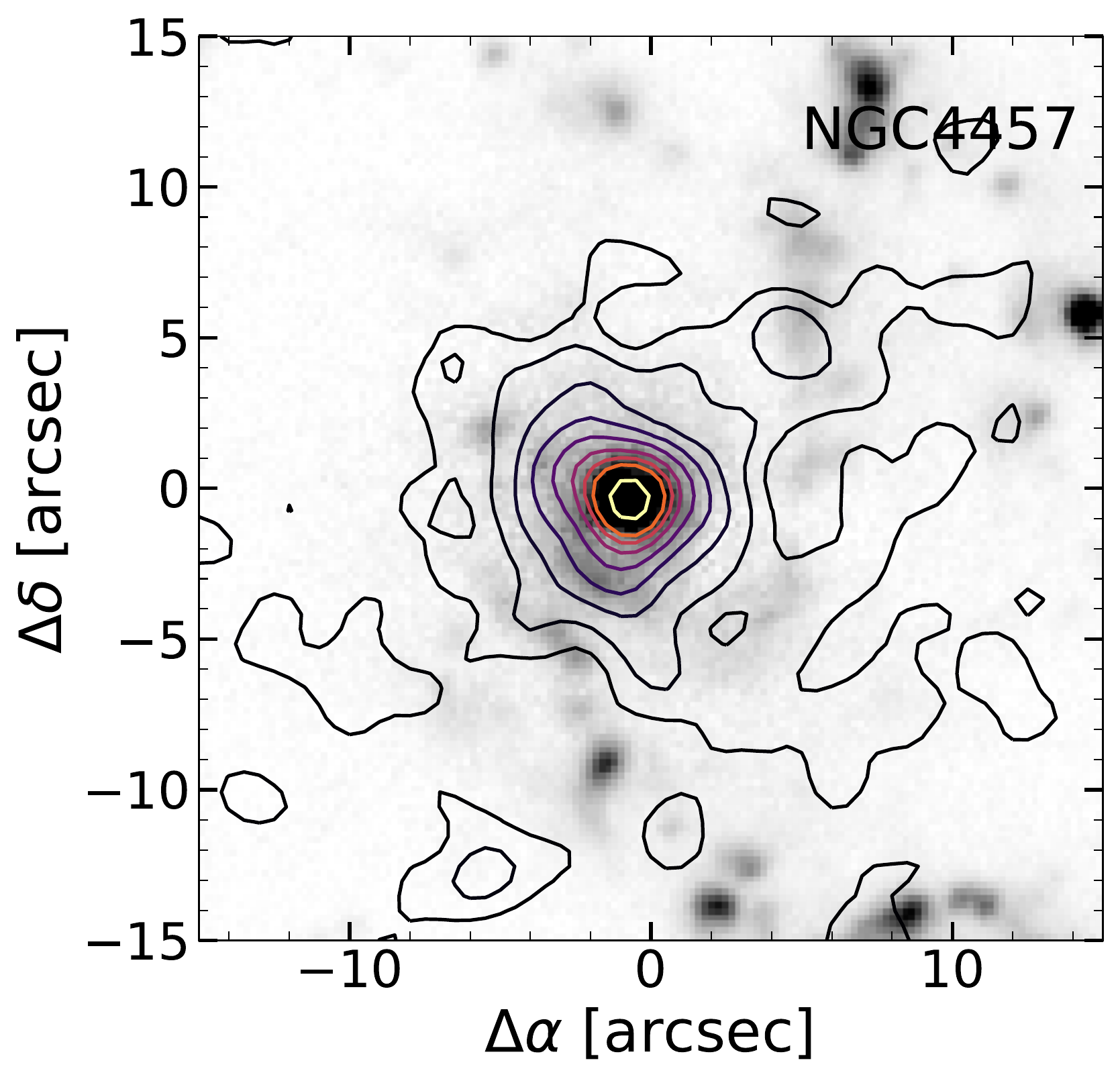}
   \includegraphics[width=0.4\columnwidth]{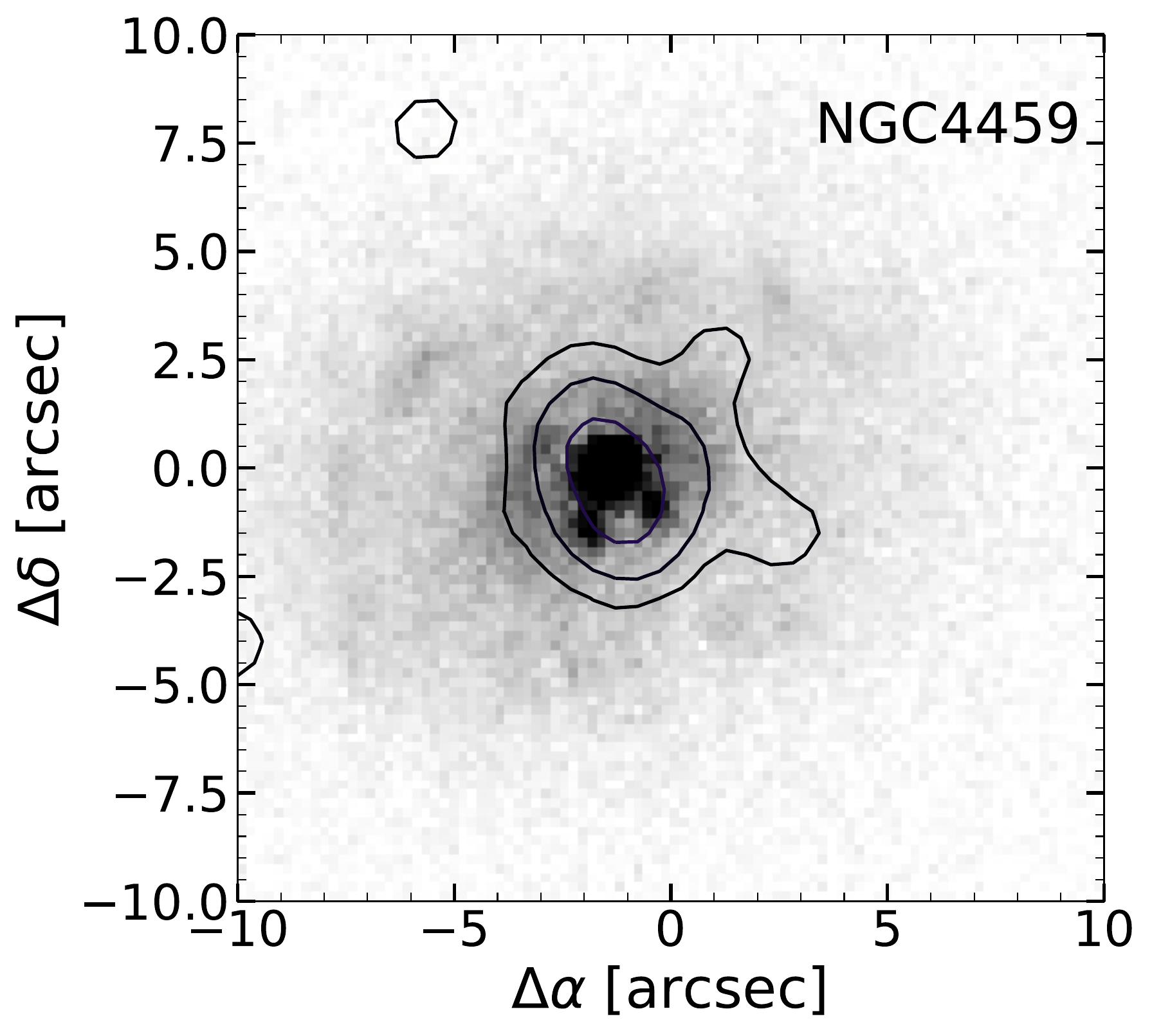}
   \includegraphics[width=0.4\columnwidth]{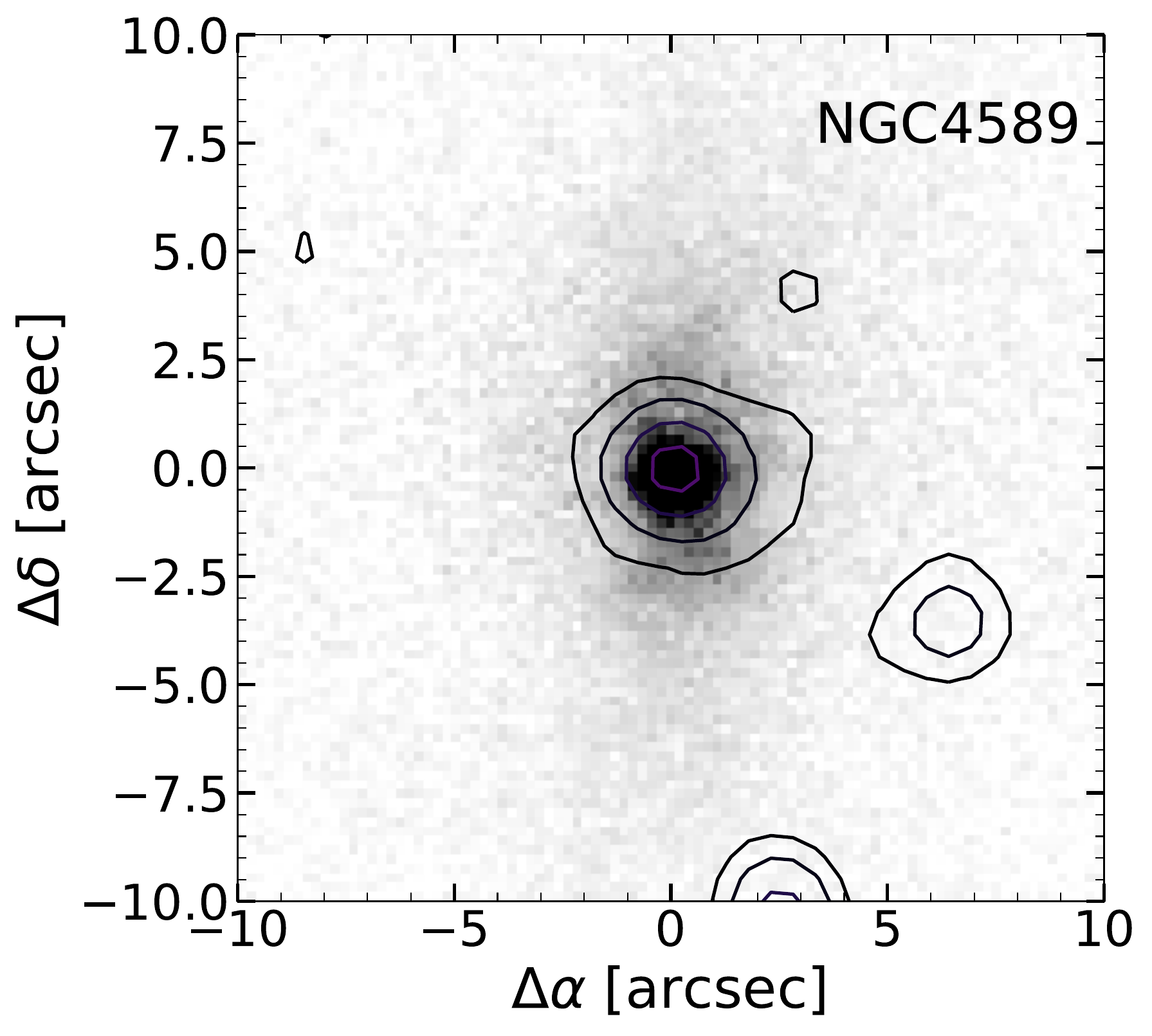}
   \includegraphics[width=0.4\columnwidth]{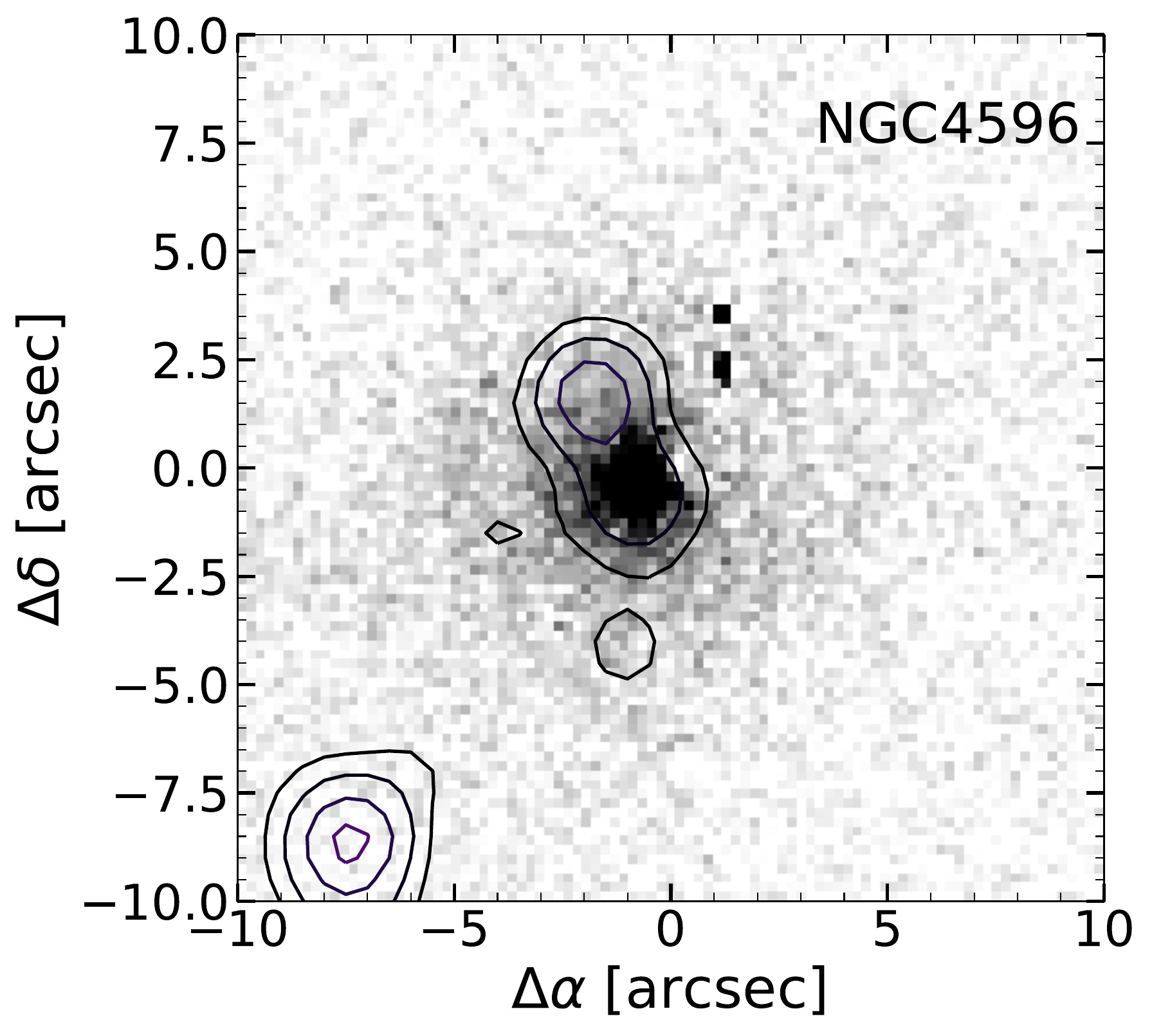}
   \includegraphics[width=0.4\columnwidth]{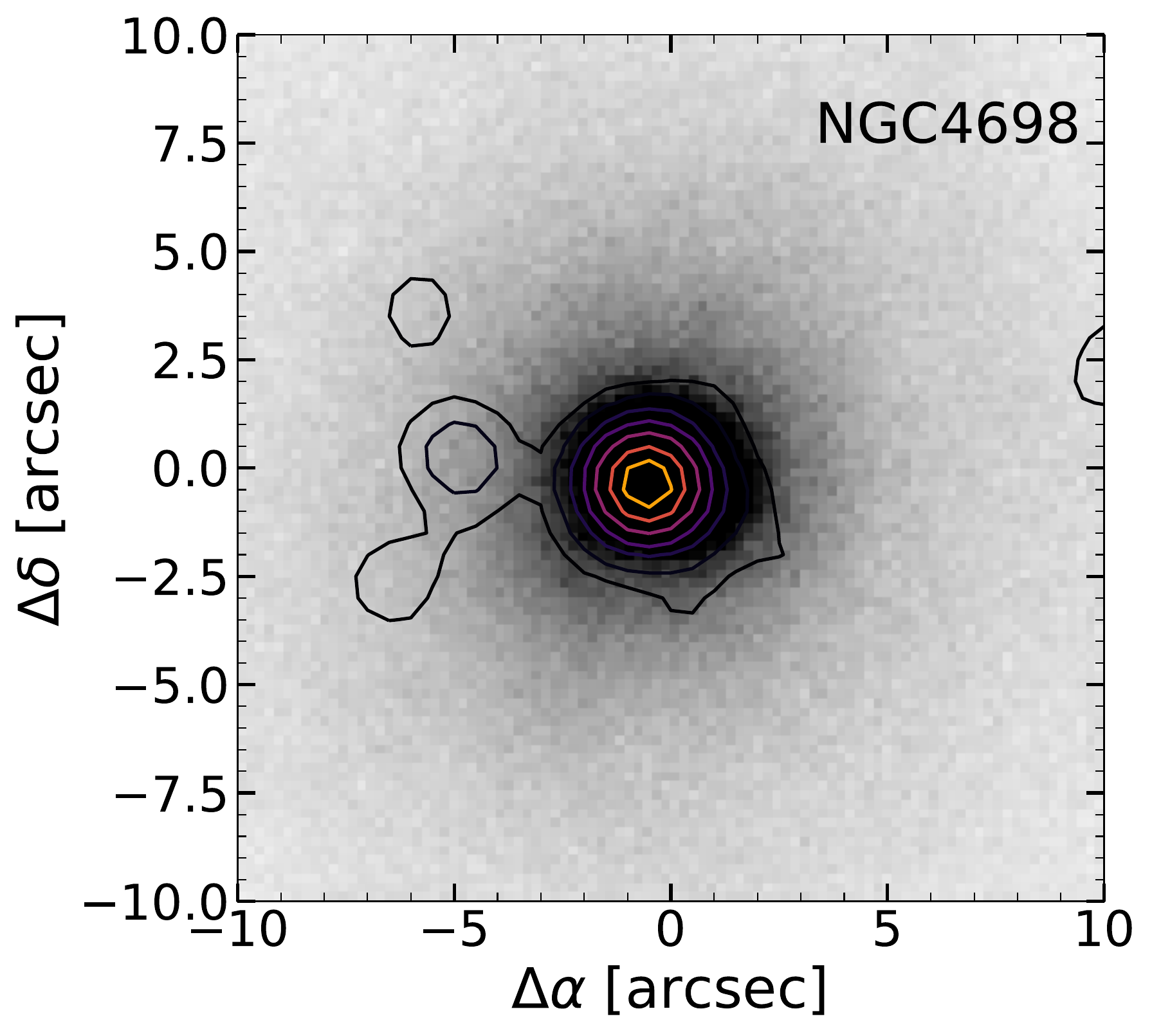}
   \includegraphics[width=0.38\columnwidth]{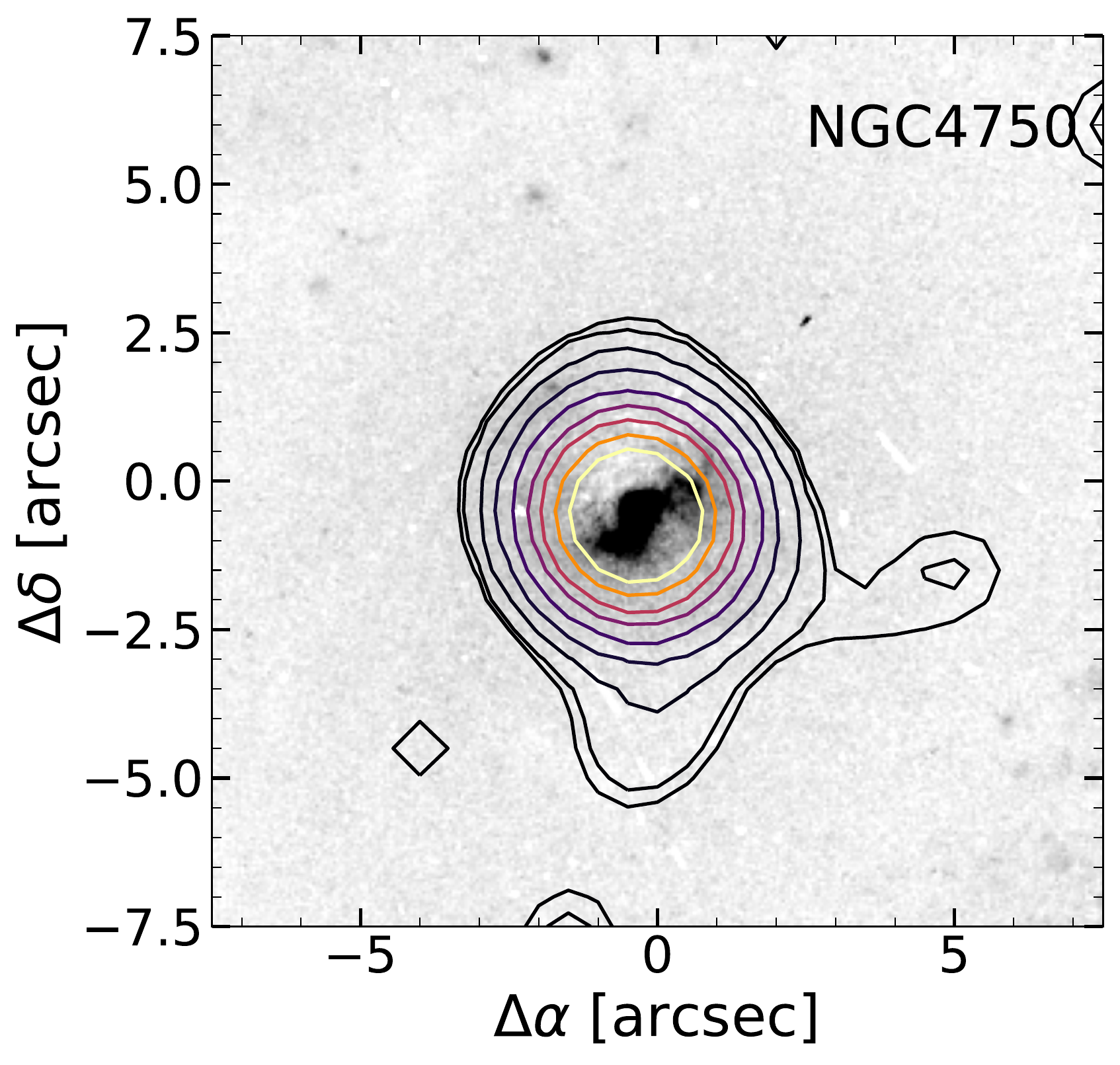}  
   \includegraphics[width=0.4\columnwidth]{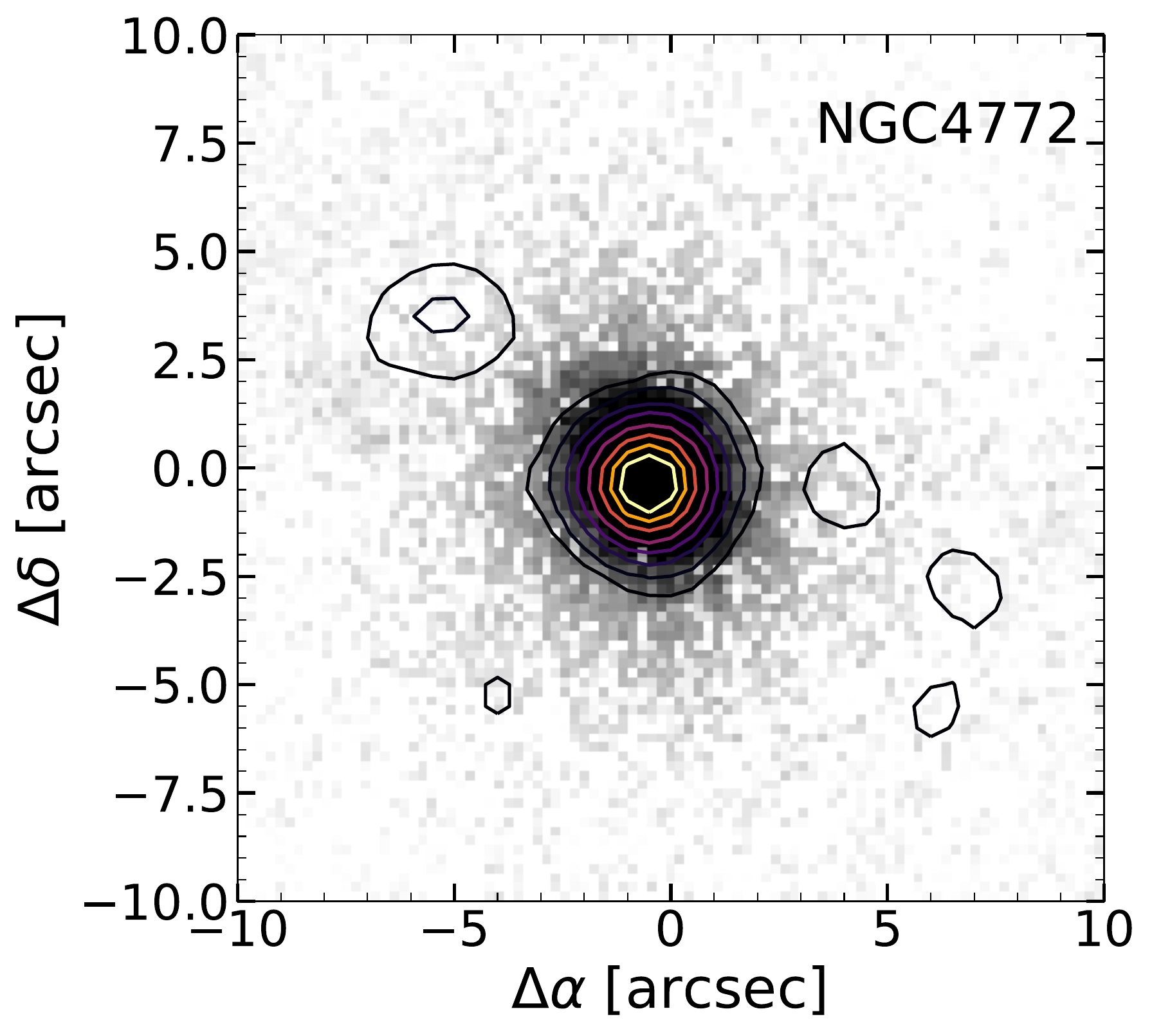}
   \includegraphics[width=0.39\columnwidth]{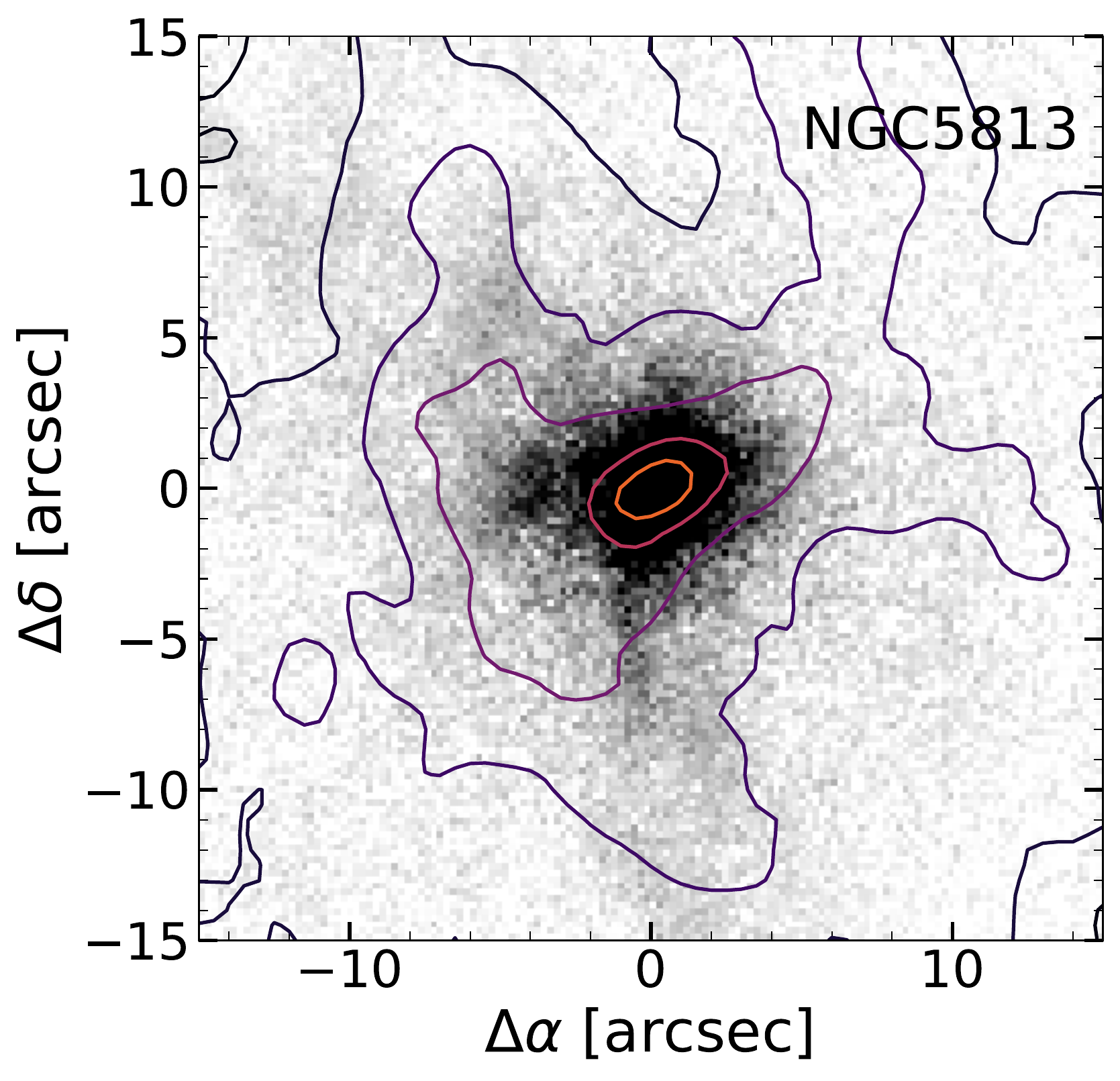}
   \includegraphics[width=0.42\columnwidth]{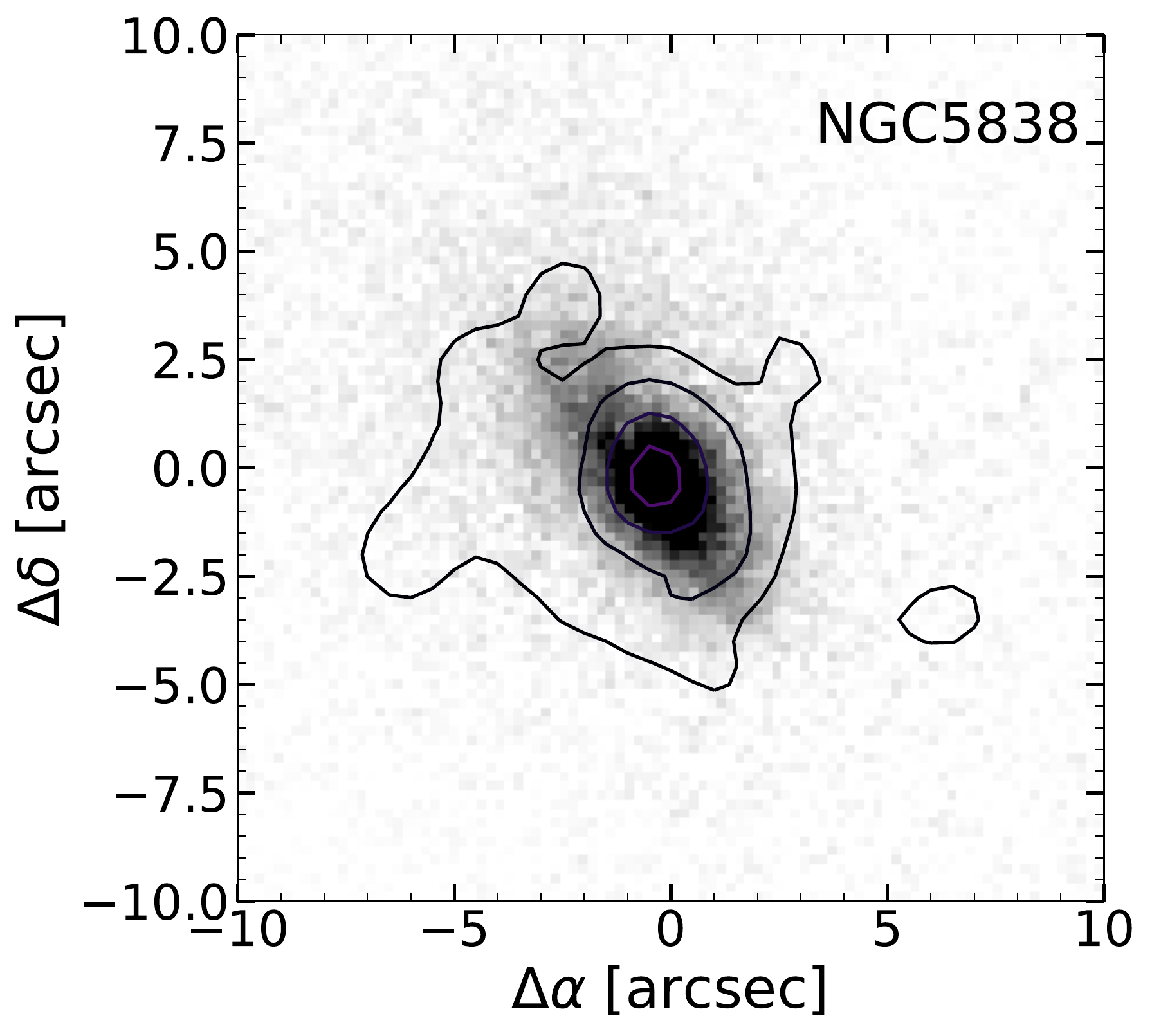}
   \includegraphics[width=0.38\columnwidth]{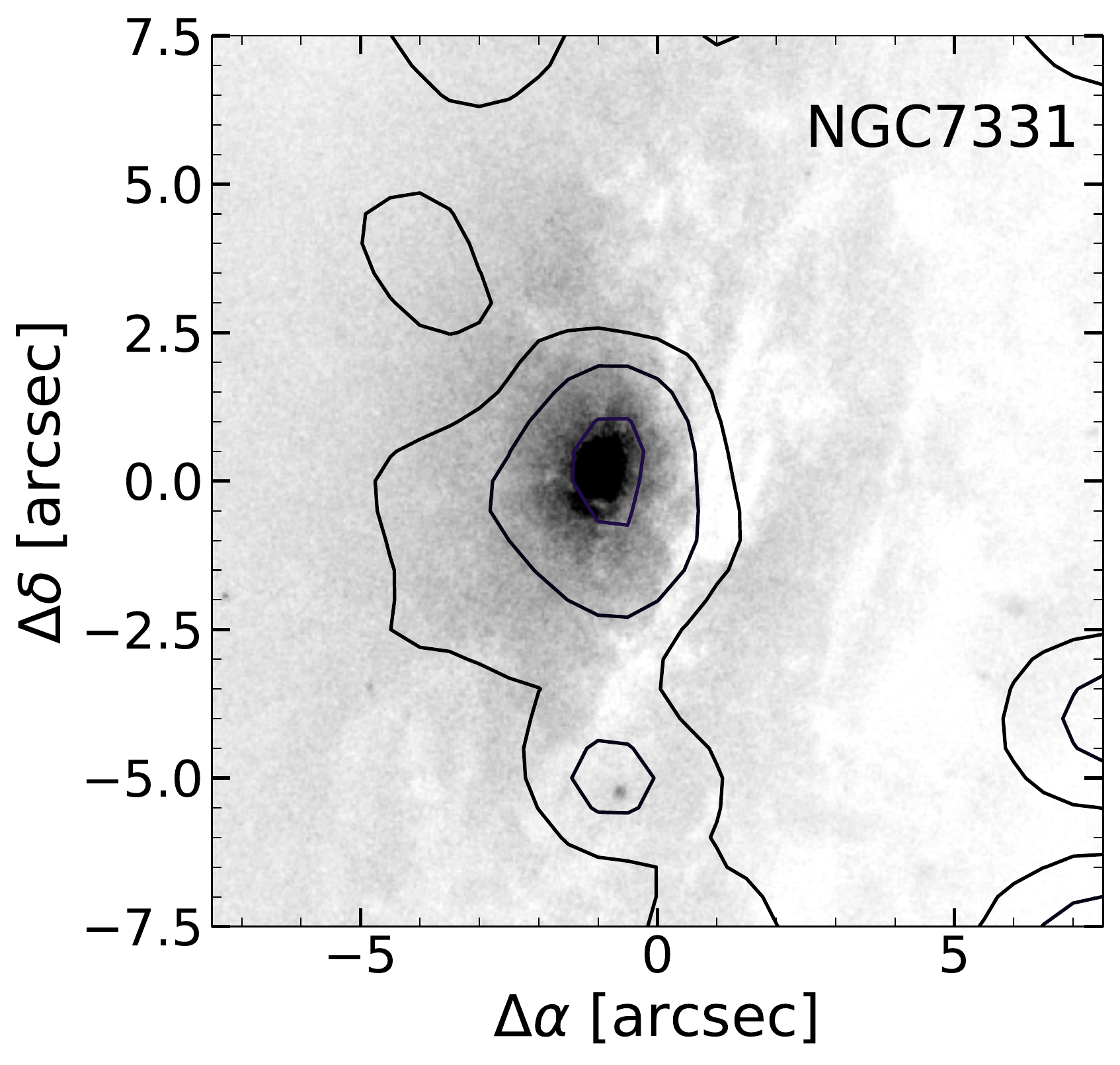}
   \caption{H$\alpha$ images with overlaid contours of the X-ray emission from Chandra. The contour levels are at 3$\sigma$ (black), 7$\sigma$ (black), 15$\sigma$ (black), 25$\sigma$ (dark-purple), 40$\sigma$ (purple), 60$\sigma$ (light-purple), 80$\sigma$ (red), 100$\sigma$ (orange) and 150$\sigma$ (yellow).} 
   \label{Figure_Xrays}
   \end{figure*}

   \noindent Some previous works have studied the correlation between the soft X-ray and ionised gas emissions for different AGN types \citep[][; M11]{Bianchi2006}, such that both emissions are believed to raise from the photoionisation of the gas around the AGN \citep{Bianchi2006,Bianchi2019}. These results are discussed in Sect.~\ref{SubSec_DiscXray}. For the new data, we searched for X-ray data with resolution comparable to that of the ALFOSC/NOT 38 H$\alpha$ images. The soft X-ray contours are shown in Fig.~\ref{Figure_Xrays}. The emission for 11 galaxies is mainly point-like around the nucleus, which would be equivalent to a \lq Core-halo\rq\, classification (see Sect.~\ref{Section_results}). We find 12 objects with an extended emission and 5 with clumpy emission. Despite the different morphologies, the soft X-ray emission is co-spatial with the ionised gas for 12 galaxies (see Table~\ref{Table_XrayMorphKin}). \\ 
   
   \noindent If we separate those galaxies in which both emissions are coincident, at least, in the direction of the emission, we end up with 12 LINERs (NGC\,0266, NGC\,0410, NGC\,3414, NGC\,3884, NGC\,3945, NGC\,4261, NGC\,4450, NGC\,4596, NGC\,4750, NGC\,4772, NGC\,5813, NGC\,5838). One example is NGC\,5813, for which the H$\alpha$ image, specially at large scales, shows an extended filamentary structure (see Fig.~\ref{Figure_largescale}), with a bubble extending up to 10\arcsec\ in the north-west direction with respect to the centre. The X-ray emission is rather complex, with the same bubble almost co-spatial with the H$\alpha$ emission (see Fig.~\ref{Figure_Xrays}). 
   
   \noindent The most notable difference in the remaining objects lies in the extension of the emission, being in some cases much more extended in the X-rays (e.g. NGC\,4278, see Fig.~\ref{Figure_AB5}), and in other cases in H$\alpha$ (e.g. NGC\,4698, see Fig.~\ref{Figure_AB7}). 
   
    \subsection{Global picture of ionised gas morphologies in LINERs}
    \label{SubSec_globalpic}

    \noindent For the new imaging data obtained with NOT and HST, as drawn by the detected H$\alpha$ emission at 3$\sigma$ level, the nuclear ionised gas in these LINERs is extended ($\geq$5\arcsec) in the majority of the cases (32 objects; $\sim$84\%). This emission is not exclusively along spiral arms or nuclear discs but also due to possible outflows (for example, as filamentary structures), or perhaps other events at large scales (i.e. $>$1\arcmin; see Fig.~\ref{Figure_largescale}), as mergers (e.g. NGC\,4125, see Appendix~\ref{Appendix_A} and Fig.~\ref{Figure_AB4}).\\

    \noindent Within our new data, as already mentioned in Sect.~\ref{SubSec_Morphology}, we have 9 objects of outflow-like morphologies in the H$\alpha$ nuclear emission, which corresponds to $\sim$25\% of the sample. The same percentage also applies for disky-like emission (9 LINERs), whereas the majority are classified as \lq Core-halo\rq\,(12 LINERS; $\sim$32\%). However, if we consider all the measurements and classifications in previous similar studies, larger statistics can be made. More specifically, in M11 they provided the final morphological classification for 30 LINERs, accounting for the data in \cite{Pogge2000} (see Sect.~\ref{Sect_sample_data}). \\
    
    \noindent In this work, we have added a total of 38 new LINERs, considering both NOT and HST data sets. Summing up, we have H$\alpha$ morphological information for a total of 70 LINERs, such that 32\% of them are classified as having outflow-like emission, 29\% show \lq Core-halo\rq\ morphology, 21\% are classified as \lq Disky\rq, 11\% as \lq Dusty\rq\ and 7\% with an \lq Unclear\rq\ morphology. These results indicate that 1 out of every 3 LINERs in the local Universe may have an outflow.

\begin{table}
        \caption{Morphological classification of the H$\alpha$ nuclear emission as in  M11. The $\dagger$ indicates the galaxies classified in M11 and \cite{Pogge2000}. The galaxies in parenthesis have an unclear classification. The * are targets analysed both with the new data and in M11.}
        \label{Table_classification}
    \centering
        \begin{tabular}{l|l|l|l}
        \hline \hline
        Core-halo & Disky & Dusty & Bubble \\ 
        (12+8) & (9+6) & (3+5) & (9+13) \\ \hline
        IC\,1459$^{\dagger}$ & NGC\,0841 & NGC\,3226$^{\dagger}$ & NGC\,0266 \\
        NGC\,0315$^{\dagger}$ & NGC\,2681$^{\dagger}$ & NGC\,3607$^{\dagger}$ & NGC\,0404$^{\dagger}$\\
        NGC\,0410 & NGC\,2841$^{\dagger}$ & NGC\,3627$^{\dagger}$ & NGC\,1052$^{\dagger}$ \\
        NGC\,2639$^{\dagger}$ & NGC\,3185 & NGC\,3628 & NGC\,2685 \\
        NGC\,2787$^{\dagger}$ & NGC\,3507 & NGC\,4125 & NGC\,3245$^{\dagger}$ \\
        NGC\,3623$^{\dagger}$ & NGC\,3608 & NGC\,4374$^{\dagger}$ & NGC\,3379*$^{\dagger}$ \\
        NGC\,3884 & NGC\,3642 & (NGC\,5363) & NGC\,3414 \\
        NGC\,3998$^{\dagger}$ & NGC\,3898 & NGC\,5746 & NGC\,3718$^{\dagger}$ \\
        NGC\,4111$^{\dagger}$ & (NGC\,4143) & NGC\,5866$^{\dagger}$ & NGC\,3945 \\
        NGC\,4261 & (NGC\,4203) &  & NGC\,4036$^{\dagger}$ \\
        NGC\,4278* & NGC\,4314$^{\dagger}$ &  & (NGC\,4143) \\
        NGC\,4450 & NGC\,4321 &  & NGC\,4192$^{\dagger}$ \\
        NGC\,4494 & NGC\,4457 &  & (NGC\,4203) \\
        NGC\,4589 & NGC\,4552$^{\dagger}$ &  & NGC\,4438$^{\dagger}$ \\
        NGC\,4698 & NGC\,4594$^{\dagger}$ &  & NGC\,4459 \\
        NGC\,4772 & NGC\,4736$^{\dagger}$ &  & NGC\,4486$^{\dagger}$ \\
        NGC\,5055$^{\dagger}$ & NGC\,5077 &  & NGC\,4579$^{\dagger}$ \\
        NGC\,5957 & (NGC\,5838) &  & NGC\,4596 \\
        NGC\,6482 & (NGC\,7331) &  & NGC\,4636$^{\dagger}$ \\
        NGC\,7743 &  &  & NGC\,4696\\
         &  &  & NGC\,4750\\
         &  &  & NGC\,5005$^{\dagger}$\\
         &  &  & (NGC\,5363) \\ 
         &  &  & NGC\,5813 \\ 
         &  &  & (NGC\,5838) \\
         &  &  & NGC\,5846$^{\dagger}$ \\
         &  &  & (NGC\,7331) \\
        \hline
        \end{tabular}
\end{table}

\section{Discussion}
\label{Section_discussion}

   \begin{figure}
   \includegraphics[width=\columnwidth]{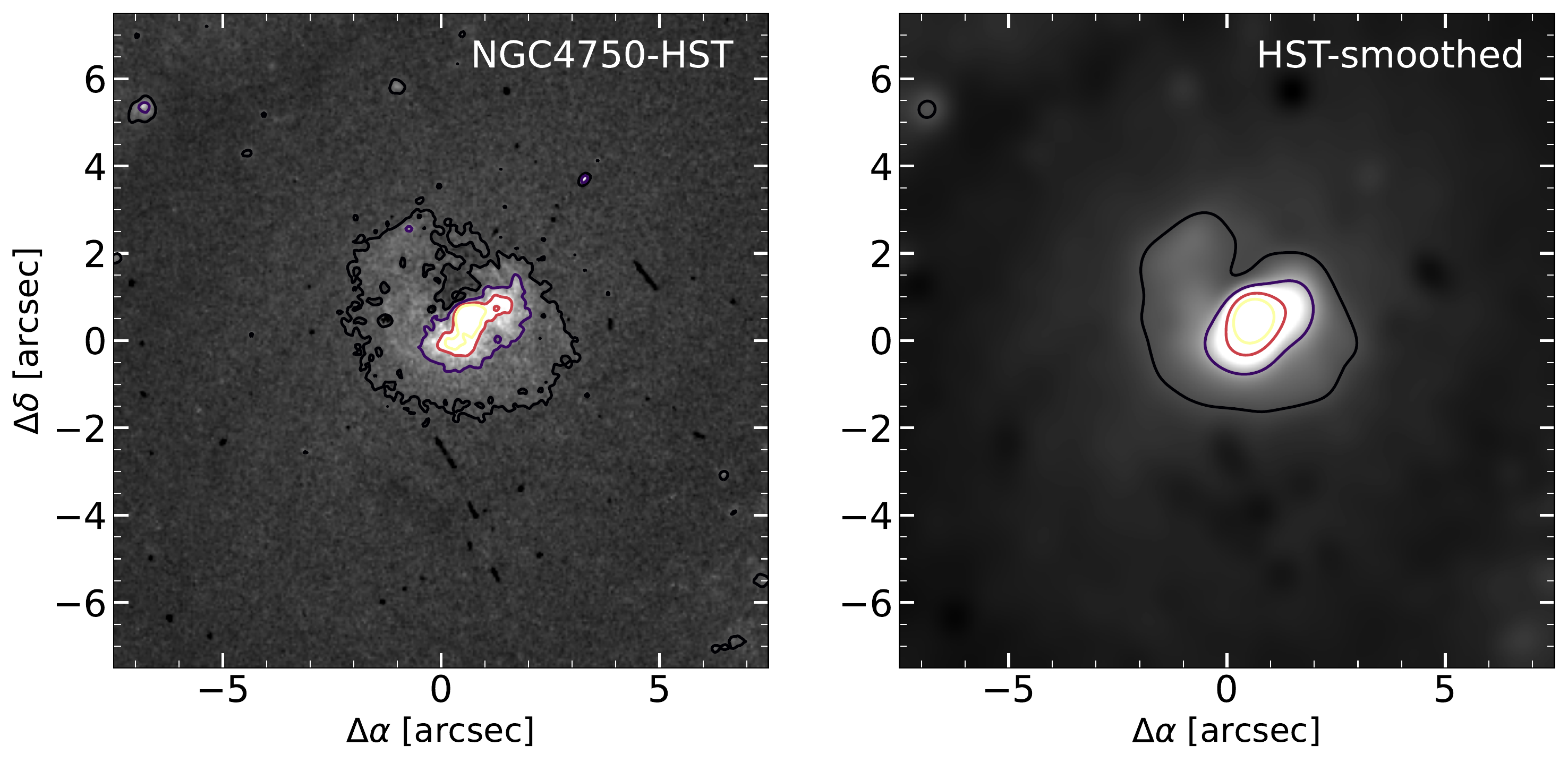}
   \caption{\textit{Left:} H$\alpha$ emission on NGC\,4750 observed with HST (M11). \textit{Right:} Smoothed version of the HST image with a Gaussian filter ($\sigma\sim 5$) to match the ALFOSC/NOT resolution. In both panels, the minimum contour is 3$\sigma$ over the background and the maximum, yellow contour of 25$\sigma$.}
   \label{Fig_compHSTNOT}
   \end{figure}

   \begin{figure}
   \includegraphics[width=\columnwidth]{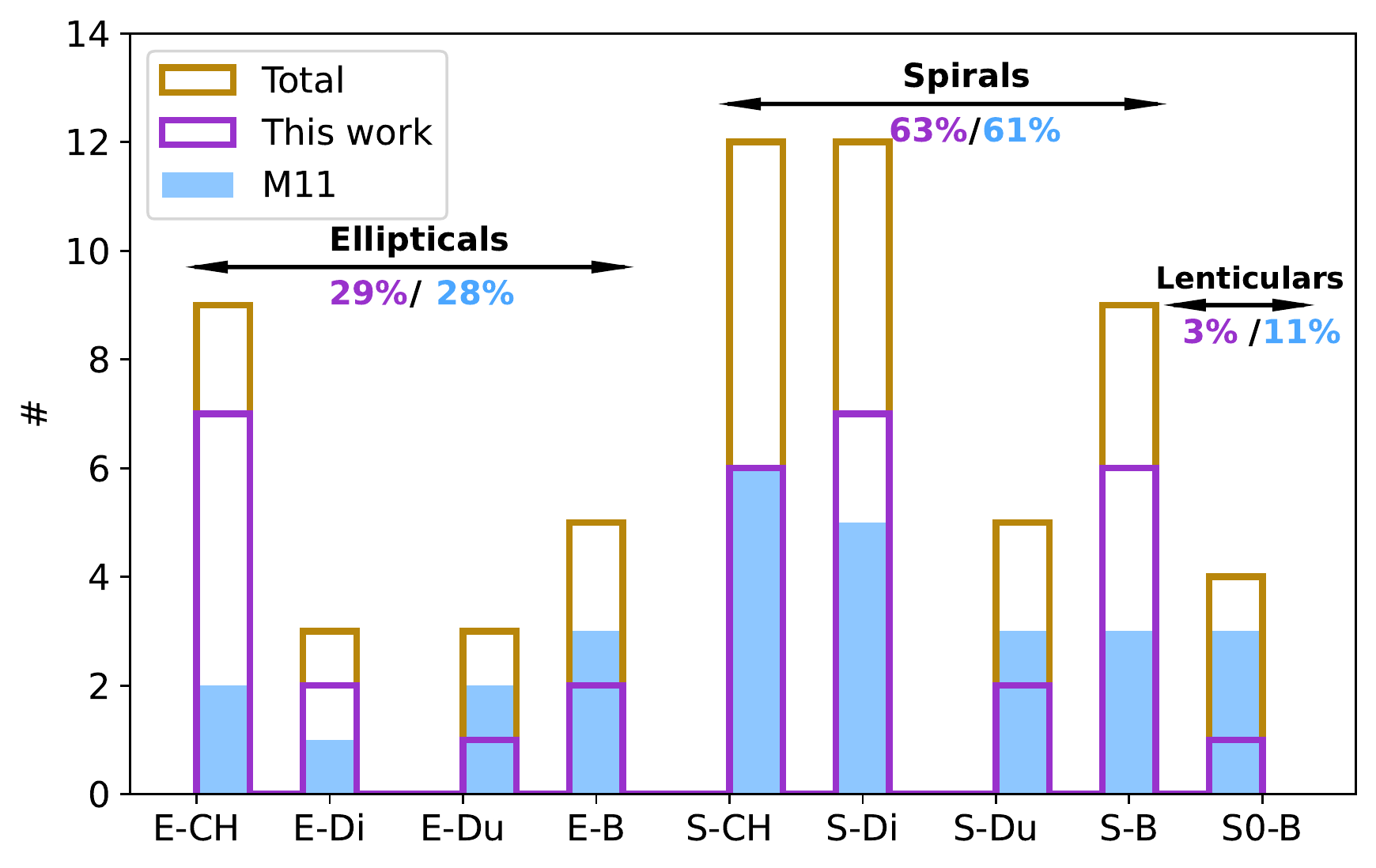}
   \caption{Histogram of the morphological type of the host galaxy and the classification of the nuclear H$\alpha$ emission. It is divided into ellipticals (E), spirals (S) and lenticulars (S0); \lq CH\rq\ stands for \lq Core-halo\rq, \lq Di\rq\ for \lq Disky\rq, \lq Du\rq\ for \lq Dusty\rq\ and \lq B\rq\ for \lq Bubble\rq. Purple lines indicates the sample from this work; blue for M11; and yellow is the combination of both samples. We have not included unclear cases or irregular galaxies in the figure, and thus the percentages do not sum up 100\%.}  
   \label{Fig_morphclass}
   \end{figure}

\noindent The combined results of the new 38 objects from this work and 32 from previous works \citep{Pogge2000,Masegosa2011} constitute the largest sample up to date of ionised gas morphological analysis of LINERs. The morphological features of the 70 galaxies indicate that a high percentage (32\%) of LINERs do have outflow-like ionised gas emission. However, we notice that a dedicated spectroscopic follow-up using IFS is needed to firmly confirm what the morphological signatures suggest and capture the full extension of the putative outflow. This specially applies to the targets with long-slit spectroscopic observations, where the outflow will not be detected if the slit is not properly oriented. 

\noindent We note that, although we classify all the images under the same criteria (see Sect.~\ref{Section_results}), both HST and ALFOSC/NOT data have different spatial resolutions. This implies a loss of detail on the possible substructures present on the images, but it should not produce any relevant change on the classification of the ionised gas morphology, as the categories defined rely on features large enough (as discs or dust lanes) to not depend on pc-scale structures (see Fig.~\ref{Fig_compHSTNOT}). Specifically, the mean extension of the H$\alpha$ emission within the HST data is 0.8 $\pm$0.3 kpc, whereas for the ALFOSC/NOT data is 1.9$\pm$1.3 kpc.\\

\noindent We compare the ionised gas morphology with the morphological type of the host galaxy (see Table~\ref{Table_galaxieslisted}) in order to see if there exists any relationship with the H$\alpha$ emission beyond the possible star-forming regions in the centres of those galaxies.
We see no correlations of the morphologies in the data from M11, in the new data, or when considering both together (see Fig.~\ref{Fig_morphclass}). We note that the H$\alpha$ nuclear emission of all the lenticular galaxies mixing both works was classified as outflow-like, although the sample is small (5 objects) to firmly state any conclusion about it. 
All the targets are early types, ranging from E to Sb. This is expected, as within the Palomar sample the largest fraction of AGNs (including higher accretion rates) were found among these morphological types \citep{Ho2008}. We find that the majority of galaxies are spirals both in M11 and in this work ($\sim$60\%; see Fig.~\ref{Fig_morphclass}). This is in contrast to \cite{Ho2008}, who stated that for LINERs approximately 35\% of the hosts galaxies were ellipticals, $\sim$45\% spirals, while lenticulars were 20\%. We selected our LINERs based on their distances (see Table~\ref{Table_galaxieslisted}) and a previous detection in X-rays that ensured their true nature as AGNs \citep[see Sect.~\ref{Sect_sample_data}][]{GM2009}. The different selection criteria of the samples may be causing the differing percentages.

   \subsection{Outflow candidates with kinematic information}
   \label{SubSec_outflows}
   
   \noindent If we take into account all the cases from the sample in this work (see Sect.~\ref{SubSec_Morphology}), M11 and \cite{Pogge2000}, 60 LINERs have spectroscopic data (at least, from long-slit spectroscopy) available in some spectral bands. This information from the literature allows us to estimate the total percentage of kinematically-confirmed (or suggested) outflows in the total sample. \\
   
    \noindent For all the targets classified as \lq Core-halo\rq, \lq Disky\rq\ and \lq Dusty\rq\ (43 targets) in our H$\alpha$ data, there is kinematic information available for 33 objects, from which there are reported outflows/inflows in 15, as specified below.
    
    \noindent From the 20 classified as \lq Core-halo\rq, 16 LINERs have kinematical information, and there are outflows reported for 8 LINERs: IC\,1459, NGC\,0315, NGC\,2787, NGC\,3884, NGC\,3998, NGC\,4278, NGC\,4450, and NGC\,7743. For NGC\,0315, NGC\,2787 \citep[see also][]{Ruschel2021}, NGC\,3884, NGC\,4278 and NGC\,4450 objects the possible outflows (inflow for NGC\,4278) were suggested by \cite{Cazzoli2018} with long-slit spectroscopy of the nuclear region. In IC\,1459 the outflow was detected using IFU data from GMOS \citep{Ricci2014,Ricci2015}, although the gas is associated to accretion from the surrounding galaxies with MUSE data in \citet{Mulcahey2021}. For NGC\,3998 there were hints of non rotational motions in \cite{Cazzoli2018}, that were compatible with the perturbed kinematics of the [O\,III] emission using IFU data of the ATLAS3D survey \citep{Boardman2017}. For NGC\,7743 there was an outflow detected in molecular gas using SINFONI \citep{Davies2014}, and several kinematical components detected in the ionised gas studied by \cite{Katkov2011}.

    \noindent In the case of the 15 LINERs classified as \lq Disky\rq, there is kinematical data for 13 targets. There are outflows/inflows reported only for 4 objects: NGC\,2841, NGC\,3642, NGC\,4594 and NGC\,5077. For NGC\,2841, \cite{Schmidt2016} detected a small inflow of HI gas, although it is not clear if is caused by gas accretion or minor mergers. For NGC\,3642 and NGC\,4594 the outflows were detected with long-slit spectroscopy by \cite{Cazzoli2018} and \cite{HM2020}, respectively. For NGC\,5077 there is MUSE IFS data studied in detail by \cite{Raimundo2021}, who reports an outflow with the morphology of a hollow cone. The extended ionised emission corresponding to the outflow in that work is similar to what we find at larger scales (see Fig.~\ref{Figure_largescale}). 
    
    \noindent From the 8 targets classified as \lq Dusty\rq, 6 of them have kinematical information in the literature. Outflows have been reported for 4 LINERs: NGC\,3226 \citep{Cazzoli2018}, NGC\,3627 \citep[inflow in][]{Casasola2011}, NGC\,3628 and NGC\,5866 \citep{Li2015}. The case of NGC\,3628 was already mentioned in Sect.~\ref{SubSec_Morphology}. There are two hints of outflows in this galaxy (see individual comments in Appendix~\ref{Appendix_A}), a kpc-scale H$\alpha$ plume (see Fig.~\ref{Figure_largescale}) detected in various wavelength bands \citep{Fabbiano1990,Tsai2012,Cicone2014,Fluetsch2019}, and a sub-kpc outflow reported in CO emission \citep{Tsai2012,Cicone2014,Roy2016}, that is not visible in our H$\alpha$ data due to obscuration from the dust lane (it is in the north direction at $\leq$\,20\arcsec from the centre). \\
    
    \noindent From the 22 LINERs classified as \lq Bubble\rq\ (see Sect.~\ref{SubSec_Morphology} and Table~\ref{Table_classification}), there is kinematical information available for 21 targets, from which 10 have kinematically detected outflows with either IFS or long-slit spectroscopy. For 13 LINERs, namely NGC\,0266, NGC\,0404, NGC\,1052, NGC\,2685, NGC\,3379, NGC\,3414, NGC\,4459, NGC\,4486, NGC\,4579, NGC\,4596, NGC\,4696, NGC\,5813 and NGC\,5846, there is IFS data available (see individual comments on each galaxy in Appendix~\ref{Appendix_A}, in \citealt{Pogge2000} and M11). The IFS data provided hints for non-rotational motions in the velocity maps which could be associated to outflows only for 6 objects, namely NGC\,0404, NGC\,1052, NGC\,3379, NGC\,4486, NGC\,4579 and NGC\,4696. NGC\,0404 images show extended H$\alpha$ emission ascribed to gas blown out possibly by starburst processes \citep{Pogge2000}, that is shown in \cite{Nyland2017} along with the central soft X-ray emission, which have similar extensions, suggesting the presence of shocks. This galaxy was suggested to have outflows associated to a small jet driven by the AGN using ALMA C0(2-1) data \citep{Nyland2017}, although the same emission was also ascribed to supernovae processes \citep{Pogge2000,Boehle2018}. NGC\,1052 was in-depth studied with IFS data, with all the works confirming an outflow in ionised gas \citep[][Cazzoli et al. in prep.]{Dopita2015,Dahmer2019}. For NGC\,3379 the outflow is detected with SAURON data in \cite{Shapiro2006}. For NGC\,4486 they are mentioned in \cite{HM2020}, although the emission is defined as ionised gas filaments with unclear nature in \cite{LopezCoba2020,Sanchez2021}. For NGC\,4579, the emission is ascribed to an outflow in several works with long-slit spectroscopy \citep[][]{Walsh2008,Davies2014,Mazzalay2014,Balmaverde2014,Molina2018}, although within the NUGA project, there was an inflow detected in molecular gas, affecting on different galactic scales \citep[][]{GarciaBurillo2009,Casasola2011}. For the remaining 7 objects, although there is disturbed gas, it is usually ascribed to past mergers or dust lanes that produce asymmetries in the gas velocities. For NGC\,4696, \cite{Canning2011} analysed VIMOS/VLT IFU data, finding a filamentary structure with at least two different kinematic components, probably caused due to interactions with the powerful radio jet and with other members of its galaxy cluster. NGC\,2685 has a complex structure: a polar ring blended with an outer ring (see Appendix.~\ref{Appendix_A}). It was studied with long-slit spectra in previous works \citep{Eskridge1997,Jozsa2009,HM2020}, but the different slits were not oriented in the direction of the possible outflow. \cite{Boardman2017} studied its [O\,III] emission with the ATLAS$^{3D}$ survey, finding perturbed kinematics likely not produced by mergers, but by gas accretion of unclear origin.
    
    \noindent For the other 8 galaxies with bubble-like morphology there is only long-slit spectroscopic data available in the literature. There are outflows reported for 4 targets: NGC\,3245 \citep[in][although not confirmed in \citealt{HM2020}]{Walsh2008}, NGC\,4036 \citep[in][although not confirmed in \citealt{Cazzoli2018}]{Walsh2008}, NGC\,4750 \citep[as an asymmetry in the emission due to dust lanes and spiral arms emerging from the centre][]{Carollo2002,Cazzoli2018} and NGC\,5005 \citep[from both space and ground-data in ][]{Cazzoli2018} using long-slit spectroscopy. For the other 4 objects there are no outflows detected (NGC\,3718, NGC\,3945, NGC\,4438 and NGC\,4636). However, NGC\,4438 has an interesting morphological structure (see Fig.~2 in M11). A clear bubble emerging from the nucleus is seen in H$\alpha$ (M11), although the slit was in a direction of the possible outflow in the study by \cite{Cazzoli2018}. The nature of this structure would be analysed with IFS data (Hermosa Mu{\~n}oz et al. in prep.). \\
    
    \noindent There are 5 galaxies whose morphologies were classified as \lq Unclear\rq. There is kinematical information in the literature for 4 of them (NGC\,4143, NGC\,4203, NGC\,5838, and NGC\,7331), where 3 have reported outflows. For NGC\,4143 there is only long-slit spectra available, from which the existence of outflows was suggested \citep{Cazzoli2018}. For NGC\,4203, NGC\,5838 and NGC\,7331, there is resolved kinematical information available in the literature (see Appendix~\ref{Appendix_A}). NGC\,4203 is included in the ATLAS$^{3D}$ survey, and the H\,I shows non-rotational motions and inner dust lanes. H\,I data from the THINGS survey for NGC\,7331 \citep{Schmidt2016} indicated the presence of inflowing/outflowing emission, also suggested in previous works \citep{Mediavilla1997,Battaner2003}. Finally, for NGC\,5838 there are not reported outflows. The IFS data comes from the SAURON survey \citep{FalconBarroso2003,Sarzi2006}, and although they detect asymmetries in the velocity maps, they ascribe them to dust in the nucleus. \\

    \noindent Summarising, the kinematic information available for 60 LINERs indicates that there are a total of 29 objects (48\%) for which the presence of outflows/inflows was either found or suggested by means of spectroscopic data. Their H$\alpha$ morphology is classified here as \lq Core-halo\rq, \lq Disky\rq, \lq Dusty\rq\ or \lq Bubble\rq, being 3 of them classified as \lq Unclear\rq\ (see Table~\ref{Table_classification} and Sect.~\ref{SubSec_Morphology}). The derived percentage is larger to the percentage derived from imaging data (32\%, see Sect.~\ref{SubSec_globalpic}). 
    In Table~\ref{Table_MorphKin} we have estimated the percentage of kinematically-identified outflows for the morphological classes derived in this work. The percentages in Table~\ref{Table_MorphKin} indicate that, on average, 1 of every 2 LINERs in the nearby Universe may host an outflow. However, we note that we lack of kinematical information for 10 LINERs in the whole sample of 70, thus the rate may actually vary from 41\% up to 56\% LINERs with outflows.
    
    \noindent Table~\ref{Table_MorphKin} shows that among the classes with the largest number of objects ($>$10), the two categories with the largest percentages are \lq Core-halo\rq\,and \lq Bubble\rq (50\% and 48\%, respectively). The objects classified as \lq Disky\rq\,show the lowest percentage within our classification (31\%); this could be a matter of the outflow orientation and the inclination of the disc, that may challenge a possible outflow detection (see Sect.~\ref{SubSec_otherAGNs}), specially in these low accretion rates where they are expected to be faint. The fraction of LINERs with kinematically confirmed outflows that are morphologically classified as \lq Bubble\rq-like emission is 34\% (10 out of 29). This implies that for the majority of the detected outflows, the ionised gas morphology does not show evident features of a bubble-like emission.
    
    \noindent If only IFS data are considered for the identification of kinematical outflows (30 objects with IFS data; see Table~\ref{Table_XrayMorphKin}), only 15 have detected outflows (50\%). Small numbers preclude a separate analysis by morphological class, since all but \lq Bubble\rq\ morphologies have IFS data for less than 10 objects. There are 13 objects with IFS information between the \lq Bubble\rq-like morphologies, among which 6 (46\%) show kinematical evidences of outflows, a slightly larger rate than that of the morphological detection (32\%).

    \begin{table}
        \caption{Percentages of the galaxies with kinematically confirmed outflows in the literature depending on their morphological classification of the H$\alpha$ nuclear emission (see Sect.~\ref{SubSec_outflows} and Appendix~\ref{Appendix_A} for the comments on individual targets).}
        \label{Table_MorphKin}
    \centering
        \begin{tabular}{l c r}
        \hline \hline
        Morphological class & Kinematical outflows & Targets \\ 
        \hline 
        Core-halo & 50\% & 8 (out of 16) \\ 
        Disky & 31\% & 4 (out of 13) \\
        Dusty & 66\% & 4 (out of 6) \\
        Bubble & 48\% & 10 (out of 21) \\
        Unclear & 75\% & 3 (out of 4) \\ 
        \hline
        All classes & 48\% & 29 (out of 60) \\
        \hline
        \end{tabular}
    \end{table}

    \subsection{Soft X-ray / Ionised-gas relationship}
    \label{SubSec_DiscXray}
    
    \noindent It is noticeable that the H$\alpha$ morphologies and the soft X-ray emission are coincident in a large fraction of the new data ($\sim$30\%), as is the case also for the targets analysed in M11. This behaviour has been seen previously for type-2 Seyferts \citep{Bianchi2006}, where the [O\,III] and the soft X-ray emission correlated. The work by \cite{Bianchi2019} suggested that Seyferts and LINERs do have similarities in their narrow line regions (see also \citealt{Pogge2000}), which could lead to the conclusion that these two emissions are produced within the same region of the AGN, and thus a correlation should be expected also in LINERs. The correspondence of the ionised gas morphological signatures, the kinematic results and the X-ray emission for all the galaxies for which we could retrieve data (56 targets: 28 from new data; see Sect.~\ref{Sect_datareduction}) is in Table~\ref{Table_XrayMorphKin}. 
    
    \noindent M11 performed a similar analysis with the objects from their sample. They gathered data for a total of 28 LINERs in both soft and hard X-rays using Chandra, 4 of which were NGC\,3379, NGC\,4278, NGC\,4676A and B (thus we consider here 24 objects). The results were that, generally, the soft X-ray emission traced that of the ionised gas, and they were correlated in all but 4 galaxies, that seemed to deviate from that trend (NGC\,3226, NGC\,4486, NGC\,5846 and NGC\,5866). From all the galaxies with X-ray data in their sample, 20 out of 24 ($\sim$83\%) showed a correlation in the emission.\\
    
    \noindent By taking into account the new data in this work (see Sect.~\ref{SubSec_ResultXrays}), from the 52 objects with soft X-ray data available in the Chandra archive, 31 of them (60\%; see Table~\ref{Table_XrayMorphKin}) do show a correlation with the ionised gas emission. According to our results (see Table~\ref{Table_XrayMorphKin}), 29 targets have reported outflows or inflows with kinematical information. Only 12 LINERs (namely IC\,1459, NGC\,2787, NGC\,2841, NGC\,3245, NGC\,3884, NGC\,3998, NGC\,4036, NGC\,4450, NGC\,4579, NGC\,4594, NGC\,4696 and NGC\,4750) have both features simultaneously, that is 23\% (12 over 52). However, the reported outflows in all the objects but 5 (IC\,1459, NGC\,2841, NGC\,3998, NGC\,4579 and NGC\,4696) come from long-slit spectroscopy. As discussed previously (see Sect.~\ref{Section_discussion}), although useful, these spectra do not fully characterise the AGN and its environment, given the spatial limitations of the technique. This combined with the small number of coincidences, do not allow to draw any firm conclusion about a possible correlation between the presence of an outflow and an extended X-ray emission.

    \begin{table*}
        \caption{Galaxies with available X-ray data and its correspondence with the morphological signatures in H$\alpha$ and the kinematically-confirmed outflows or inflows in the literature (see Sect.~\ref{SubSec_outflows}, Sect.~\ref{SubSec_DiscXray} and Fig.~\ref{Figure_Xrays}). In column (2) we include the morphological classification according to Sect.~\ref{Section_results}. Column (3) indicates whether the spectral information was obtained with long-slit spectroscopy (LSS) or IFS. In column (4) and (5): \lq Y\rq\,stands for Yes and \lq N\rq\,for No. Column (4) indicates if the X-ray emission is co-spatial with the H$\alpha$ morphology (see Sect.~\ref{SubSec_ResultXrays}). Targets from \cite{Pogge2000} and \cite{Masegosa2011} are marked with $\dagger$.}
        \label{Table_XrayMorphKin}
    \centering
        \begin{tabular}{llccc|llccc}
        \hline \hline
        Galaxy & Morphology & Spectra & Outflow/Inflow & X-ray & Galaxy & Morphology & Spectra & Outflow/Inflow & X-ray \\ 
        (1) & (2) & (3) & (4) & (5) & (1) & (2) & (3) & (4) & (5) \\ 
        \hline 
        NGC\,0266 & Bubble & IFS & N & Y              & NGC\,4698 & Core-halo & IFS & N & N \\  
        NGC\,0404$^{\dagger}$ & Bubble & IFS & Y & -  & NGC\,4772 & Core-halo & IFS & N & Y \\
        NGC\,1052$^{\dagger}$ & Bubble & IFS & Y & -  & NGC\,5055$^{\dagger}$ & Core-halo & IFS & N & Y \\
        NGC\,2685 & Bubble & IFS & N & -              & NGC\,5957 & Core-halo & - & - & - \\  
        NGC\,3245$^{\dagger}$ & Bubble & LSS & Y & Y  & NGC\,6482 & Core-halo & LSS & N & - \\   
        NGC\,3379 & Bubble & IFS & Y & N              & NGC\,7743 & Core-halo & IFS & Y & - \\ 
        NGC\,3414 & Bubble & IFS & N & Y              &  &  &  &  & \\
        NGC\,3718$^{\dagger}$ & Bubble & LSS & N & -    & NGC\,0841 & Disky & LSS & N & - \\
        NGC\,3945 & Bubble & LSS & N & Y              & NGC\,2681$^{\dagger}$ & Disky & LSS & N & Y \\   
        NGC\,4036$^{\dagger}$ & Bubble & LSS & Y & Y  & NGC\,2841$^{\dagger}$ & Disky & IFS & Y & Y \\    
        NGC\,4192$^{\dagger}$ & Bubble & - & - & -    & NGC\,3185 & Disky & - & - & - \\
        NGC\,4438$^{\dagger}$ & Bubble & LSS & N & Y    & NGC\,3507 & Disky & IFS & N & N \\ 
        NGC\,4459 & Bubble & IFS & N & N              & NGC\,3608 & Disky & IFS & N & N \\   
        NGC\,4486$^{\dagger}$ & Bubble & IFS & Y & N  & NGC\,3642 & Disky & LSS & Y & N \\   
        NGC\,4579$^{\dagger}$ & Bubble & IFS & Y & Y  & NGC\,3898 & Disky & LSS & N & N \\    
        NGC\,4596 & Bubble & IFS & N & Y              & NGC\,4314$^{\dagger}$ & Disky & - & - & Y \\    
        NGC\,4636$^{\dagger}$ & Bubble & LSS & N & Y    & NGC\,4321 & Disky & IFS & N & N \\   
        NGC\,4696$^{\dagger}$ & Bubble & IFS & Y & Y  & NGC\,4457 & Disky & IFS & N & N \\   
        NGC\,4750 & Bubble & LSS & Y & Y              & NGC\,4552$^{\dagger}$ & Disky & LSS & N & Y \\   
        NGC\,5005$^{\dagger}$ & Bubble & LSS & Y & -  & NGC\,4594$^{\dagger}$ & Disky & LSS & Y & Y \\ 
        NGC\,5813 & Bubble & IFS & N & Y              & NGC\,4736$^{\dagger}$ & Disky & LSS & N & Y \\ 
        NGC\,5846$^{\dagger}$ & Bubble & IFS & N & N  & NGC\,5077 & Disky & IFS & Y & - \\
         &  &  &  &                                       &  &  &  & \\ 
        IC\,1459$^{\dagger}$ & Core-halo & IFS & Y & Y    & NGC\,3226$^{\dagger}$ & Dusty & LSS & Y & N \\ 
        NGC\,0315$^{\dagger}$ & Core-halo & LSS & Y & N   & NGC\,3607$^{\dagger}$ & Dusty & - & - & Y \\
        NGC\,0410 & Core-halo & IFS & N & Y               & NGC\,3627$^{\dagger}$ & Dusty & IFS & Y & - \\ 
        NGC\,2639$^{\dagger}$ & Core-halo & - & - & -     & NGC\,3628 & Dusty & IFS & Y & N \\ 
        NGC\,2787$^{\dagger}$ & Core-halo & LSS & Y & Y   & NGC\,4125 & Dusty & - & - & N \\ 
        NGC\,3623$^{\dagger}$ & Core-halo & - & - & -     & NGC\,4374$^{\dagger}$ & Dusty & LSS & N & Y \\ 
        NGC\,3884 & Core-halo & LSS & Y & Y               & NGC\,5746 & Dusty & IFS & N & - \\
        NGC\,3998$^{\dagger}$ & Core-halo & IFS & Y & Y   & NGC\,5866$^{\dagger}$ & Dusty & LSS & Y & N \\ 
        NGC\,4111$^{\dagger}$ & Core-halo & - & - & Y     &  &  &  &  & \\
        NGC\,4261 & Core-halo & IFS & N & Y               & NGC\,4143 & Unclear & LSS & Y & N \\ 
        NGC\,4278 & Core-halo & LSS & Y & N               & NGC\,4203 & Unclear & IFS & Y & N \\
        NGC\,4450 & Core-halo & LSS & Y & Y               & NGC\,5363 & Unclear & - & - & - \\
        NGC\,4494 & Core-halo & IFS & N & -               & NGC\,5838 & Unclear & IFS & N & Y \\ 
        NGC\,4589 & Core-halo & LSS & N & N               & NGC\,7331 & Unclear & IFS  & Y & N \\ 
        \hline
        \end{tabular}
    \end{table*}

    \subsection{Comparison with other active galaxies}
    \label{SubSec_otherAGNs}
    
    \noindent So far, there are two systematic searches of outflows in LINERs with both ground and space-based long-slit spectroscopic data. These were done by \cite{Cazzoli2018} and \cite{HM2020}, who analysed 22 (type-1) and 9 (type-2) LINERs, finding an outflow detection-rate of 41\% and 22\%, respectively. Considering the sample number, the rate of outflows we report in this study (48\%) is in agreement with that by \cite{Cazzoli2018}.
    
    \noindent There are several works searching for outflows in the nearby Universe for different active galaxies in different wavelength ranges \citep{Cazzoli2016,Fiore2017,Fluetsch2019,Fluetsch2021,Ruschel2021}, and also at higher redshifts \citep[e.g.][0.6$\leq$\,z\,$\lesssim$\,1.7, with $\sim$50\% kinematic detections of ionised outflows]{Harrison2016}. Within the kinematical data from the literature gathered in this work with LINERs in the Local Universe, not all the targets with kinematically-identified outflows are based on ionised gas (19 out of 29 in ionised gas). There are evidences for extended outflows on different spatial scales (from sub-kpc to several kpc), associated to both AGNs and starburst activity \citep[][e.g. M\,82, see \citealt{Leroy2015}]{Veilleux2005}. \cite{Luo2021} studied the ionised emission for 40 nearby (z$<$0.1) type-2 AGNs, finding a correlation between the outflow sizes and the [O\,III] luminosities. The typical sizes of the kinematically-identified outflows varied from 0.9 to 4 kpc (median 2 kpc), which are similar sizes to those quoted by previous works \citep[e.g. mean of 1.8 kpc for luminous z$<$1 AGNs in Bae et al. 2017; 0.1 to 3 kpc for 6 Seyfert-2 galaxies in][]{Revalski2021}. In this work we focus on the circumnuclear emission of the galaxies as it is expected that the possible outflows are less intense in LINERs. However, the H$\alpha$ emission in our images has similar extensions, with a mean observed size of 1.9 kpc, ranging in all morphologies from 0.2 to 6 kpc (0.5 to $\sim$2 kpc for only the bubble-like morphologies).

    \noindent \cite{Concas2019}, by exploiting SDSS spectra, suggested that although the inclination of the host galaxy is crucial to detect the outflowing emission, if the ionised outflows are AGN-driven and the accretion disc is not connected to the host disc, then the outflow may be launched in another direction, not perpendicular to the host. \cite{Fischer2013} obtained the inclination of the AGN in 17 Seyfert galaxies, finding that the biconical outflows were more easily detected in those that were type-2 AGNs (i.e. oriented edge-on).
    From our imaging data and the kinematic analysis from the literature, we find that inclination could be important, as the lowest percentages of outflows are in \lq Disky\rq\ systems (e.g. NGC\,4321). Given that the intensity of the outflowing gas scales with the activity of the AGN \citep{Fluetsch2019,Revalski2021}, and that their extension is typically $\sim$ 1~kpc (one of the largest seen with our images is NGC\,3945 $\sim$1.5\,kpc; see Fig.~\ref{Figure_AB4}), we could be missing outflows due to inclination effects. As already noticed by \cite{Fischer2013}, these effects are potentially more relevant for close to face-on targets in which the outflow has a cone-like shape, as the inclination effects may lead to misleading morphological classifications. Within our sample, between the kinematically-detected outflows there are 19 type-1 (5 IFS, 14 LSS) and 10 type-2 LINERs (8 IFS, 2 LSS)\footnote{The distribution of type-1 and type-2 LINERs in the complete sample is 38\% and 62\% respectively \citep[based on][]{Ho1997,Cazzoli2018,HM2020}}. Those detected with long-slit spectroscopy are only nuclear outflows. Curiously, for the IFS outflowing detections in type-1s, the typical sizes are larger than 1\,kpc (i.e. $\sim$1\,kpc for NGC\,1052, $\sim$2.3\,kpc for NGC\,4696, $\sim$5\,kpc for IC\,1450), whereas for type-2s the sizes are usually detected on a sub-kpc scale (i.e. $\sim$40\,pc NGC\,0404 or $\sim$700\,pc for NGC\,7743). This would be expected based on the unified model, as type-2 LINERs are oriented edge-on and thus outflows oriented in the polar direction would be more easily detected than on more face-on type-1 LINERs.

\section{Summary and conclusions}
\label{summary_conclusions}

\noindent In this work we present a morphological analysis of 70 nearby LINERs with both ALFOSC/NOT and HST narrow band imaging corresponding to the H$\alpha$ emission. We have obtained new data for a total of 38 galaxies, whereas the remaining 32 come from previous works \citep[][M11]{Pogge2000}. The H$\alpha$ imaging data indicate that among these AGNs there is a variety of morphologies within the ionised gas, that can be classified in 4 different groups, as in M11, plus an additional class for objects with an ambiguous classification.

\noindent The statistics derived from the total sample of 70 LINERs suggest that $\sim$32\% of LINERs have bubble-like ionised gas morphologies, $\sim$29\% show compact circumnuclear emission (\lq Core-halo\rq) and $\sim$21\% have emission associated to the disc or the spiral arms of the host galaxy. For $\sim$11\%\,of the sample we cannot state anything on the morphologies, as there is nuclear dust preventing from obtaining a complete vision of the ionised gas. Additionally, the complex distribution of the gas in $\sim$7\% of the sample lead us to an unclear classification of the sources based exclusively on imaging information. However, there is kinematical information in the literature for 80\% (4 out of 5) of these sources with \lq Unclear\rq\ morphology that provides more evidences on the possible origin of the emission (see Sect.~\ref{SubSec_outflows}). \\

\noindent Considering all the kinematic information available, accounting for both long-slit and IFS data, we have data for 60 LINERs. The statistics evidence that 48\% of the objects do show outflow kinematic signatures (see Sect.~\ref{SubSec_outflows}). This percentage is somewhat larger than that derived from the morphological signatures (32\%). Since we still miss kinematic information for 10 targets of the 70 LINERs, the total rate could vary from 41\% up to 56\% LINERs with outflows. With the complete morphological and kinematical information combined, we find that at least 1 out of every 3 nearby (z\,$<$\,0.025) LINERs may have a galactic outflow, probably associated to the activity of the super massive black hole, given that they are mainly detected in ionised gas.

\noindent The morphological signatures combined with the kinematical information suggest that, considering our number statistics, it is equally probable to find an outflow in a LINER in which the ionised gas morphology is bubble-like than if it is core-halo-like (see Sect.~\ref{SubSec_outflows}). \\

\noindent We found evidences that the soft X-ray emission follows that from the ionised gas in 31 LINERs ($\sim$60\% above the total with X-ray data) from the galaxies analysed in this work. This means that for the majority of targets there is a correlation between the ionised H$\alpha$ gas and the soft X-ray, as it was proposed in previous works for higher-luminosity AGNs \citep{Bianchi2006,Bianchi2019}. Both emissions are then expected to be raised in the same spatial region of the AGN. We note though that the results may vary and that we could find (or not) a strong correlation between the emissions as we still lack soft X-ray data for a 25\% of the sample (i.e. 18 LINERs). Specifically, the correlation percentage may vary up to 70\% (44\%) if both emissions are (not) co-spatial for all the remaining targets. We do not see any correlation between the spectroscopically-confirmed outflows and the co-spatiality of these emissions (ionised gas and soft X-rays). Therefore, with the data that we have, we can say that the soft X-ray emission is not a strong predictor of the presence of kinematically evident outflows. As for the H$\alpha$ imaging, considering all the kinematical information available, our results suggest that those objects with Disky-like structures have less probability of hosting an outflow, probably due to the orientation, whereas for the remaining morphologies, the probability of detection is 55\%.

\begin{acknowledgements}
We thank the anonymous referee for his/her very constructive comments that have helped to improve the paper. 
We acknowledge financial support from the Spanish Ministerio de Ciencia, Innovaci{\'o}n y Universidades (MCIU) under the grants AYA2016-76682-C3 and PID2019-106027GB-C41. Authors acknowledge financial support from the State Agency for Research of the Spanish MCIU through the \lq Centre of Excellence Severo Ochoa\rq\ award to the Instituto de Astrof{\'i}sica de Andaluc{\'i}a (SEV-2017-0709). LHM acknowledge financial support under the grant BES-2017-082471.\\
The data presented here were obtained in part with ALFOSC, which is provided by the Instituto de Astrof{\'i}sica de Andaluc{\'i}a (IAA) under a joint agreement with the University of Copenhagen and NOTSA. This work is based on observations made with the NASA/ESA Hubble Space Telescope, and obtained from the Hubble Legacy Archive, which is a collaboration between the Space Telescope Science Institute (STScI/NASA), the Space Telescope European Coordinating Facility (ST-ECF/ESA) and the Canadian Astronomy Data Centre (CADC/NRC/CSA).
This research has made use of data obtained from the Chandra Data Archive and software provided by the Chandra X-ray Centre in the application package CIAO.\\
This research has made use of the NASA/IPAC Extragalactic Database (NED), which is operated by the Jet Propulsion Laboratory, California Institute of Technology, under contract with the National Aeronautics and Space Administration.
We acknowledge the usage of the HyperLeda database (http://leda.univ-lyon1.fr).\\
This work has made extensive use of IRAF (v2.16) and Python (v3.8.10), particularly with \textsc{astropy} \citep[v4.2, \nolinkurl{http://www.astropy.org};][]{astropy:2013, astropy:2018}, \textsc{matplotlib} \citep[v3.4.1;][]{Hunter:2007}, \textsc{photutils} \citep[v1.1.0;][]{LarryBradley_2020} and \textsc{numpy} \citep[v1.19.4;][]{Harris2020}.\\
Authors acknowledge M. Guerrero and B. P{\'e}rez for their help in some observing runs, and the support astronomers at the NOT telescope that made the observations during service time.
\end{acknowledgements}

%
   \bibliographystyle{aa} 
   \bibliography{bibliography.bib} 

%
%

\begin{appendix}

\section{Individual comments on galaxies}
\label{Appendix_A}

\begin{figure*}
   \includegraphics[width=\textwidth]{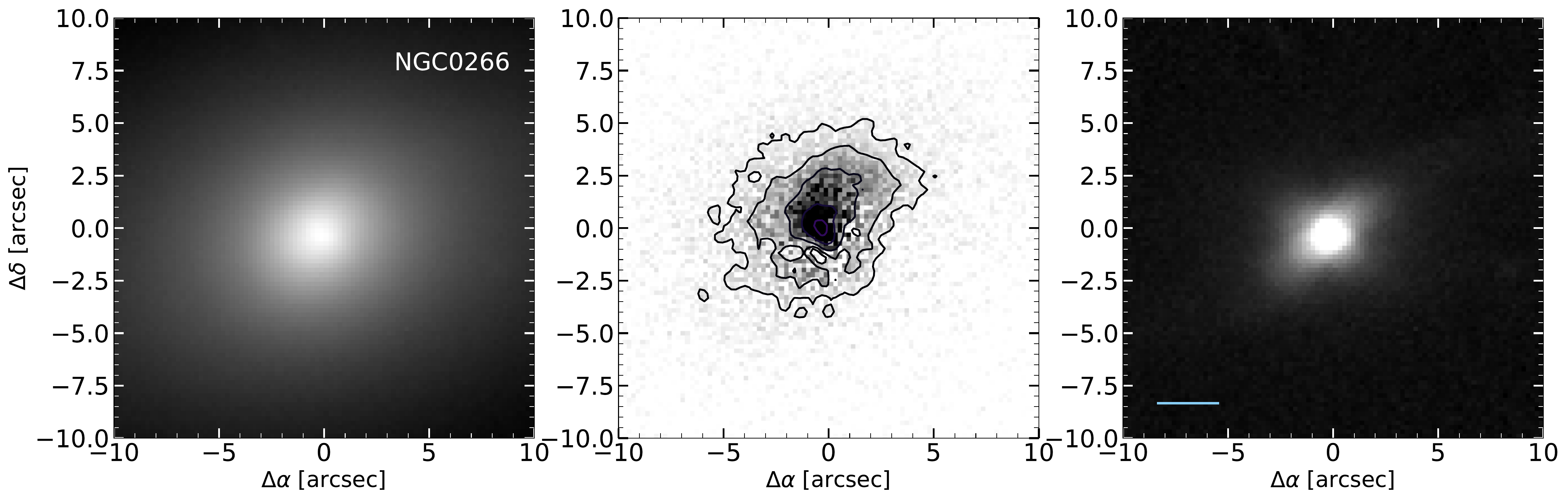}
   \includegraphics[width=\textwidth]{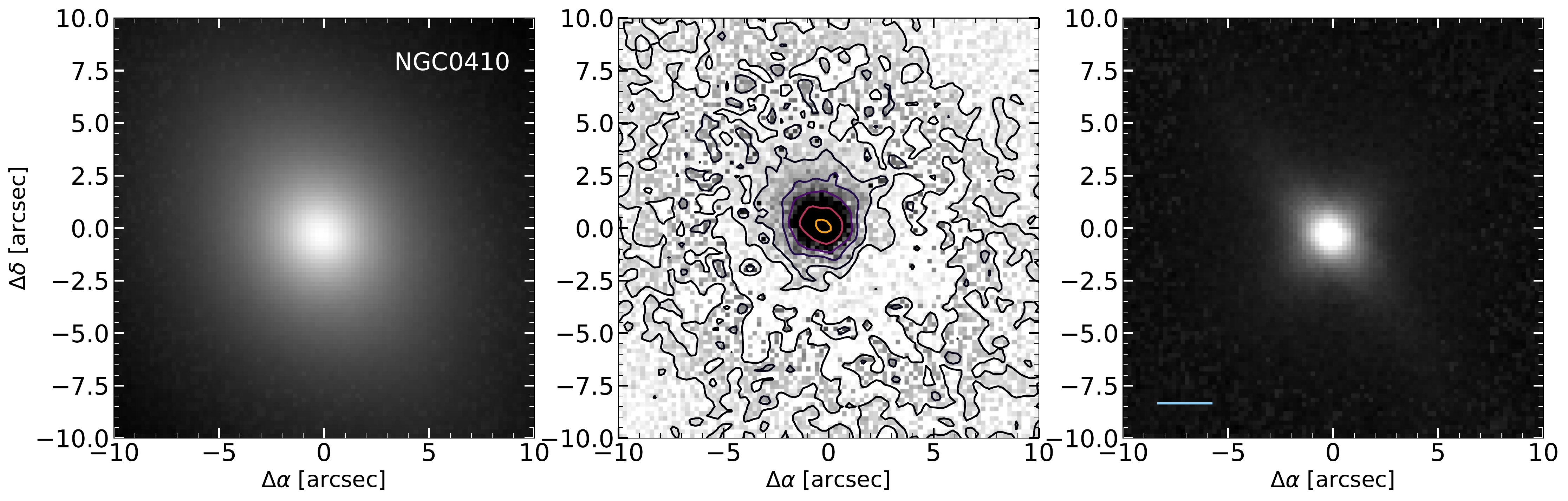}
   \includegraphics[width=\textwidth]{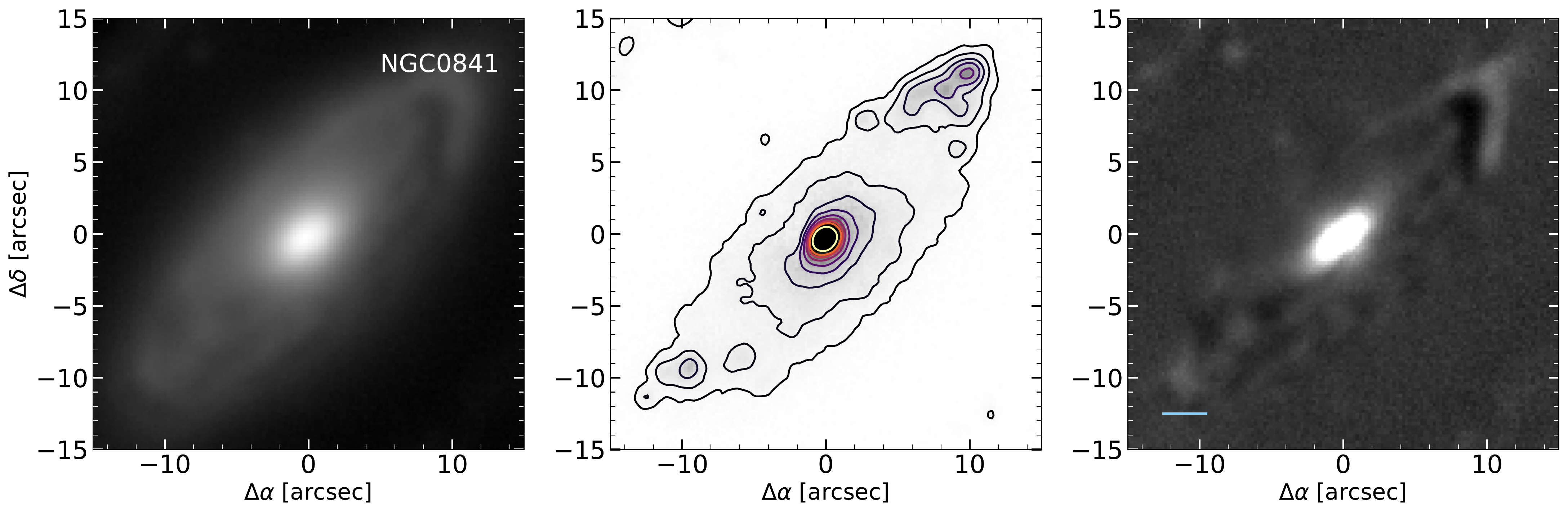}
   \includegraphics[width=\textwidth]{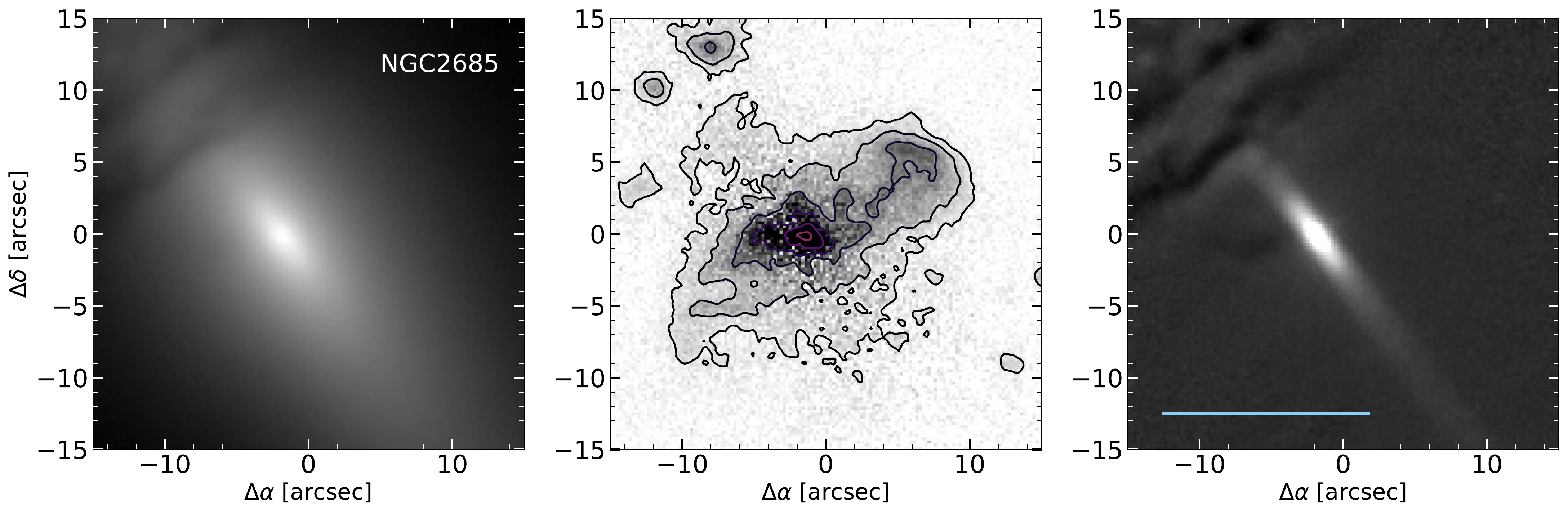}
   \caption{\textit{Left:} Original NF image. \textit{Middle:} H$\alpha$ image of NGC\,0266, NGC\,0410, NGC\,0841 and NGC\,2685 with contours at 3$\sigma$ (black), 7$\sigma$ (black), 15$\sigma$ (black), 25$\sigma$ (dark-purple), 40$\sigma$ (purple), 60$\sigma$ (light-purple), 80$\sigma$ (red), 100$\sigma$ (orange) and 150$\sigma$ (yellow) levels. \textit{Right:} Sharp-divided BF image. The blue line indicates the 1 kpc scale.}  
   \label{Figure_AB1}
   \end{figure*}

\noindent \textit{NGC\,0266} (see Fig.~\ref{Figure_AB1}): The H$\alpha$ emission of this barred galaxy was studied by \citet{Epinat2008}. These authors obtained the velocity maps with data from the GHASP survey, finding gas with disturbed morphology, that could be consistent with past interactions. Considering all the regions with H$\alpha$ emission of NGC\,0266 (out to $\sim$2\arcmin), we only find emission along the spiral arms and extended in the centre (see upper panel of Fig.~\ref{Figure_AB1}). In the work by \cite{Epinat2008}, they detected additional gas distributed along the major axis of the galaxy, following the spiral arms but also randomly distributed, with large voids of gas near the centre. \\

\noindent \textit{NGC\,0410} (see Fig.~\ref{Figure_AB1}): It belongs to a small group of galaxies together with NGC0407 and NGC0414 (non-active systems). We have classified the emission of this object as \lq Core-halo\rq\,(see Table~\ref{Table_classification}). No extended H$\alpha$ emission has been reported previously \citep{Lakhchaura2018}. This galaxy is also included in the MASSIVE survey \citep{Ma2014}, but only the stellar kinematics is shown \citep{Ene2020}. \\

\noindent \textit{NGC\,0841} (see Fig.~\ref{Figure_AB1}): \cite{GD2008} characterised this galaxy as hosting a nuclear spiral and a dust lane using F547M HST imaging. Our sharp-divided image also suggests a ring-like nuclear structure. The H$\alpha$ image shows strong emission in the nucleus and clumpy emission regions along the disc, probably associated to multiple star forming regions. Thus, we classified it as \lq Disky\rq.
The HI data of this system show a perturbed morphology indicative of its interaction with NGC\,0834 (not in this sample), located south-east of our galaxy, although both systems are optically undisturbed \citep{Kuo2008}. In fact, the ionised and neutral gas of this galaxy follow a rotation pattern \citep{Cazzoli2018}. \\

\noindent \textit{NGC\,2685} (see Fig.~\ref{Figure_AB1}): This galaxy shows a complex structure, hosting an inner, polar ring and an outer ring, proposed to correspond to a warped disc, oriented as the galaxy disc \citep{Jozsa2009}. The H\,I gas morphology is disturbed with the presence of non-circular motions (likely due to the interaction between these two different rings), although its rotation curve is nearly that of a spiral \citep{Jozsa2009}. \cite{Boardman2017} studied the [O\,III] kinematics for this galaxy within the ATLAS$^{3D}$ survey, finding a disturbed velocity map, probably ascribed to gas accretion. In \cite{HM2020} the long-slit spectrum analysed shows very narrow line profiles ($\sigma\sim$65 kms$^{-1}$) in both space- and ground-based spectra, although neither was oriented in the direction in which we see an extended ionised emission (PA$\sim$120\degr; see middle-down panel in Fig.~\ref{Figure_AB1}). This asymmetric emission coming from the nucleus, extending up to 10 arcsec, was also pointed out by \cite{Ulrich1975} using long-slit spectroscopy around the [O\,II]$\lambda$3727\AA\,line. \cite{Eskridge1997} obtained optical spectroscopy to study the H\,II regions of the galaxy and H$\alpha$ imaging data where they traced the nuclear emission, stating that the asymmetry probably is produced due to extinction effects. In our work we ascribe this emission to the possible presence of an outflow, probably not detected kinematically due to the orientation of the slit (e.g. PA$=$38\degr\,and 54\degr\,in \citealt{HM2020}). \\

\begin{figure*}
   \includegraphics[width=\textwidth]{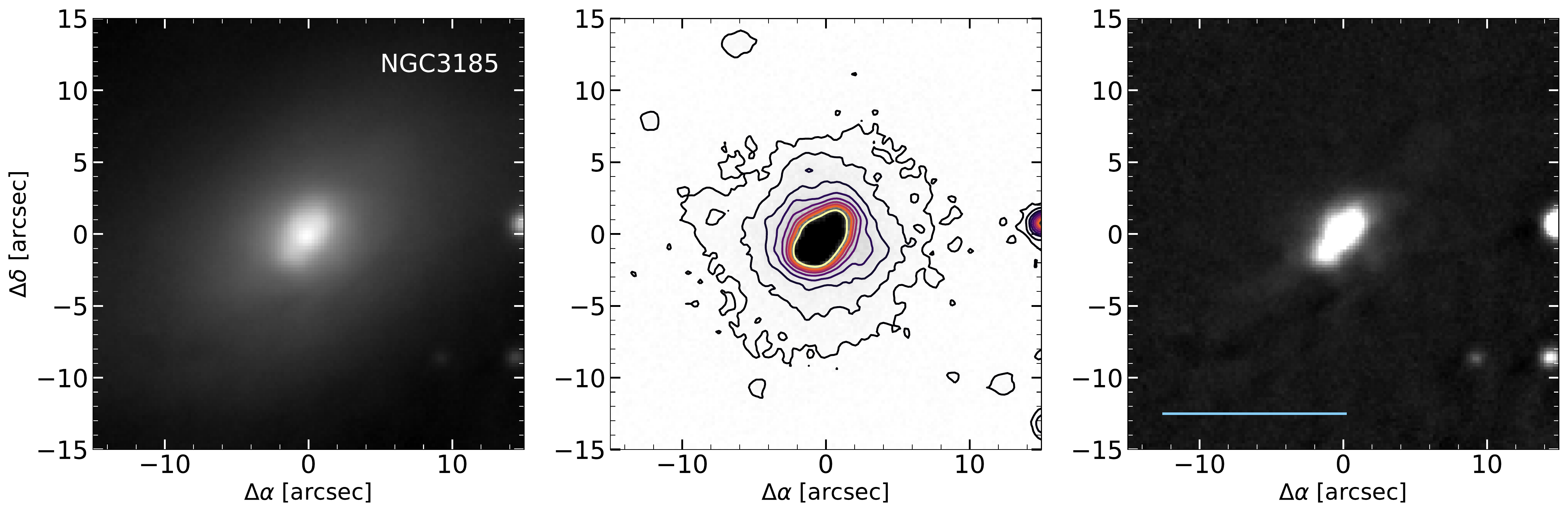}
   \includegraphics[width=\textwidth]{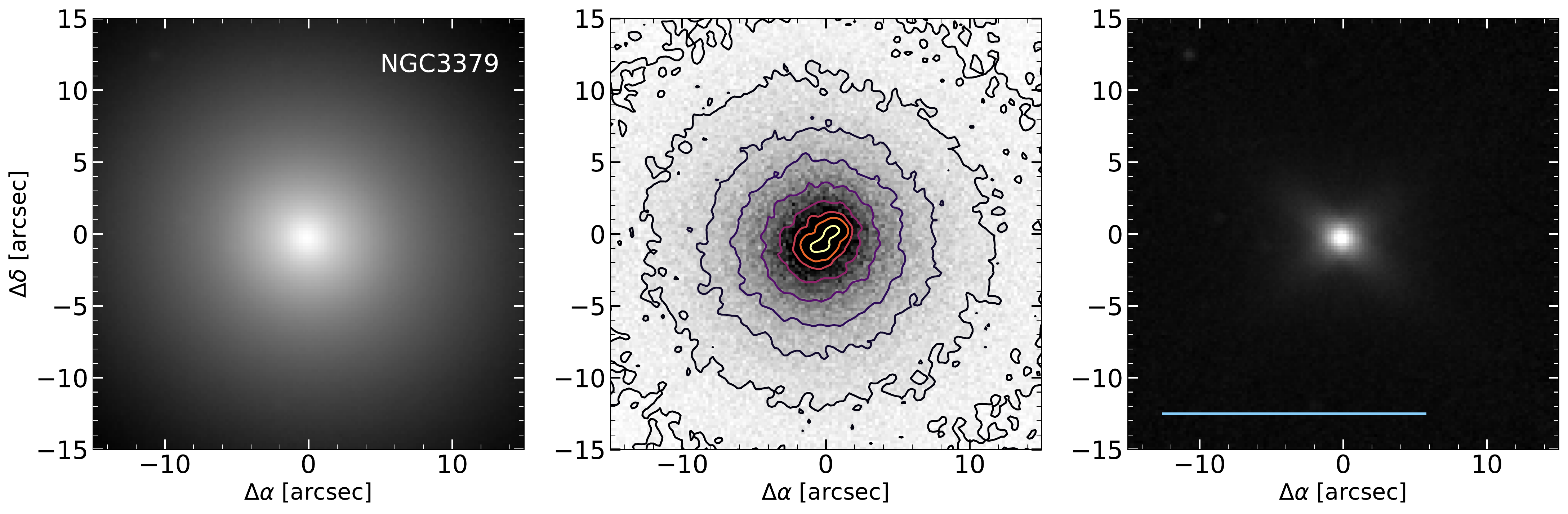}
   \includegraphics[width=\textwidth]{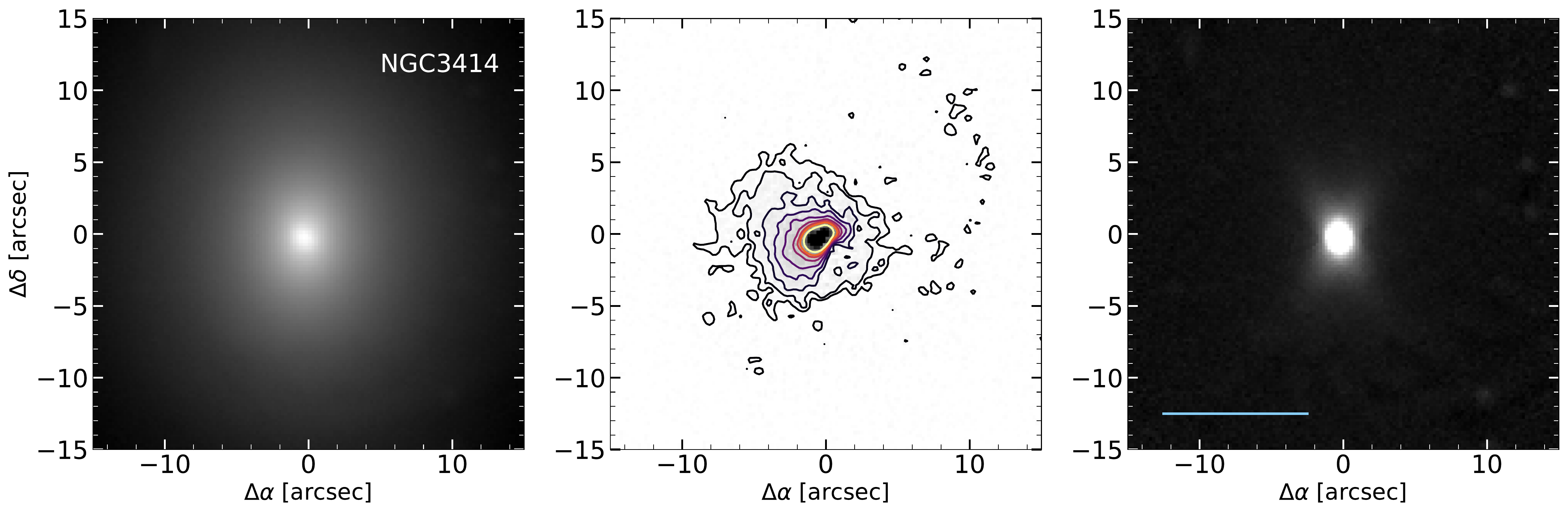}
   \includegraphics[width=\textwidth]{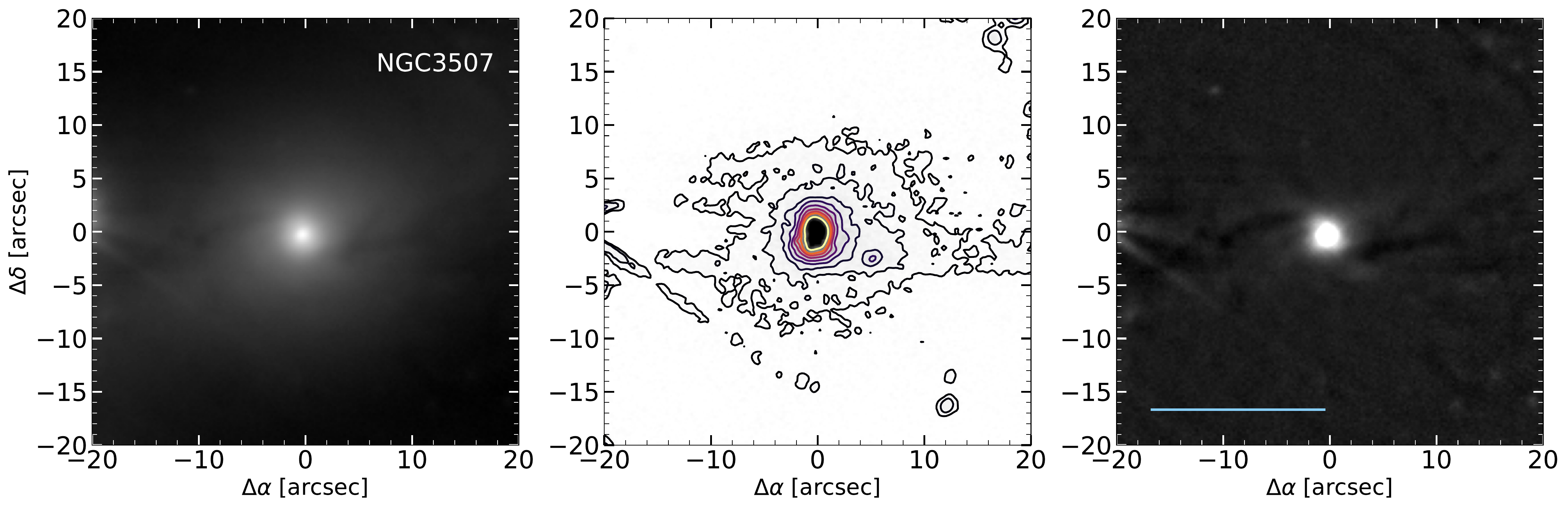}   
   \caption{NGC\,3185, NGC\,3379, NGC\,3414 and NGC\,3507 H$\alpha$ emission. The complete description is in Fig.~\ref{Figure_AB1}.}  
   \label{Figure_AB2}
   \end{figure*}

\noindent \textit{NGC\,3185} (see Fig.~\ref{Figure_AB2}): We have classified the ionised gas emission of this galaxy as \lq Disky\rq, with a clear unresolved nuclear structure that was proposed to be a ring in several works \citep[][]{James2016,Laurikainen2017}, coincident with the radio emission in \cite{Chiaraluce2019}. In these works the nature of the ring is ascribed to a bar that is driven star formation in the nucleus. \cite{Laurikainen2017} also supports that view, finding an \lq X-shape\rq\, in their image, usually present in barred galaxies. On the contrary, \cite{DiazGarcia2020} defines the nuclear enhanced star forming region not as a ring, but as a circumnuclear starburst for NGC\,3185. \\

\noindent \textit{NGC\,3379} (see Fig.~\ref{Figure_AB2}): We classified the ionised gas in this galaxy as \lq Bubble\rq, as we found evidence for extended H$\alpha$ emission with an elongated shape in the innermost parts of the image ($\sim$4\arcsec). This substructure (PA$\sim$41\degr) is spatially coincident with a nuclear dust lane at a PA$\sim$50\degr \citep{Masegosa2011}. \cite{Watkins2014} also found extended emission when studying this galaxy. The H$\alpha$ emission shows no particular asymmetric structures and no evidences of any previous merger interactions. When they apply a elliptical fitting to the brightness profile, their residuals show an \lq X\rq-like shape, just as seen in our SD image (see Fig.~\ref{Figure_AB2}), usually present in barred systems. \cite{Trinchieri2008} discovered the presence of outflowing emission near the centre with a width of $\sim$\,800 pc (larger than the apparent width in our H$\alpha$ image $\sim$300 pc) with deep soft X-ray Chandra observations. Using kinematic data, \cite{Shapiro2006} obtained information within the SAURON sample of the stellar and gaseous component of this galaxy. They found a regular rotating gas disc and also evidences of non-rotational motions. \\

\noindent \textit{NGC\,3414} (see Fig.~\ref{Figure_AB2}): This is a peculiar S0 galaxy, with a faint disc over a prominent bulge \citep{Koopmann2006}. In our images, the H$\alpha$ morphology is classified as an \lq Bubble\rq. Our results show the nuclear emission with a non-symmetric shape that could be due to the presence of dust in the south-west direction, as previously reported. Specifically, the analysis in the R band with HST data done by \cite{Rest2001} suggest the presence of dust in the nucleus. This is also confirmed by \cite{Koopmann2006}, as they found centrally concentrated ionised gas distorted by the presence of dust. \cite{Sarzi2006} studied this galaxy using SAURON data, finding a spiral pattern in the gas distribution, that was generally rotating perpendicularly to the stellar component. \\

\noindent \textit{NGC\,3507} (see Fig.~\ref{Figure_AB2}): We have classified the ionised gas morphology of this galaxy as \lq Disky\rq. Previous H$\alpha$ imaging from the literature \citep{SanchezGallego2012} trace the large scale ionised gas distribution that is also seen in our images, but with worse seeing (theirs 1.3\arcsec\ vs ours 1.0\arcsec). The CO map from the PHANGS-ALMA survey, which provides the CO(2-1) line emission at about 1\arcsec\ resolution \citep{Leroy2021}, traces molecular gas along the dust-lanes seen in the sharp-divided images, broadly resembling the ionised gas emission. A detailed analysis of the CO kinematics would reveal whether the atomic gas is participating from the putative outflow indicated by the H$\alpha$ morphology. \\

\begin{figure*}
   \includegraphics[width=\textwidth]{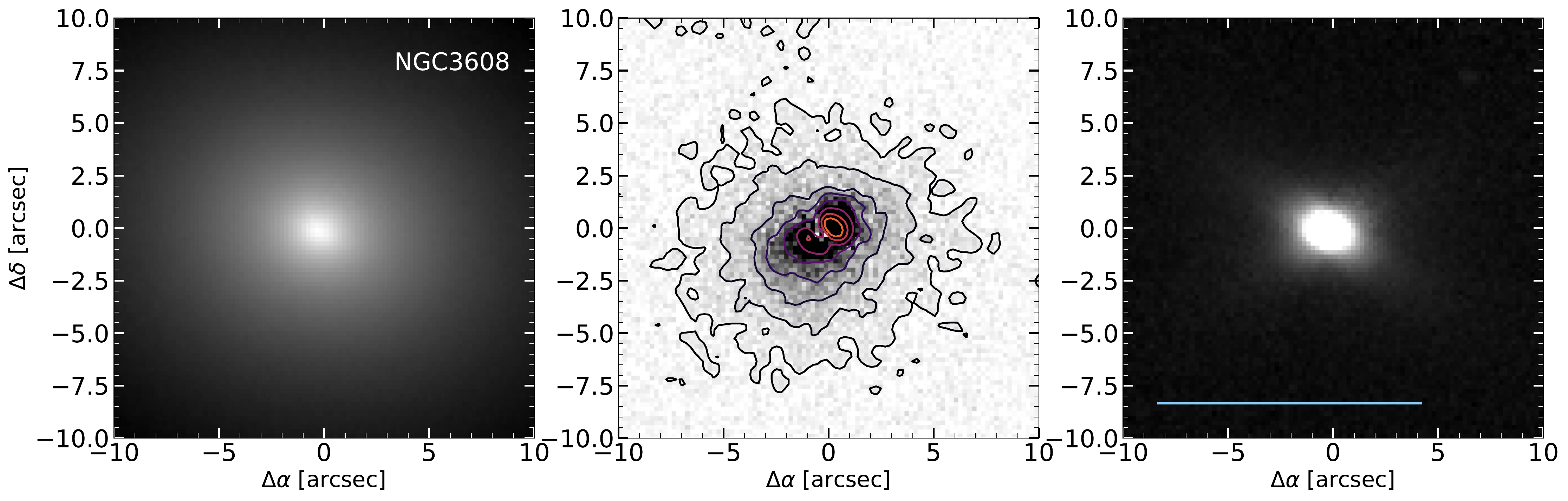}
   \includegraphics[width=\textwidth]{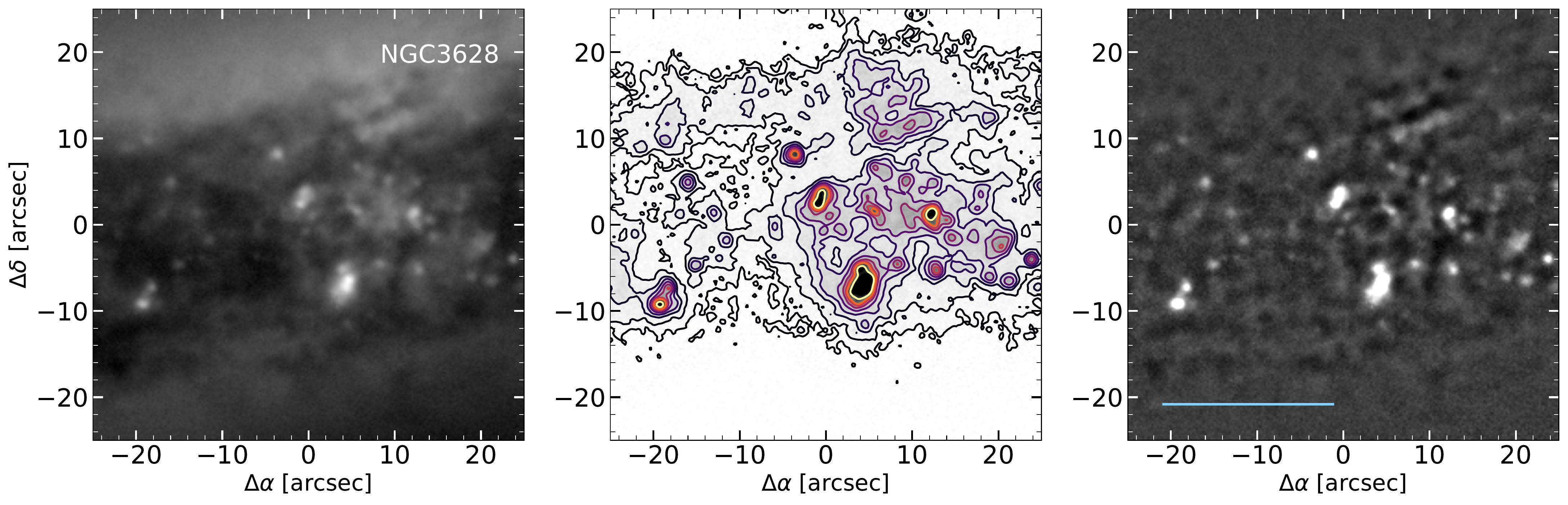}   
   \includegraphics[width=\textwidth]{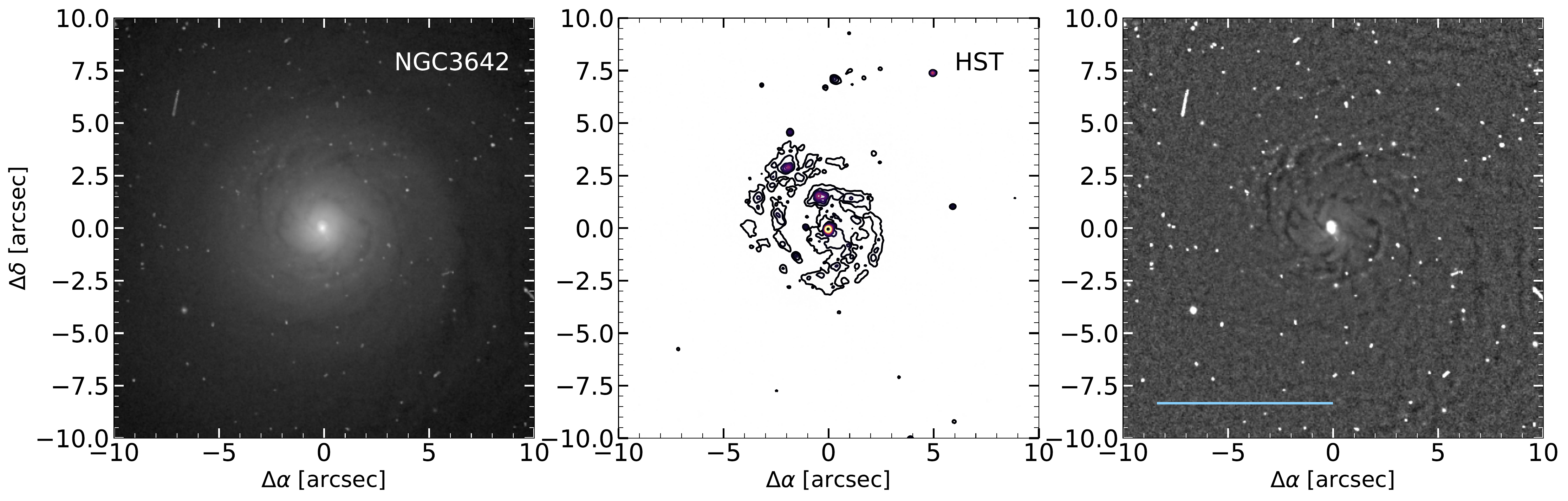}   
   \includegraphics[width=\textwidth]{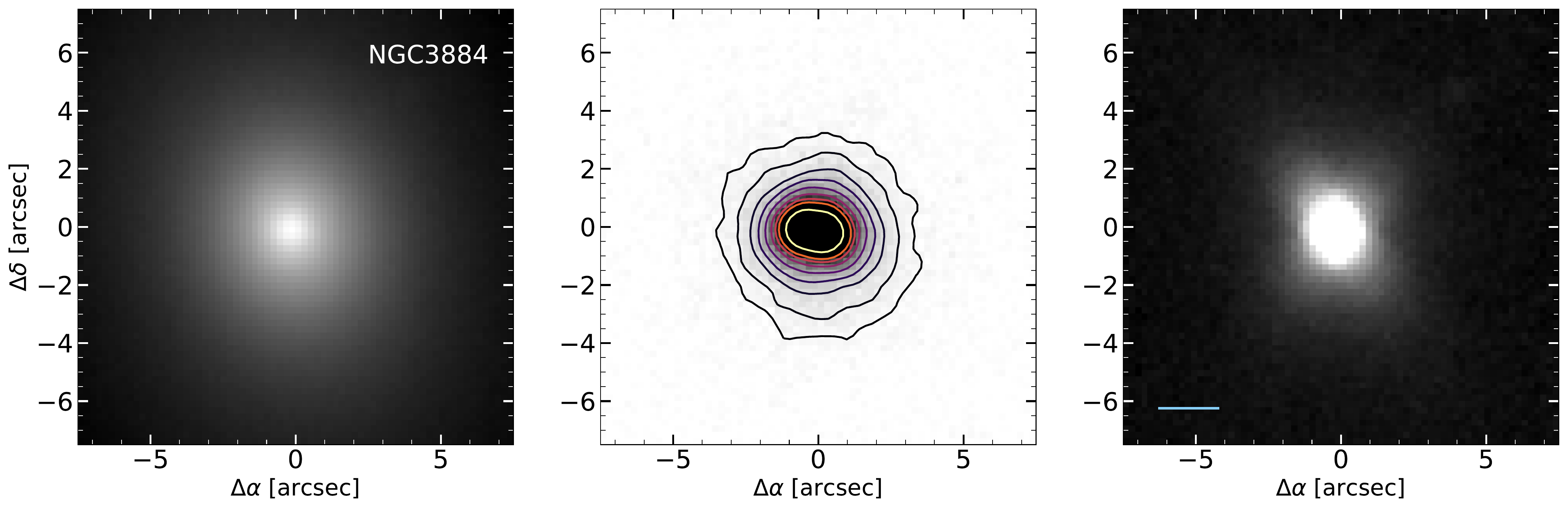}
   \caption{NGC\,3608, NGC\,3628, NGC\,3642 and NGC\,3884 H$\alpha$ emission. The complete description is in Fig.~\ref{Figure_AB1}.}  
   \label{Figure_AB3}
   \end{figure*}

\noindent \textit{NGC\,3608} (see Fig.~\ref{Figure_AB3}): We have classified the H$\alpha$ morphology of this galaxy as \lq Disky\rq. Similarly, \cite{Goudfrooij1994b} found the distribution of ionised gas to be nearly face-on and symmetric, with no evidences of dust absorption. \cite{Afanasiev2007} found evidences of the ionised gas rotating perpendicular to the stars, with complex stellar kinematics (counter rotating components). They found no evidences of outflows, although they do see a strong variation of the gas photometric and kinematic major axes at $\sim$ 1-4 arcsec from the nucleus, that would indicate the presence of polar gas rotation. They claim the existence of a inclined gaseous ring of $\sim$ 200 pc, which would be coincident with the inner contours of our H$\alpha$ image. This galaxy was observed within the ATLAS3D sample, but the weakness of the measured emission lines avoid any kinematical analysis. \\

\noindent \textit{NGC\,3628} (see Fig.~\ref{Figure_AB3}): The nuclear emission in this galaxy is classified as \lq Dusty\rq\,given the large amount of dust that prevents from seen the galactic centre. At larger scales (up to 100\arcsec) \cite{Fabbiano1990} reported for the first time the presence of an outflow in this galaxy, based on H$\alpha$ observations, reporting that it is coincident with an X-ray plume. This work was latter supported by several other studies in other wavelength bands. Specifically, \cite{Tsai2012} reported the existence of is a sub-kpc scale outflow north of the disc, detected with CO(1-0) emission, expanding at 50\,km\,s$^{-1}$ outwards of the galaxy. This was confirmed by \cite{Cicone2014} using also molecular gas observations, and they mention that this outflow is spatially coincident with a large scale plasma outflow ($\sim$ 10 kpc) that was detected with soft X-ray observations with Chandra. The outflow studied by \cite{Tsai2012} is described as being a weak bubble coming through a central and larger outflow \citep{Roy2016}. \cite{Sharp2010} observed this galaxy with the AAOmega IFS, finding filamentary emission and outflowing gas, with evidences of large-scale shocks. We also see signatures of a large scale outflow in our data, out of the dust lane (up to $\sim$80\arcsec), seen in Fig.~\ref{Figure_largescale}.\\

\noindent \textit{NGC\,3642} (see Fig.~\ref{Figure_AB3}): As in our image, \cite{Pogge2000} detect a strong nuclear source surrounded by some diffuse circumnuclear H$\alpha$ emission. We have classified this emission as \lq Disky\rq, concentrated along the spiral arms, with no outflowing signatures. \cite{Scarlata2004} describe the nuclear region of this galaxy as being elongated, probably due to a dust lane, from where dusty spiral arms emerge. The kinematic study performed by \cite{Cazzoli2018} suggest the presence of two kinematic components in the nuclear spectrum, being the broadest component a candidate signature of a nuclear outflow. \\

\noindent \textit{NGC\,3884} (see Fig.~\ref{Figure_AB3}):  The H$\alpha$ morphology of our image shows a rather elongated structure along the galaxy disc (see last row of Fig.~\ref{Figure_AB3}); we classified it as \lq Core-halo\rq. The soft X-ray emission is coincident with the ionised gas (see Fig.~\ref{Figure_Xrays}). Despite the simple distribution of the ionised gas, in \cite{Cazzoli2018} they modelled the emission lines of the ionised gas with two kinematic components. The broadest component is blueshifted and was interpreted as a possible nuclear outflow.\\

\begin{figure*}
   \includegraphics[width=\textwidth]{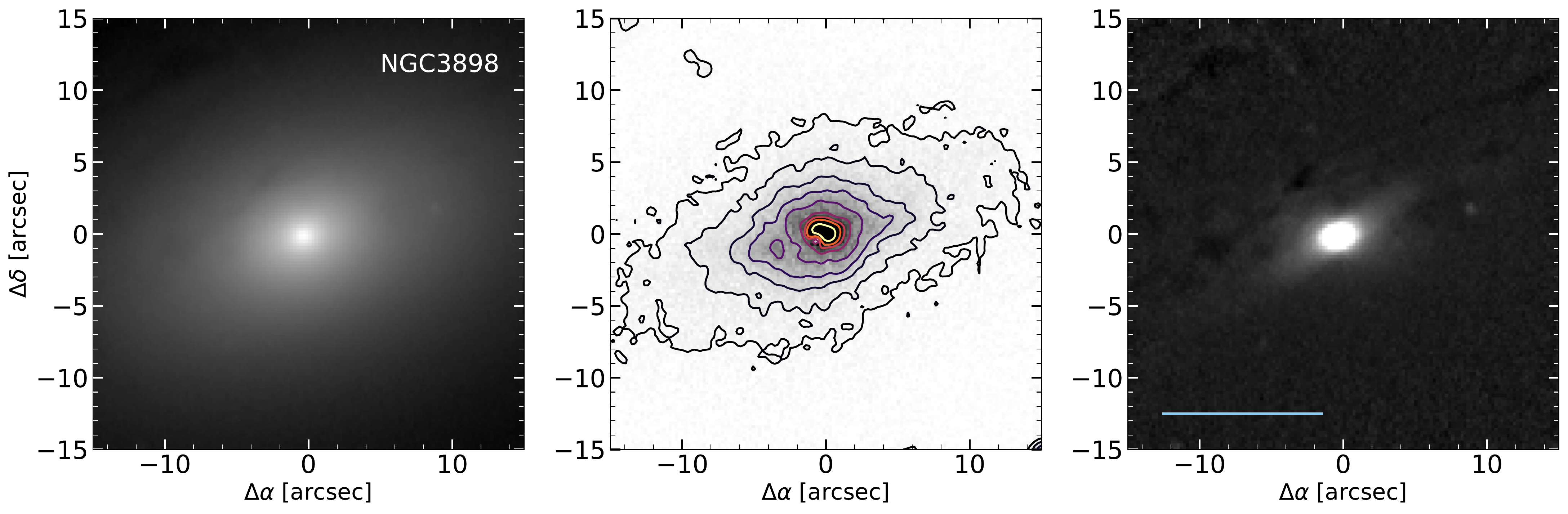}
   \includegraphics[width=\textwidth]{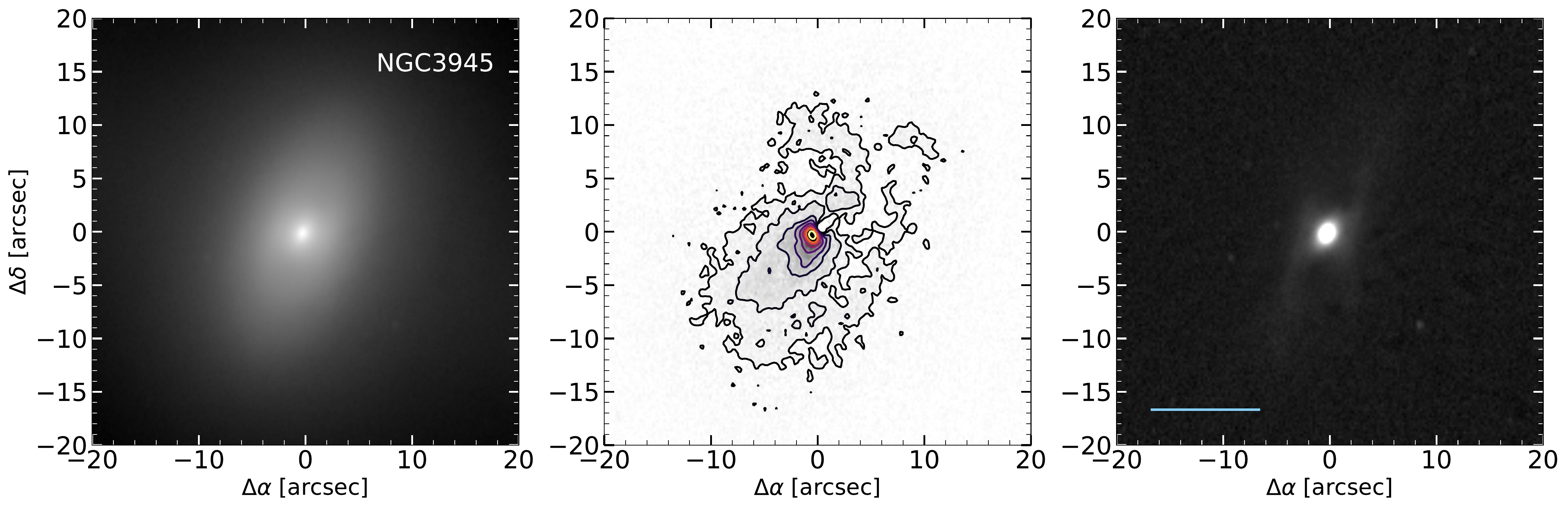}
   \includegraphics[width=\textwidth]{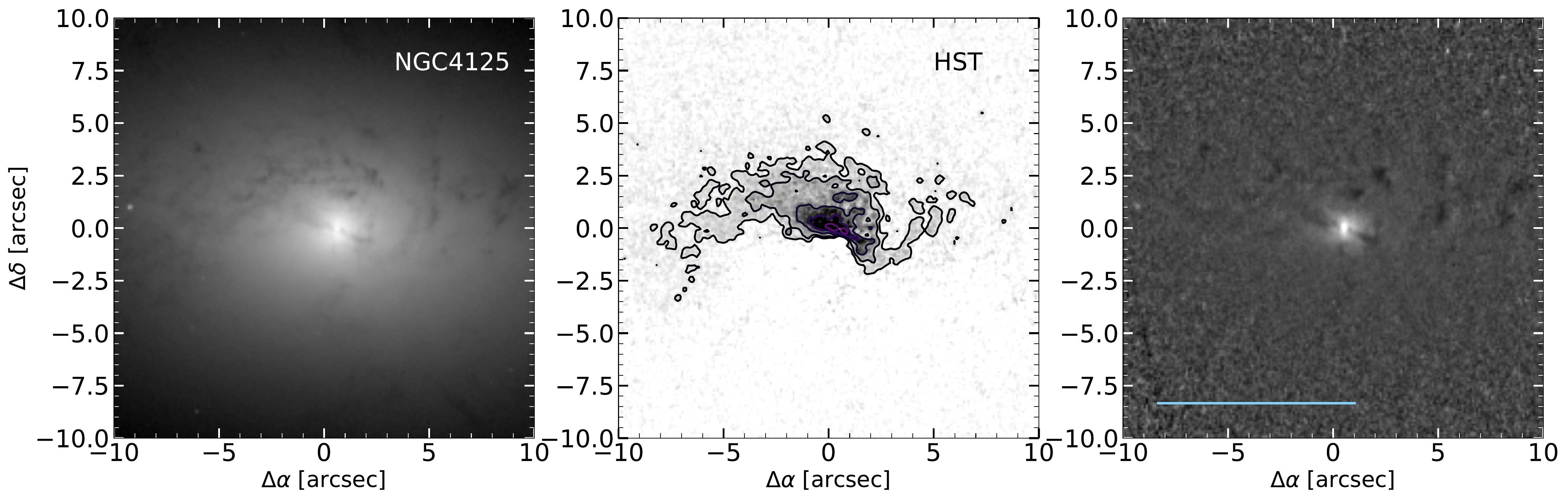}
   \includegraphics[width=\textwidth]{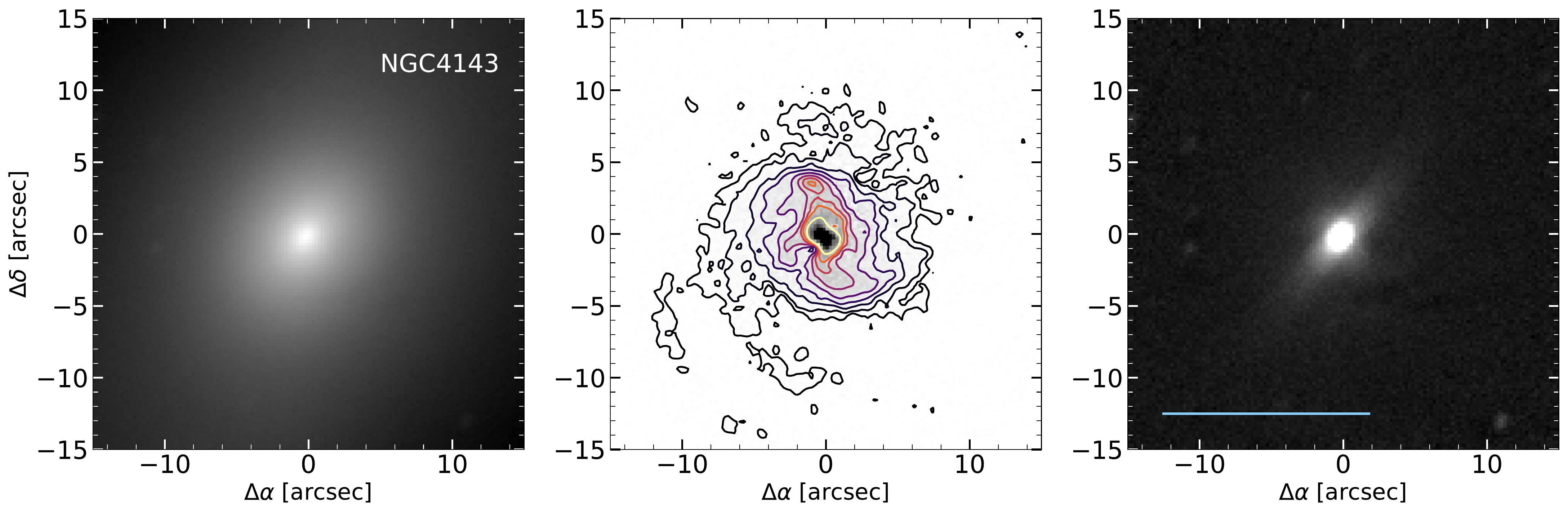}
   \caption{NGC\,3898, NGC\,3945, NGC\,4125 and NGC\,4143 H$\alpha$ emission. The complete description is in Fig.~\ref{Figure_AB1}.}  
   \label{Figure_AB4}
   \end{figure*}

\noindent \textit{NGC\,3898} (see Fig.~\ref{Figure_AB4}): In our image we find that the distribution of the H$\alpha$ gas is extended and Disky. This is consistent with previous works as \cite{Hameed2005} with data from the CCD imager of the Kitt Peak National Observatory, or \cite{Pignatelli2001}, using imaging data from the Vatican Advanced Technology Telescope and spectra from the Isaac Newton Telescope. This latter work found that that the distribution of the ionised gas is smooth within the bulge of the galaxy, but rather clumpy within the disc, due to the presence of H\,II regions. From this work, we can infer from the distribution of these regions that both the gas and stellar components are coincident in the same plane. On the contrary \cite{Carollo2002} reported an irregular nuclear emission with strong dust component with data from NICMOS/HST. They found red nuclear emission with elongated features, surrounded by a large scale disc or bar-like structure connected with a spiral feature north-east from the nucleus. \\

\noindent \textit{NGC\,3945} (see Fig.~\ref{Figure_AB4}): In our H$\alpha$ image we see clear hints of filamentary emission (up to $\sim$15\arcsec), which we have classified as \lq Bubble\rq. This emission could be co-spatial with some internal features mentioned in previous works such as a pseudo-bulge \citep[R$\sim$4\arcsec][]{Kormendy1979}, two bars \citep[R$\sim$2-3\arcsec][]{Erwin1999} or a lens which is coincident with the larger bar \citep[R$\sim$26\arcsec][]{Dullo2016}. \cite{Laurikainen2011} identified the barlens for the first time oriented along the minor axis of the galaxy, however \cite{Erwin2017} stated that instead that feature resembles to be associated to a larger inner/nuclear disc. \\

\noindent \textit{NGC\,4125} (see Fig.~\ref{Figure_AB4}): This elliptical galaxy contains a disc \citep[e.g.][]{vandenBosch1994,Goudfrooij1994a} and filamentary nuclear dust \citep[e.g.][]{Goudfrooij1994b,Braun2007}, visible in our sharp-divided image, that lead us to a \lq Dusty\rq\,morphological classification. This galaxy is not detected with radio emission\citep{Filho2002}, H\,I or CO gas \citep{Wilson2013}. The soft-X rays are much more extended and not correlated with the ionised gas, that is however correlated with the PAH emission \citep{Kaneda2011}. \cite{Pu2010} obtained the long slit spectrum of the nucleus, finding evidences of rotation along both the major and minor axes of the galaxy, and young stellar populations in the centre, which could indicate the existence of material coming from a past merger. They found the ionised gas to be oriented along the major axis of the galaxy (as we see it in our H$\alpha$ image), following non-rotational motions, no ascribed to outflows. \\

\noindent \textit{NGC\,4143} (see Fig.~\ref{Figure_AB4}): Our H$\alpha$ morphology is classified as \lq Unclear\rq, given that the image shows extended emission along the galaxy disc, but part of the gas (or even the whole emission) could be associated to a possible outflow \citep[see][]{Cazzoli2018}. The nuclear spectra of this galaxy was studied by \cite{Cazzoli2018} based on TWIN/CAHA observations. They found two kinematic components in the ionised gas, with the broadest component consistent with an outflow. \\

\begin{figure*}
   \includegraphics[width=\textwidth]{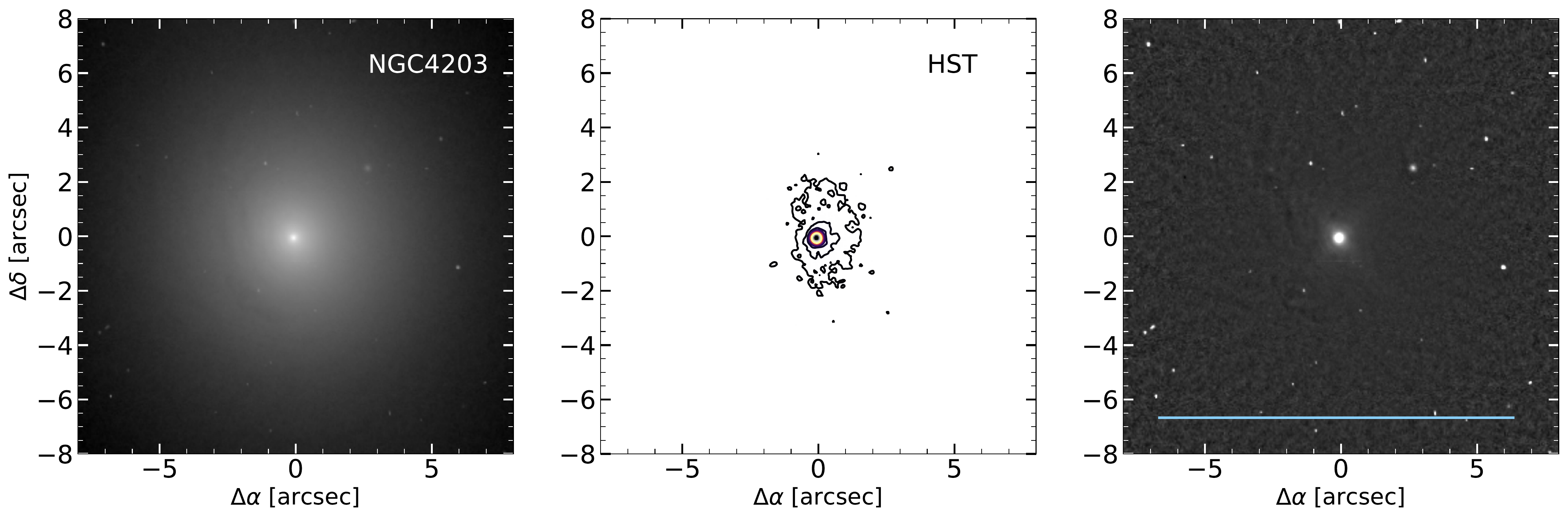}
   \includegraphics[width=\textwidth]{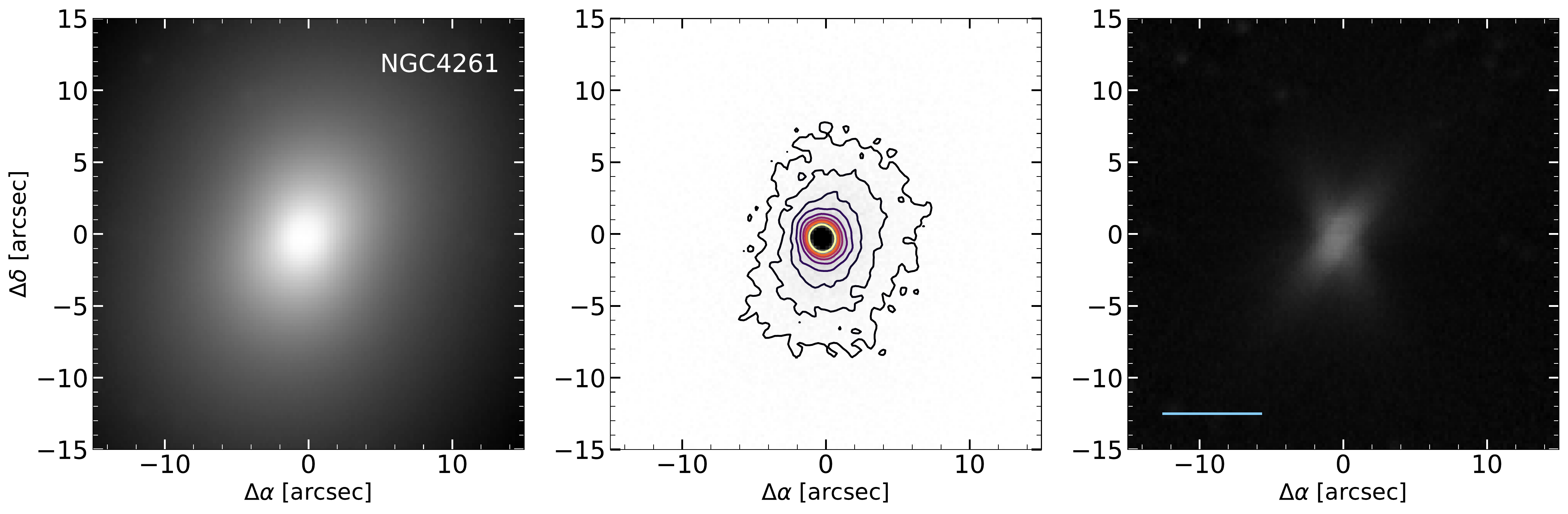}
   \includegraphics[width=\textwidth]{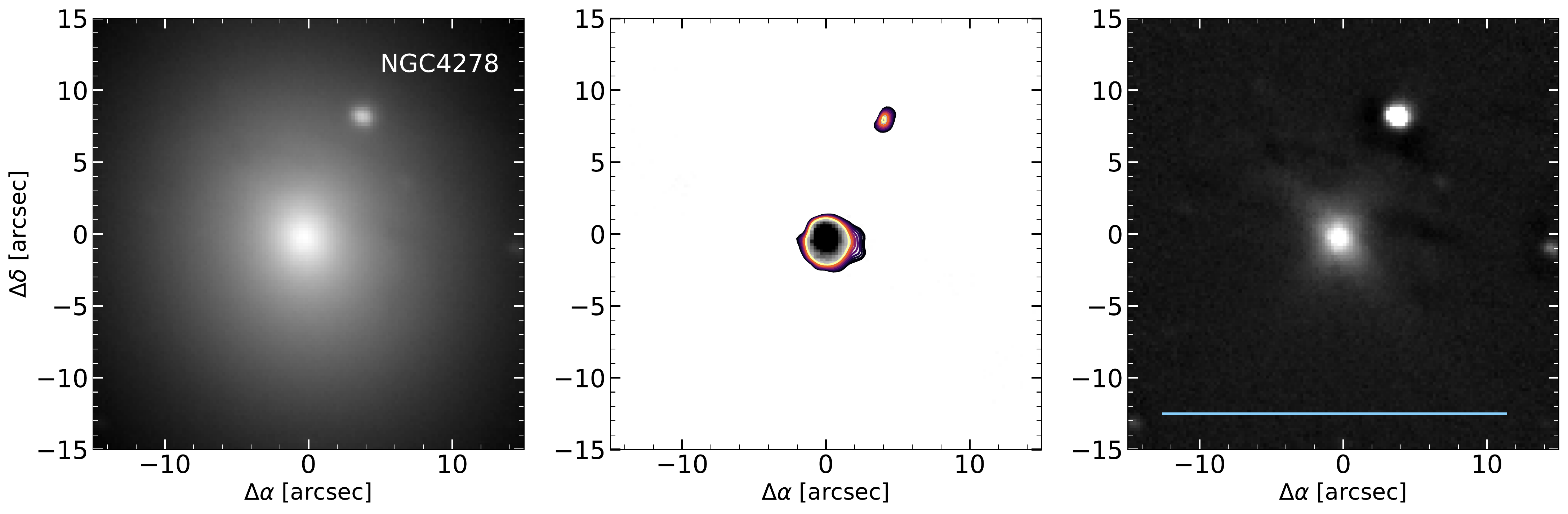}
   \includegraphics[width=\textwidth]{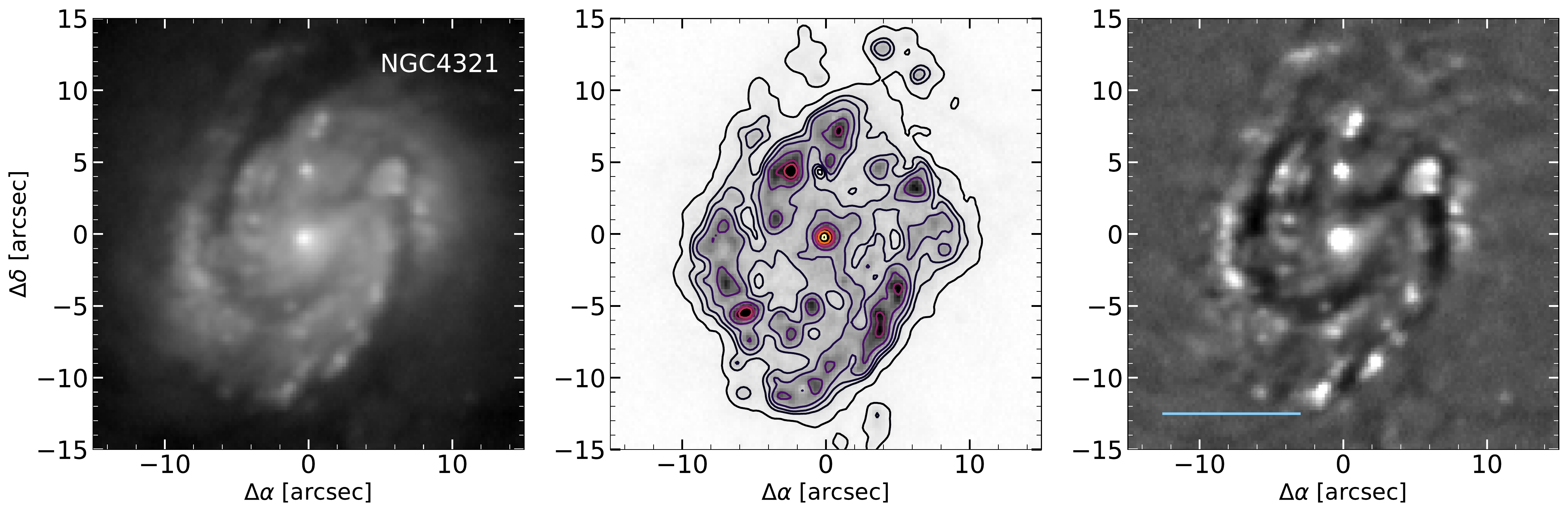}
   \caption{NGC\,4203, NGC\,4261, NGC\,4278 and NGC\,4321 H$\alpha$ emission. The complete description is in Fig.~\ref{Figure_AB1}.}  
   \label{Figure_AB5}
   \end{figure*}

\noindent \textit{NGC\,4203} (see Fig.~\ref{Figure_AB5}): The ionised gas morphology in our data set is classified as \lq Unclear\rq, as there is some extended emission that cannot be classified as coming from the disc or from outflowing emission exclusively from imaging data. \cite{Yildiz2020} characterised the object as a spiral-like, HI-rich, dusty galaxy (their Fig.~C1), and gave an extinction map oriented east-west, in contrast to the north-south orientation of the H$\alpha$ emission in our image. The nuclear spectrum of this galaxy has complicated profiles with several components \citep{Storchi2017,Cazzoli2018}. Stellar and gas kinematics from IFU spectroscopy were obtained by \cite{Boardman2017} with the Mitchell spectrograph. The derived gas kinematics, even with sparser spatial resolution, seem to indicate that non-rotational motions, associated to outflows, could be present roughly along the north-south direction. \\

\noindent \textit{NGC\,4261} (see Fig.~\ref{Figure_AB5}): The H$\alpha$ emission is concentrated near the nucleus, thus we classified it as \lq Core-halo\rq. We find a correlation in the orientation of the soft X-ray emission and the ionised gas, although the first is much more extended ($>$15\arcsec from the centre). \cite{Baldi2019} published H$\alpha$ and [O\,III] images obtained with HST, showing a notably less extended H$\alpha$ emission (maximum 3\arcsec vs our $\geq$5\arcsec) and more asymmetric (more emission to the west) than what we find. \cite{Boizelle2021} used high resolution (about 0.2\arcsec) CO observations with ALMA, finding that the molecular gas extends 2\arcsec on both sides along the north-south axis. The molecular gas kinematics can be globally reproduced by a rotating disc model.\\

\noindent \textit{NGC\,4278} (see Fig.~\ref{Figure_AB5}): This HI-rich \citep{Yildiz2020}, elliptical galaxy hosts a regular disc \citep{Sarzi2006}. The ionised gas morphology is classified here as \lq Core-halo\rq\,as it is centrally concentrated \citep[see also ][]{Masegosa2011}. In \cite{Cazzoli2018} the emission lines of the nuclear spectrum show two kinematic components, one of them consistent with an inflow. \\

\noindent \textit{NGC\,4321} (see Fig.~\ref{Figure_AB5}): This galaxy has been studied in different wavelength bands. Specifically, the ionised gas emission \citep[e.g.][]{Arsenault1990,Cepa1990,Knapen2000}, the molecular gas \citep[e.g.][]{Sakamoto1995,GarciaBurillo1998} and the HI emission, which is coincident with the optical disc \citep{Knapen1993}. Our H$\alpha$ morphological image is classified as \lq Disky\rq\,as it clearly follows the disc of the galaxy, the separation of the 4 spiral arms and several star forming regions \citep[also visible in molecular gas][]{GarciaBurillo1998}. The central region was analysed with SAURON IFS data \citep{Allard2005}, that reported an enhancement of the H$\beta$ emission in the ring as well as two dust lanes at the end of the bar of the galaxy. The existence of dust in the disc plane makes the definition of the spiral arms difficult \citep{Scarlata2004}. \\

\begin{figure*}
   \includegraphics[width=\textwidth]{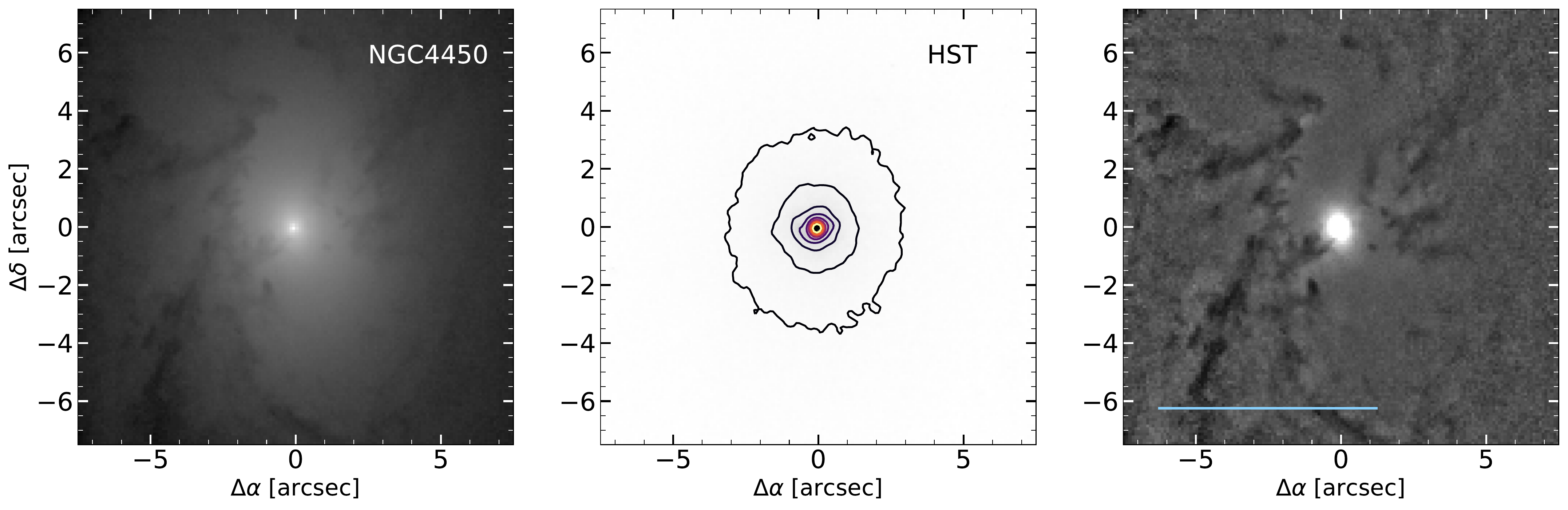}
   \includegraphics[width=\textwidth]{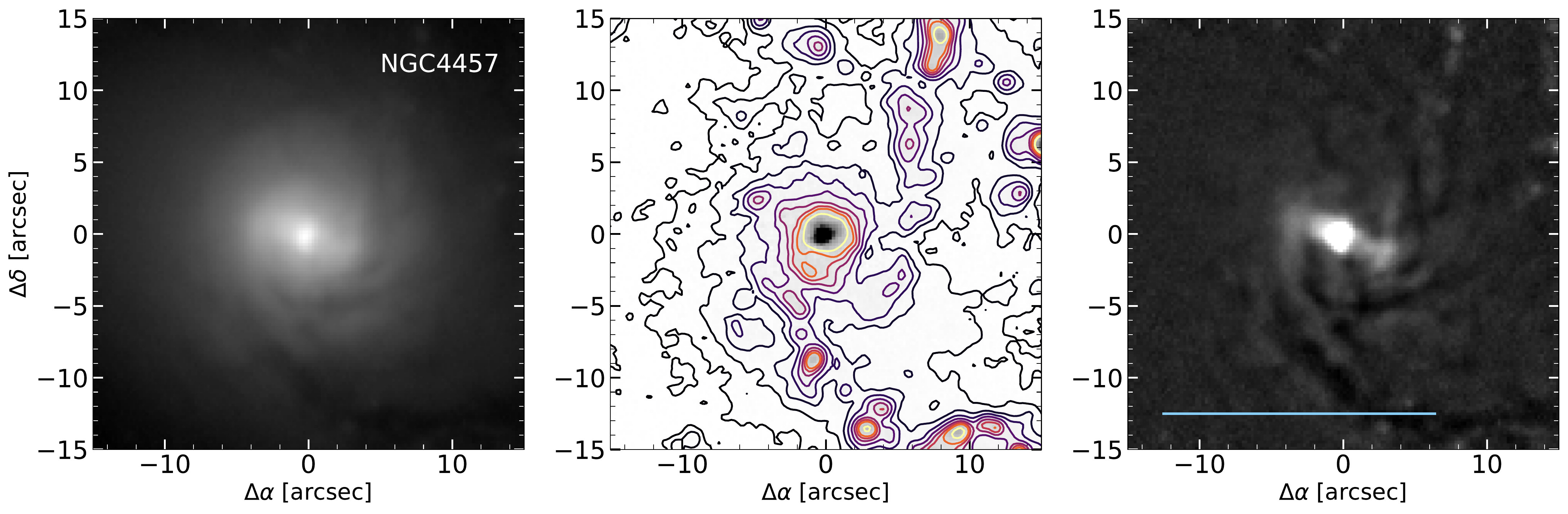}
   \includegraphics[width=\textwidth]{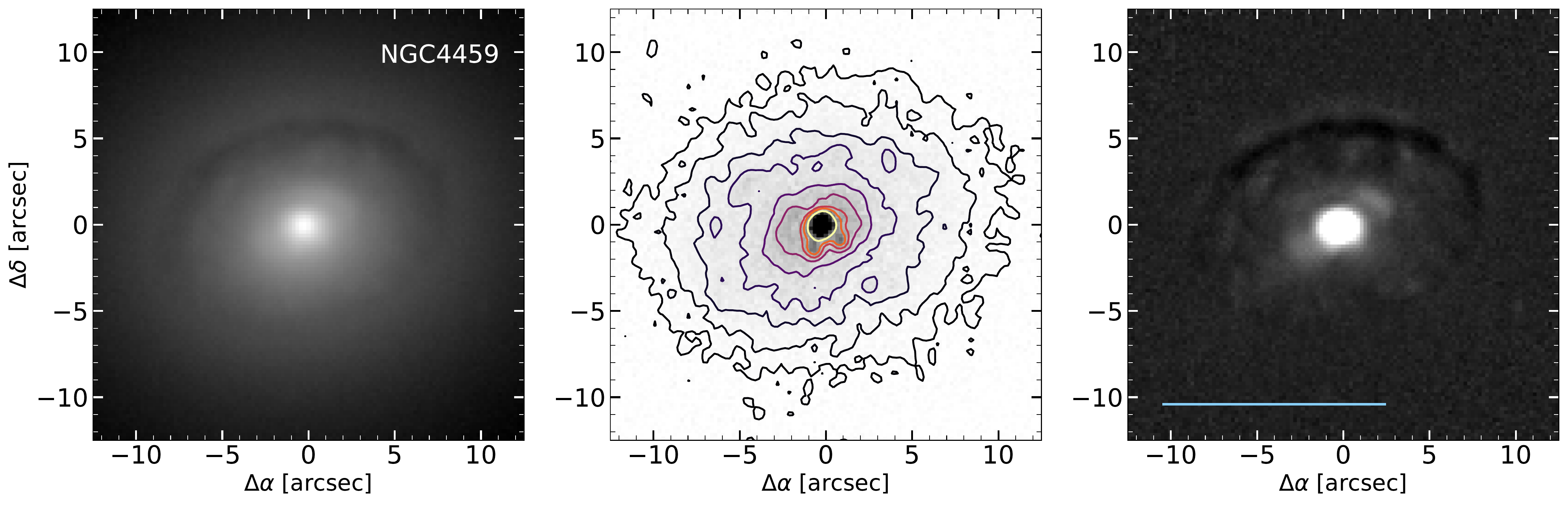}
   \includegraphics[width=\textwidth]{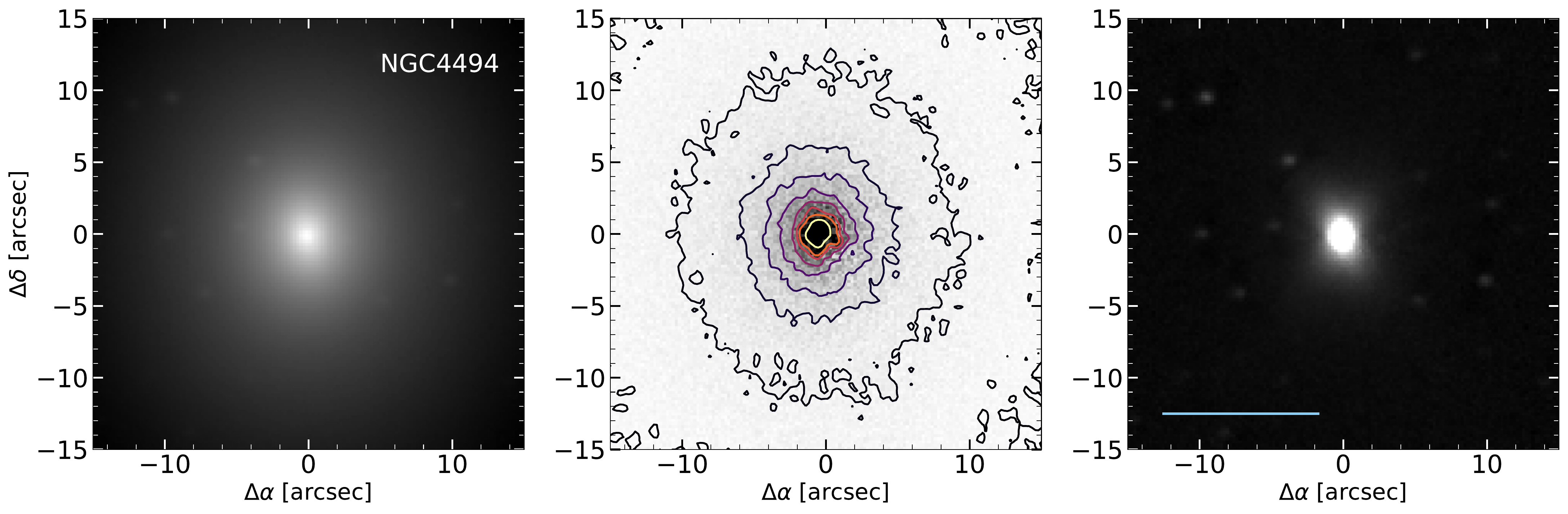}
   \caption{NGC\,4450, NGC\,4457, NGC\,4459 and NGC\,4494 H$\alpha$ emission. The complete description is in Fig.~\ref{Figure_AB1}.}  
   \label{Figure_AB6}
   \end{figure*}

\noindent \textit{NGC\,4450} (see Fig.~\ref{Figure_AB6}): This barred spiral galaxy was classified as hosting a truncated star forming disc at $\sim$60\arcsec\,from the nucleus, given the non-continuous H$\alpha$ emission at large scales \citep{Koopmann2001, Chemin2006}. In the innermost parts of the galaxy (i.e. 5\arcsec\,from the centre) we find the H$\alpha$ emission concentrated in a \lq Core-halo\rq\,morphology. This is in agreement with the image on \cite{Koopmann2001}, where the nuclear emission is not extended further out to the galaxy ($<$10\arcsec), except for some sparse regions at larger scales ($\sim$30\arcsec, i.e.$\sim$3~kpc). \cite{Cortes2015} obtained the velocity field for this galaxy, finding two plateaus at a distance $\sim$5\arcsec\,from the centre, that suggested the presence of an additional rotating component. The [O\,III] emission has an offset of 25\degr\,with respect to the main stellar component, that was ascribed to the recent acquisition of gas into the galaxy \citep{Cortes2015}. This additional component is also confirmed by \cite{Cazzoli2018}, who detected a secondary component in the forbidden lines, although it was interpreted as outflows/inflows. \\

\noindent \textit{NGC\,4457} (see Fig.~\ref{Figure_AB6}): This face-on galaxy has many star forming regions, clearly visible in our H$\alpha$ image (classified as \lq Disky\rq), found only in the inner 30\arcsec\,\citep{Cortes2015}. This galaxy is known to host a radio jet, although the PA is not reported in the literature \citep[see][and references therein]{Nemmen2014}. Both the stellar and the ionised gas components are rotating, but non-circular motions have been detected. This is seen in the gas velocity map, that is more disturbed than that of the stars, specially along the stellar kinematic minor axis \citep{Cortes2015}. In our image it is clearly visible an H$\alpha$ arm in the approaching side of the galaxy, as in previous works \citep{Chemin2006,Cortes2015}. The gas kinematic centre has the most blue-shifted velocities \citep{Chemin2006}, and its centre is not spatially coincident with the stellar kinematic centre \citep{Cortes2015}. In the work by \cite{Cortes2015}, they suggest that the arm and the velocity maps of the ionised gas component are indicative of ram pressure stripping effects. \\

\noindent \textit{NGC\,4459} (see Fig.~\ref{Figure_AB6}): This unbarred lenticular galaxy has an extended dust disc up to $\sim$8\arcsec\ from the nucleus \citep[][and references therein]{Pagotto2019}. The ionised gas morphology is classified here as \lq Bubble\rq, as it shows two small blobs south from the nucleus, in the direction of the molecular gas kinematic minor axis \citep{Young2008}. The blobs are not seen in previous kinematic maps of the ionised gas \citep[H$\beta$ line in][]{Sarzi2006}. The H$\alpha$ emission is extended up to $\sim$ 10\arcsec\, from the nucleus in the west-east direction \citep[in agreement with][]{Koopmann2001}, concentrated in a disc as the H$\beta$ emission \citep{Sarzi2006,Sarzi2010}. Various works have studied the molecular gas traced by the CO in the galaxy \citep{Combes2007,Young2008,Davis2017} which is detected along an oval shape, similar to that of the H$\alpha$, coincident with dust lanes (visible in our sharp-divided image, see third, right panel in Fig.~\ref{Figure_AB6}). The concentrated molecular gas suggests the existence of ongoing circumnuclear star formation \citep{Sarzi2006}. The kinematic information of the CO (2-1) suggest a regular rotating disc \citep{Davis2017}. \\

\noindent \textit{NGC\,4494} (see Fig.~\ref{Figure_AB6}): This elliptical galaxy is characterised by an inner edge-on dusty ring of star formation \citep{Forbes1996}. So far, our study is the first one focused on the ionised gas of this galaxy. In our image the H$\alpha$ emission shows a \lq Core-halo\rq\, morphology, extended up to 40\arcsec\, from the nucleus (with a 3$\sigma$-detection). \cite{Foster2011} found a double structure in the global kinematics derived from individual globular clusters, which they ascribe to a recent gas-rich merger, although we do not see any feature of this in the ionised gas. The stellar kinematics was further studied within the ATLAS3D survey \citep{Krajnovic2015} and the SLUGGS survey \citep{Brodie2014}, in which they found a regular rotation pattern, particularly flat in the outer parts of the galaxy.\\ 

\begin{figure*}
   \includegraphics[width=\textwidth]{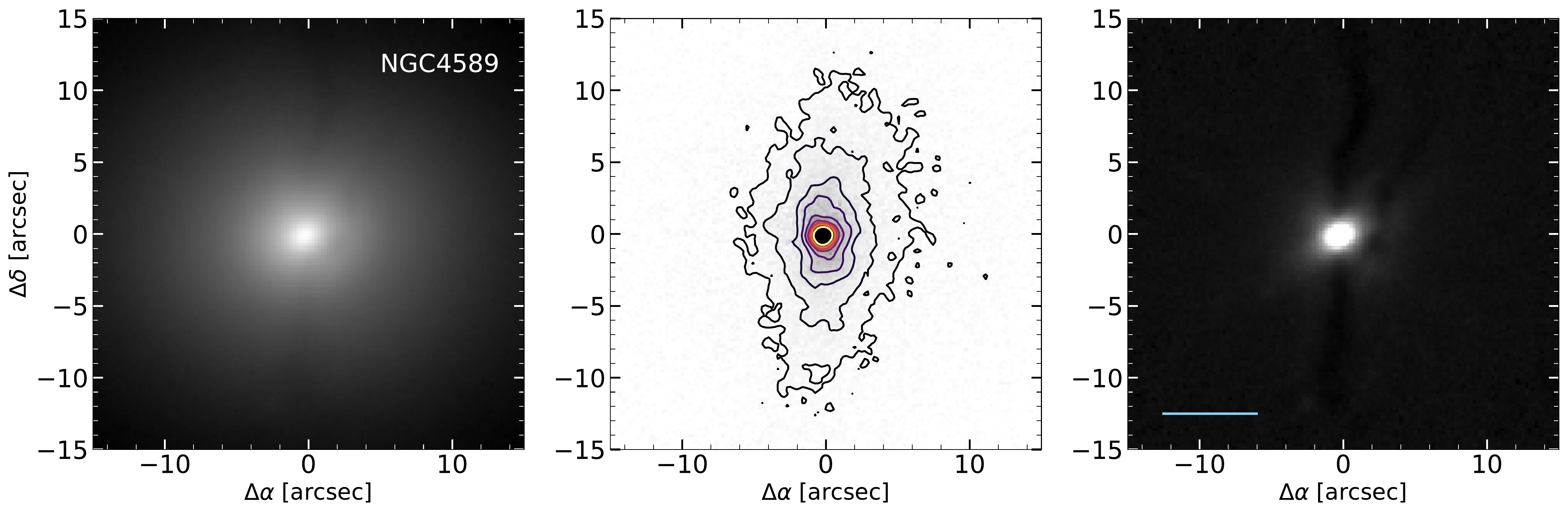}
   \includegraphics[width=\textwidth]{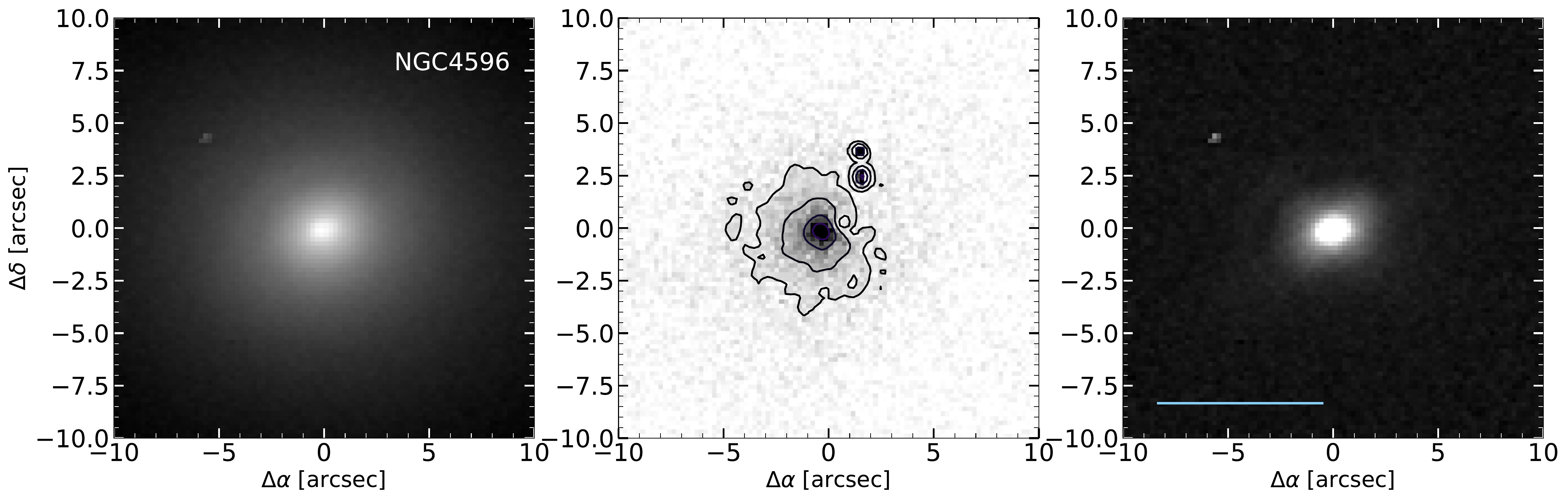}
   \includegraphics[width=\textwidth]{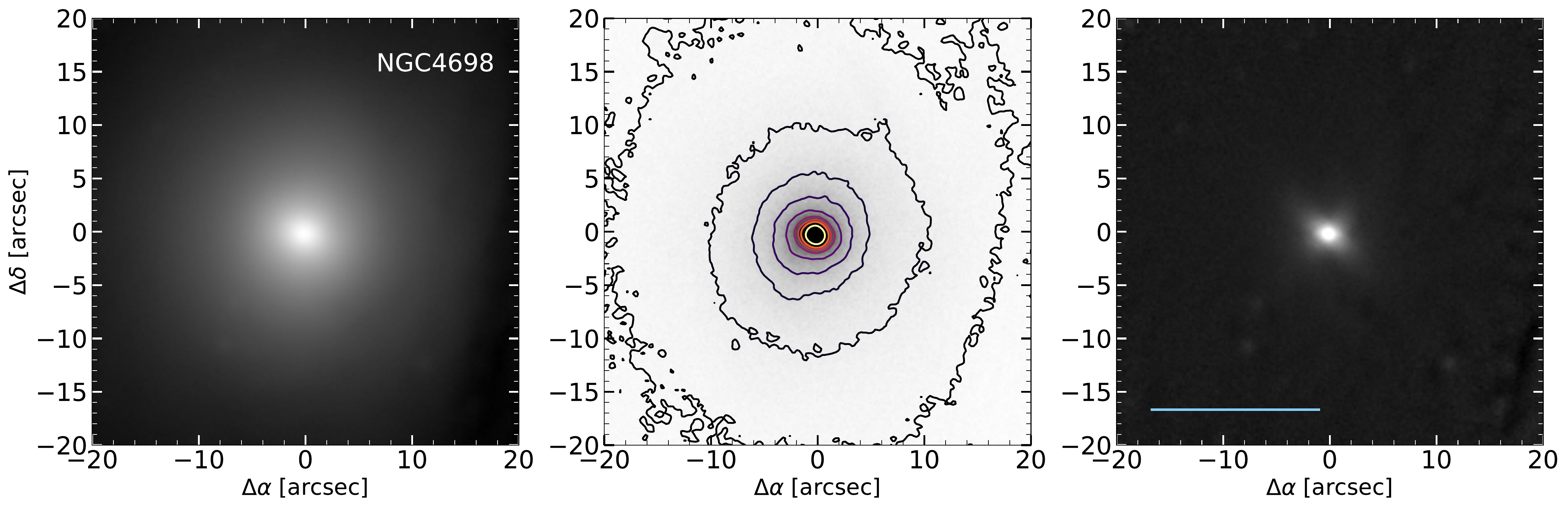}
   \includegraphics[width=\textwidth]{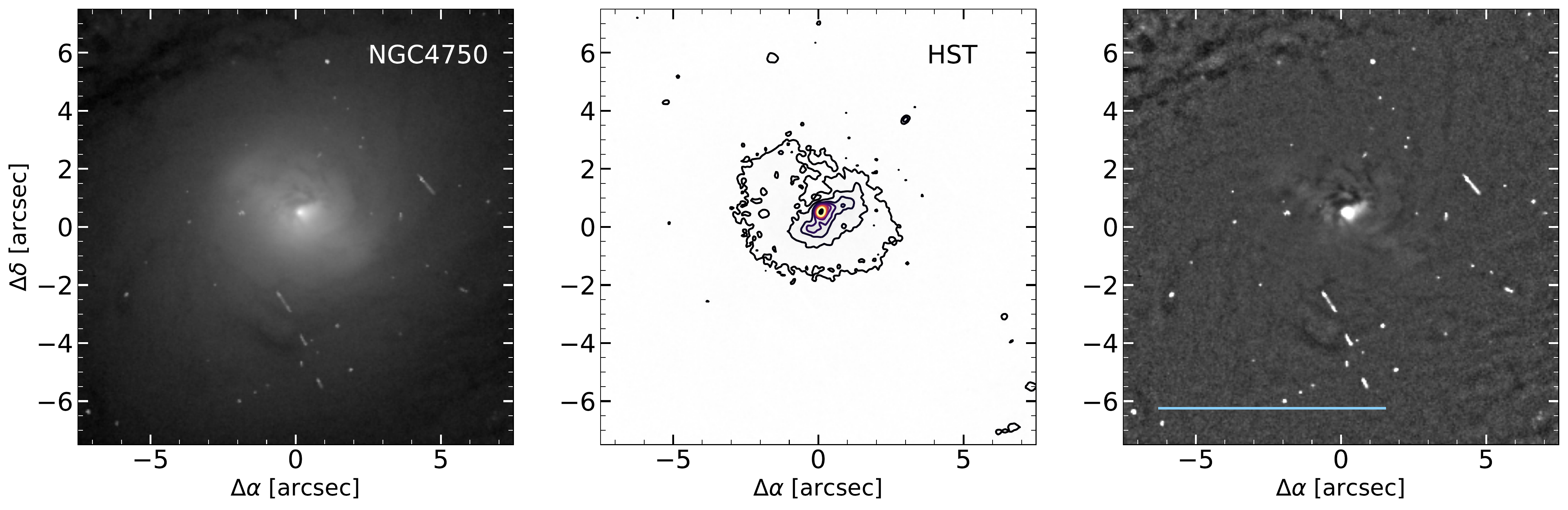}
   \caption{NGC\,4589, NGC\,4596, NGC\,4698 and NGC\,4750 H$\alpha$ emission. The complete description is in Fig.~\ref{Figure_AB1}.}  
   \label{Figure_AB7}
   \end{figure*}

\noindent \textit{NGC\,4589} (see Fig.~\ref{Figure_AB7}): The ionised gas in this elliptical galaxy shows a \lq Core-halo\rq\, morphology with a dust lane along the minor axis also seen in previous works \citep{Moellenhoff1989,Goudfrooij1994b}. \cite{Moellenhoff1989} performed a spectroscopic study of the object that revealed complex gas and stellar kinematics, with gas moving along several position angles. They associated these motions to a previous merger. This latter event may also have caused the minor-axis dust lane by accretion of external material. The ionised gas is co-spatial with the dust lane as it already settled down, which is consistent with being produced in an old merger \citep{Moellenhoff1989}. \\

\noindent \textit{NGC\,4596} (see Fig.~\ref{Figure_AB7}): It is a strong barred galaxy \citep{Kent1990, Gerssen1999,Laurikainen2017} with a faint dust spiral and a compact source in the nucleus \citep{GD2008}. Our image of the H$\alpha$ emission reveals an outflow-like morphology, with an asymmetric profile with respect to the nucleus. \cite{FalconBarroso2006} studied this galaxy within the SAURON sample \citep{Bacon2001}, mentioning the misalignment in the kinematic and photometric axes of this galaxy, due to a bar. The stellar and gaseous component are aligned, both with a regular rotation pattern. In the SAURON map, the H$\beta$ gas is centrally concentrated \citep{FalconBarroso2006}, similar to what we see for H$\alpha$ in our image. \\

\noindent \textit{NGC\,4698} (see Fig.~\ref{Figure_AB7}): Our H$\alpha$ image has a morphology that we classified as \lq Core-halo\rq. \cite{ErrorFerrez2013} included this object in their sample, where they also obtained H$\alpha$ imaging with ALFOSC/NOT. They were only interested in the double-ring structure seen at large scales, hence the comparison of the nuclear region with the published H$\alpha$ image is not possible. 
\cite{Bertola1999} reported that the nuclear disc of gas and stars are rotating perpendicularly with respect to the galaxy main disc, most probably as a result of acquisition of external gas \citep{Corsini2012}. \cite{Cortes2015} spatially studied resolved stellar and ionised gas kinematics, and suggested that this Virgo cluster galaxy is the product of an ancient merger. They also concluded that the ionised gas kinematics (traced with [O\,III] in the central 5\arcsec) is non-planar. In \cite{HM2020} the line profiles are narrow ($\sigma\sim$90 kms$^{-1}$) for both space- and ground-based spectroscopic data. \\

\noindent \textit{NGC\,4750} (see Fig.~\ref{Figure_AB7}): The morphology of the H$\alpha$ gas of this galaxy, \lq Bubble\rq, is correlated with the soft X-ray emission (see Sect.~\ref{SubSec_DiscXray}). \cite{Carollo2002} reported the presence of spiral features emerging from the nucleus, stronger in the north-east direction, using NICMOS/HST images. They found evidence of strong dust features and star formation. In our data, specially in the sharp-divided image, we also detect the spiral arms and the presence of dust. However, the gas seems to be outflowing from the nucleus rather than being distributed along the spiral arms, and we do not detect star-forming regions. In \cite{Cazzoli2018} the nuclear spectrum show features of an outflow (broad component in forbidden lines and narrow H$\alpha$). Indeed, the slit is oriented towards the outflow detected by our imaging data (PA = 231\degr). \\

\begin{figure*}
   \includegraphics[width=\textwidth]{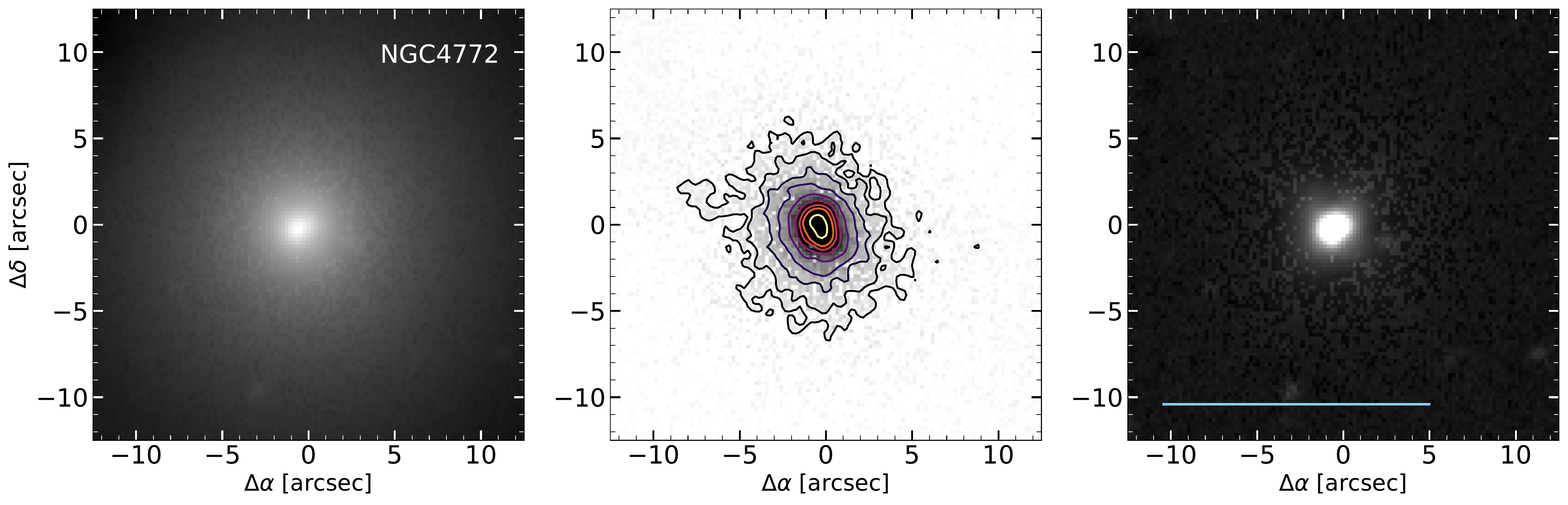}
   \includegraphics[width=\textwidth]{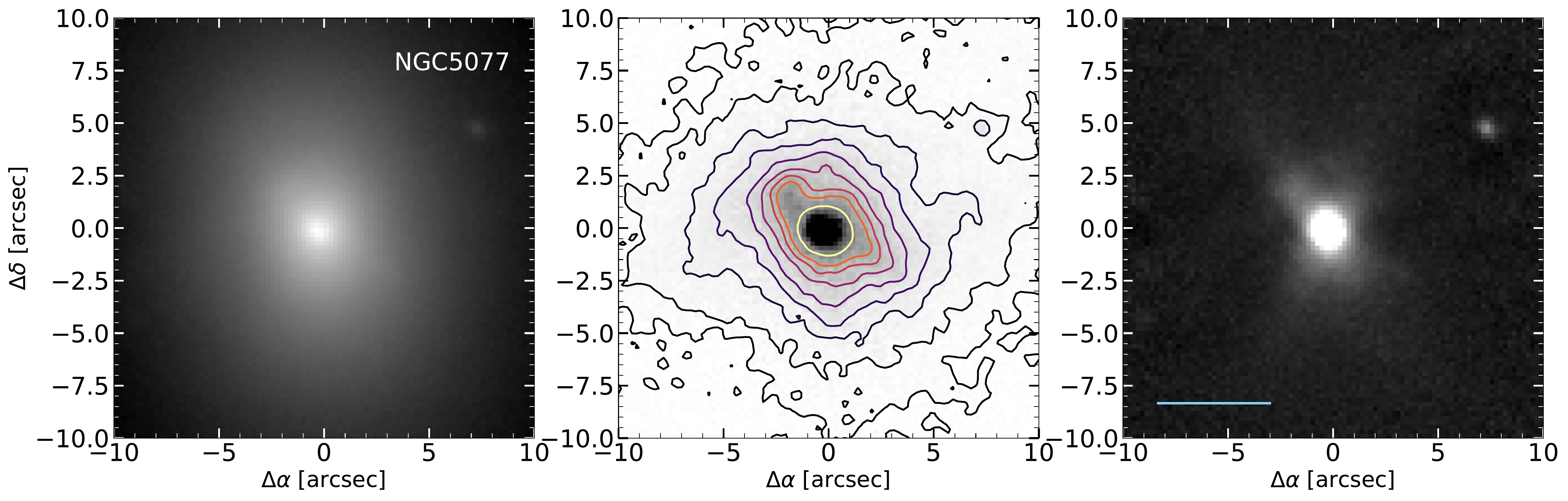}
   \includegraphics[width=\textwidth]{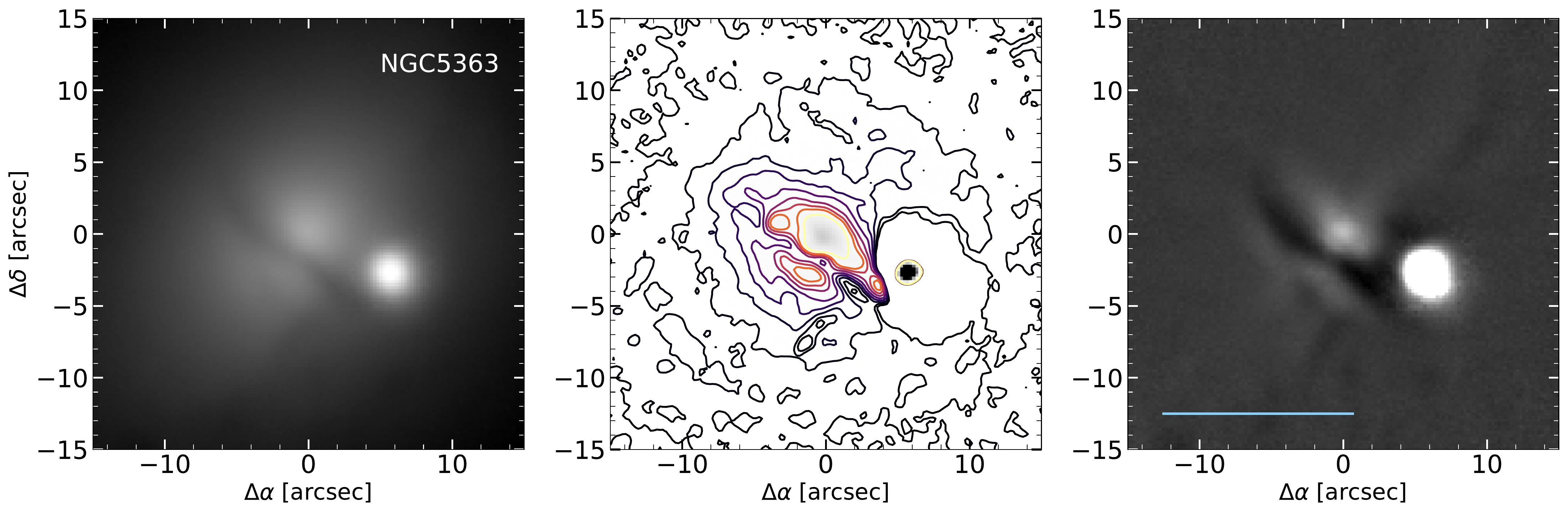}
   \includegraphics[width=\textwidth]{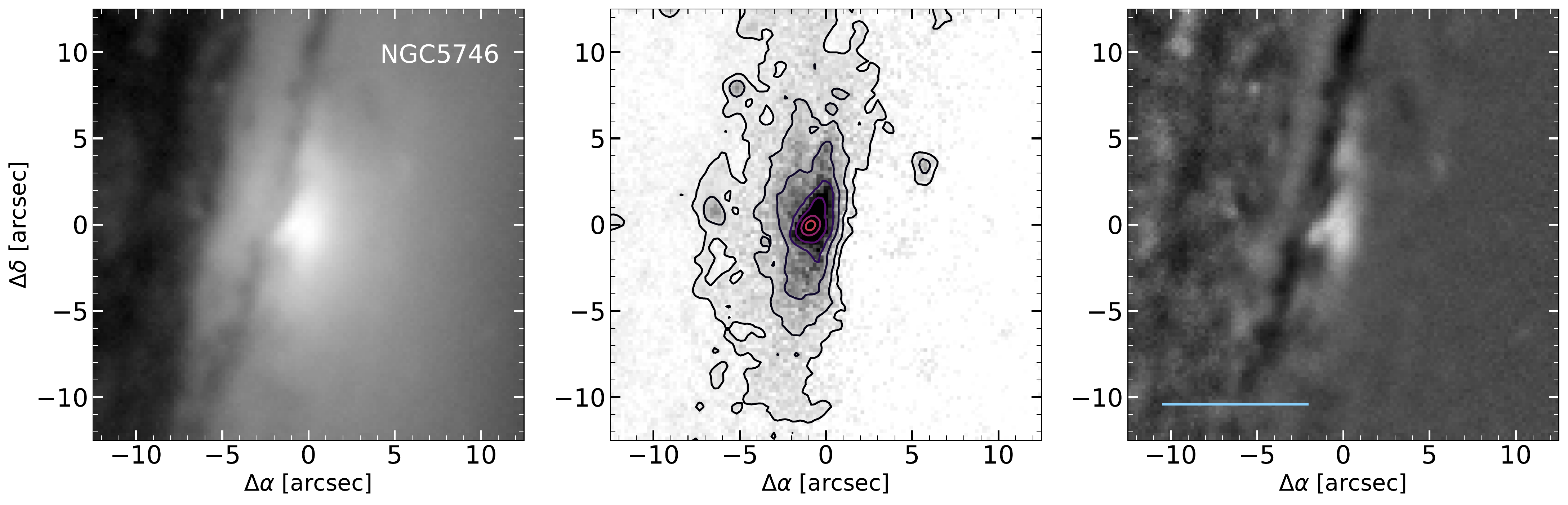}
   \caption{NGC\,4772, NGC\,5077, NGC\,5363 and NGC\,5746 H$\alpha$ emission. The complete description is in Fig.~\ref{Figure_AB1}.}  
   \label{Figure_AB8}
   \end{figure*}

\noindent \textit{NGC\,4772} (see Fig.~\ref{Figure_AB8}): The H$\alpha$ emission of this galaxy is classified here as \lq Core-halo\rq. \cite{Haynes2000} reported a centrally peaked emission surrounded by diffuse gas affected by the presence of a dust lane, coincident with HI gas. They suggest that this galaxy may have gone through a merger that spread out the gas through the disc. \citet{FalconBarroso2006} detected ionised gas with SAURON data that was oriented in the north-west and south-west direction, possibly coming from a ring out of the main galactic plane. Although there could be other interpretations as the main stellar and gas components of this galaxy are known to co-rotate. In our image we detect a nuclear source both in the ionised and soft X-ray emissions, that is diffuse and extended for H$\alpha$ in the same direction as in \cite{FalconBarroso2006}. \\

\noindent \textit{NGC\,5077} (see Fig.~\ref{Figure_AB8}): In our image the H$\alpha$ gas seems to follow the galactic spiral arms or disc; thus we classified it as a \lq Disky\rq\,morphology. The analysis of MUSE IFS data by \cite{Raimundo2021} confirmed the presence of a stellar distinct, counter-rotating core with complex gas dynamics. In fact, our H$\alpha$ distributions are very similar. She reports the discovery of a nuclear outflow, consistent with a hollow cone intersecting the plane of the sky. \\

\noindent \textit{NGC\,5363} (see Fig.~\ref{Figure_AB8}): The galaxy is classified as I0 in NED database, although the images may contradict this classification due to an inner spiral structure \citep{Finkelman2010}. In our H$\alpha$ image this internal structure is seen, as in \cite{Finkelman2010}, although is not detected in the R broad-band image. This galaxy shows an internal dust lane that obscures the nuclear emission, and extends further out from the nucleus \citep{Finkelman2010}. No outflow/inflow signatures are seen neither in the sharp-divided image nor in the pure H$\alpha$ image in the nuclear region. However, at larger scales in our image there is extended emission, whose nature is not clear. This lead to our classification as \lq Unclear\rq. \\

\noindent \textit{NGC\,5746} (see Fig.~\ref{Figure_AB8}): This is an edge-on, quiescent spiral galaxy, with no evidences of recent mergers \citep{Barentine2012,Mosenkov2020}. Our image shows a heavily obscured nucleus due to a dust lane that prevents from seeing the nuclear H$\alpha$ emission. Integral field spectroscopy is reported only for the extraction of the stellar kinematics, from which the presence of a bar in determined \citep{Molaeinezhad2016, Peters2017}.\\

\begin{figure*}
   \includegraphics[width=\textwidth]{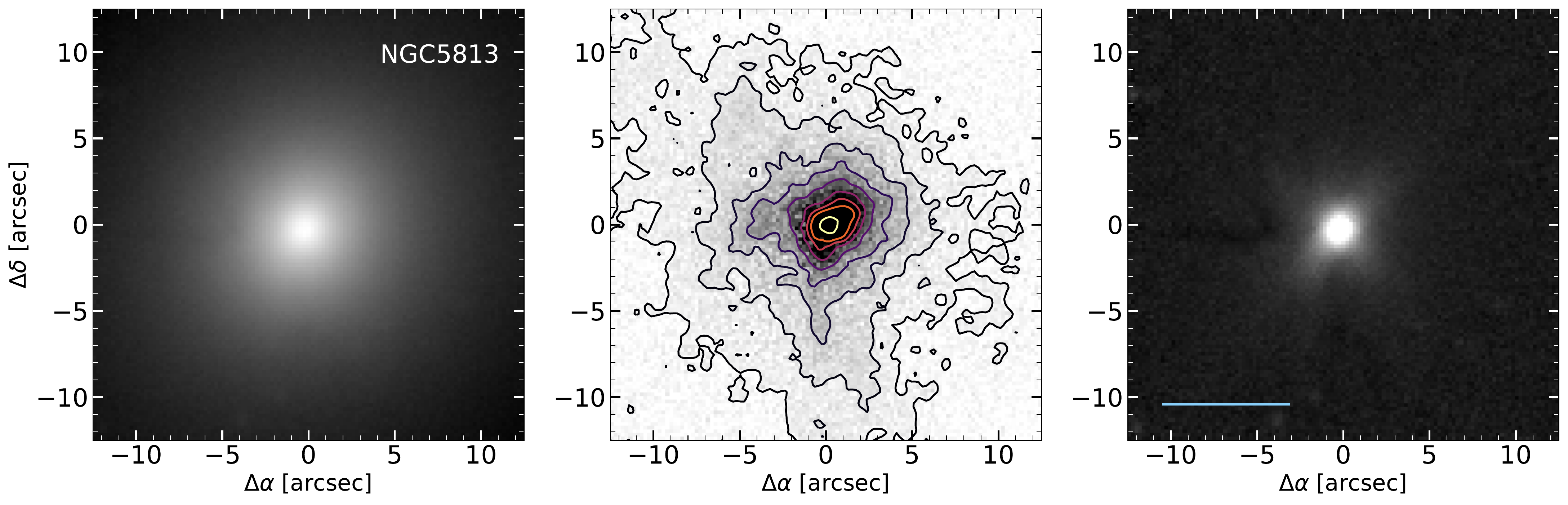}
   \includegraphics[width=\textwidth]{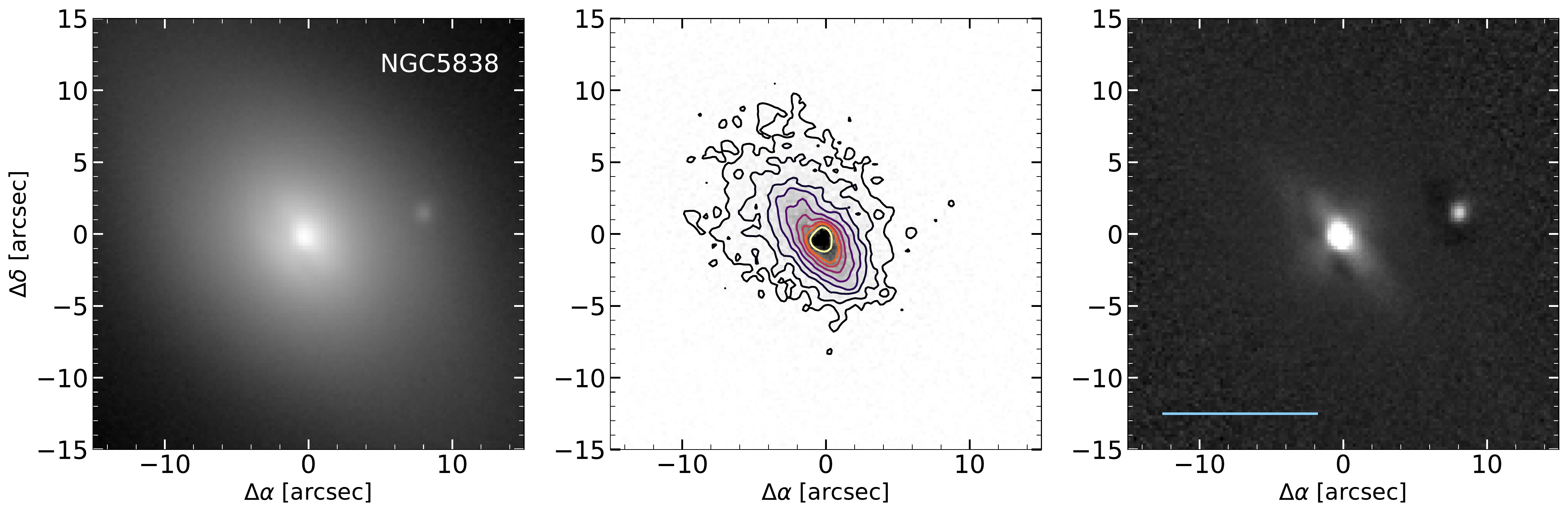}
   \includegraphics[width=\textwidth]{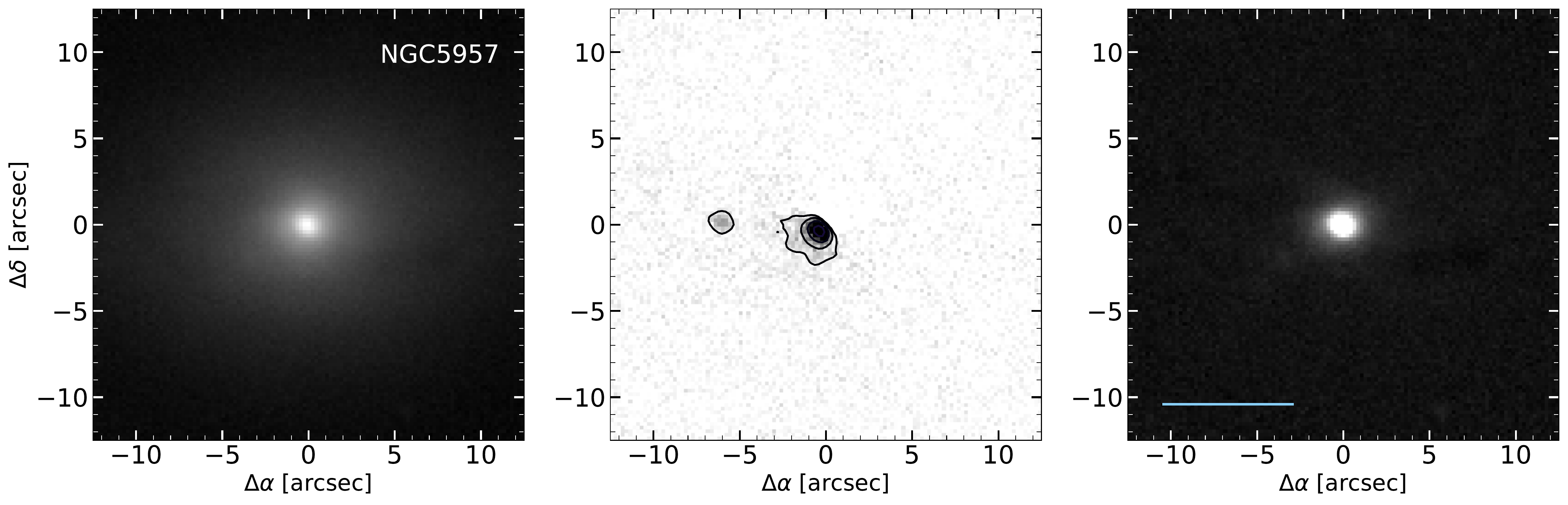}
   \includegraphics[width=\textwidth]{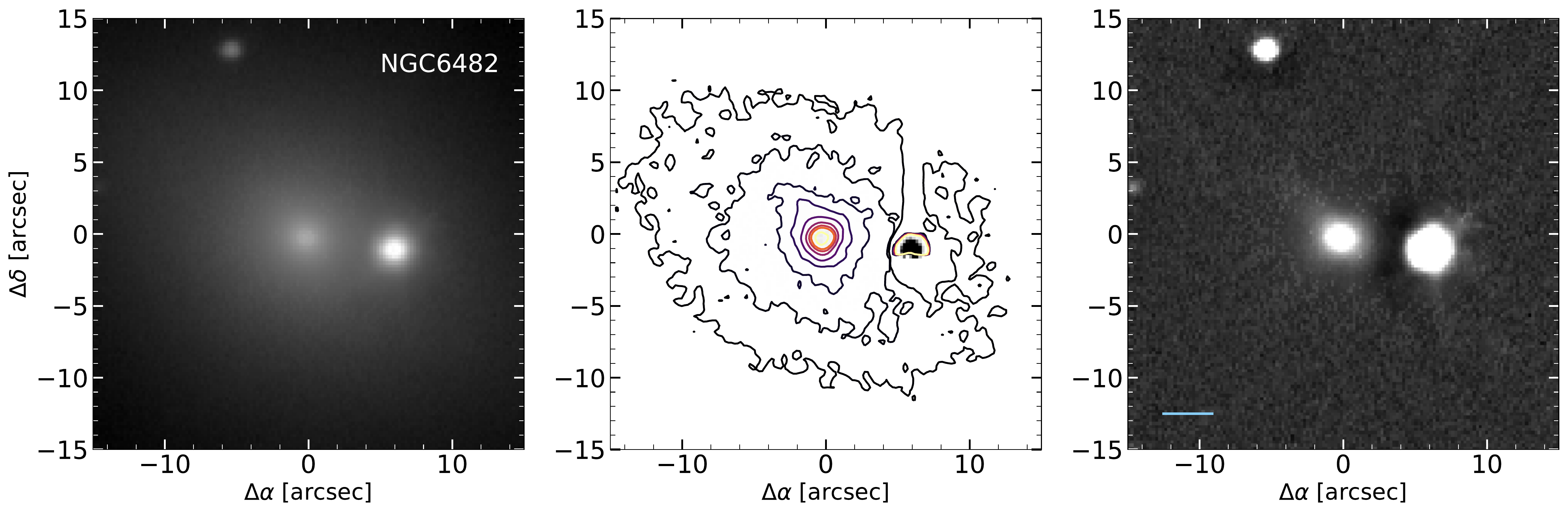}
   \caption{NGC\,5813, NGC\,5838, NGC\,5957 and NGC\,6482 H$\alpha$ emission. The complete description is in Fig.~\ref{Figure_AB1}.}  
   \label{Figure_AB9}
   \end{figure*}

\noindent \textit{NGC\,5813} (see Fig.~\ref{Figure_AB9}): The most notable feature of this elliptical is that it hosts a kinematically decoupled core \citep{Kormendy1984,Krajnovic2015}. \cite{Carollo1997} reported the presence of a dust lane along the major axis of the galaxy, stating that the dust itself may be suggesting the existence of the two reported cores. Our image of the ionised gas reveals that is very extended emission (see Fig.~\ref{Figure_largescale}) with filamentary structures that resemble an outflow. In previous H$\alpha$ observations with the Imager on the SOAR telescope, \cite{Randall2011} reported the presence of filaments co-spatial with the radio emission and with cool gas. They also reported that part of the central H$\alpha$ emission is anti-correlated with some inner X-rays cavities, although we do find a rough correlation with the soft X-ray emission. With MUSE IFU data, \cite{Krajnovic2015} identified two counter-rotating components (visible in the emission lines) and a disturbed velocity dispersion in the stellar component. The H$\alpha$-[N\,II] emission revealed the presence of knots and filaments along the polar direction of the galaxy, also visible in our image. Both lines have a uniform velocity dispersion and both approaching and receding velocities with respect to the systemic velocity of the galaxy \citep{Krajnovic2015}. The origin of this emission is explained in \cite{Krajnovic2015} as being related to the X-rays cavities and the jet activity, due to the interaction of the gas filaments with the plasma, being a gas reservoir of the galaxy rather than an inflow or an outflow. Despite its complex kinematics and its double nucleus, this galaxy is not believed to have gone through a major merger \citep{Randall2015}.\\

\noindent \textit{NGC\,5838} (see Fig.~\ref{Figure_AB9}): This is an inclined lenticular galaxy (\textit{i}\,=\,72\degr, see Table.~\ref{Table_galaxieslisted}), with a thin bar \citep{Molaeinezhad2016,Laurikainen2017}, fast, regular disc rotation \citep{Emsellem2004}. There is reported minor axis rotation, probably produced by asymmetry effects in the velocity maps due to the presence of a nuclear dust disc \citep{FalconBarroso2003}. This dust disc is subtle, but visible in our broad-band image. Our narrow-band image precludes to interpret if the gas is lying in the galaxy disc or it may be produced by outflowing processes given its asymmetry. \cite{Sarzi2006} found a regular dust disc and a relatively small scale of the gas distribution (maximum extension up to $\sim$10\arcsec\ north-east from the nucleus), being H$\beta$ similar to our H$\alpha$ distribution. These authors remarked its similarity with NGC\,4459, both in the gas distribution and in the kinematic maps, whose ionised gas we classified here as an \lq Bubble\rq (see individual comments for that galaxy). \\

\noindent \textit{NGC\,5957} (see Fig.~\ref{Figure_AB9}): This galaxy has previous observations with the NOT telescope in the broad R filter, studied in the work by \cite{Erwin2008}. However, they could not determine all their properties due to problems with the sky subtraction. Similarly, our observations of this galaxy suffered from poor sky conditions and high clouds (see Sect.~\ref{Sect_sample_data}), which is translated into a flux loss; we cannot state anything about the ionised gas morphology beyond the pure nuclei, due to the low signal-to-noise. \\

\noindent \textit{NGC\,6482} (see Fig.~\ref{Figure_AB9}): This galaxy is the brightest object of its fossil group \citep{Corsini2018}. The ionised gas in our image has a \lq Core-halo\rq\,morphology. The object was studied in the context of the MASSIVE survey \citep{Ma2014}, were \cite{Ene2018} found a regular rotation pattern in the stellar component, well aligned with the photometric axis. They found that it is one of the galaxies with more warm gas from the sample, extended up to 16 kpc, in contrast to the ionised gas that goes up to 4 kpc. They associated it to cooling flows that may be connected to the X-ray emission in the system \citep{Pandya2017}. \cite{Corsini2018} analysed the stellar and gaseous components of the galaxy via long-slit spectroscopy, finding a young stellar population that may have been produced by a recent merger or AGN feedback processes. \\

\begin{figure*}
   \includegraphics[width=\textwidth]{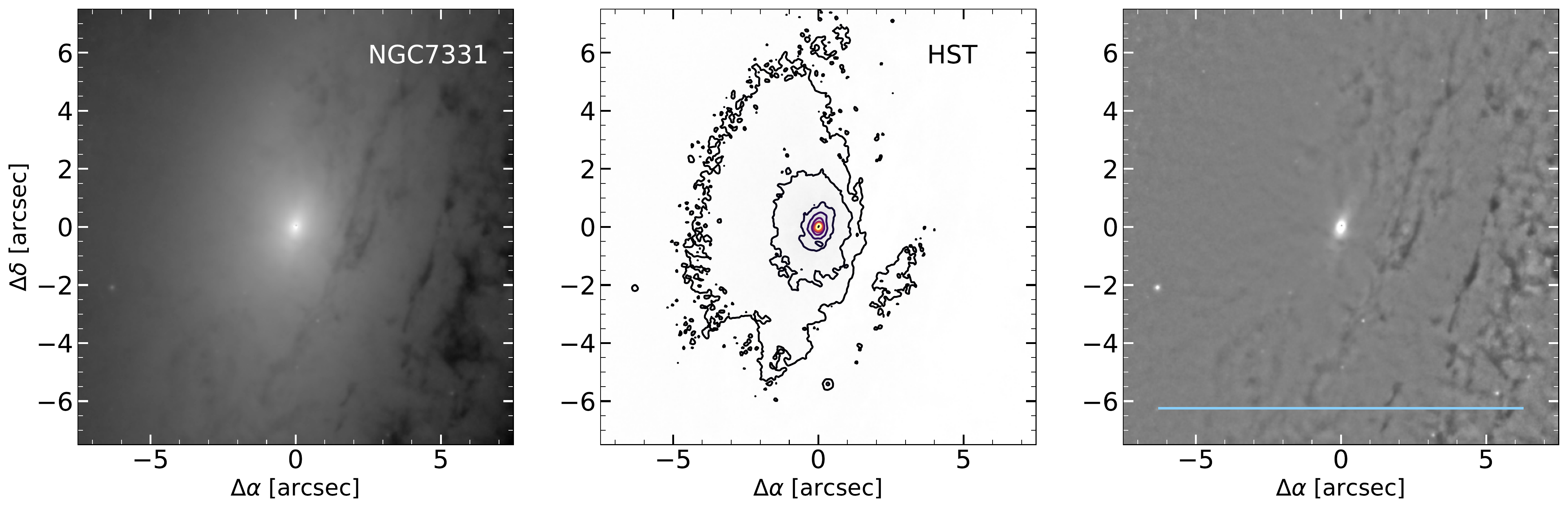}
   \includegraphics[width=\textwidth]{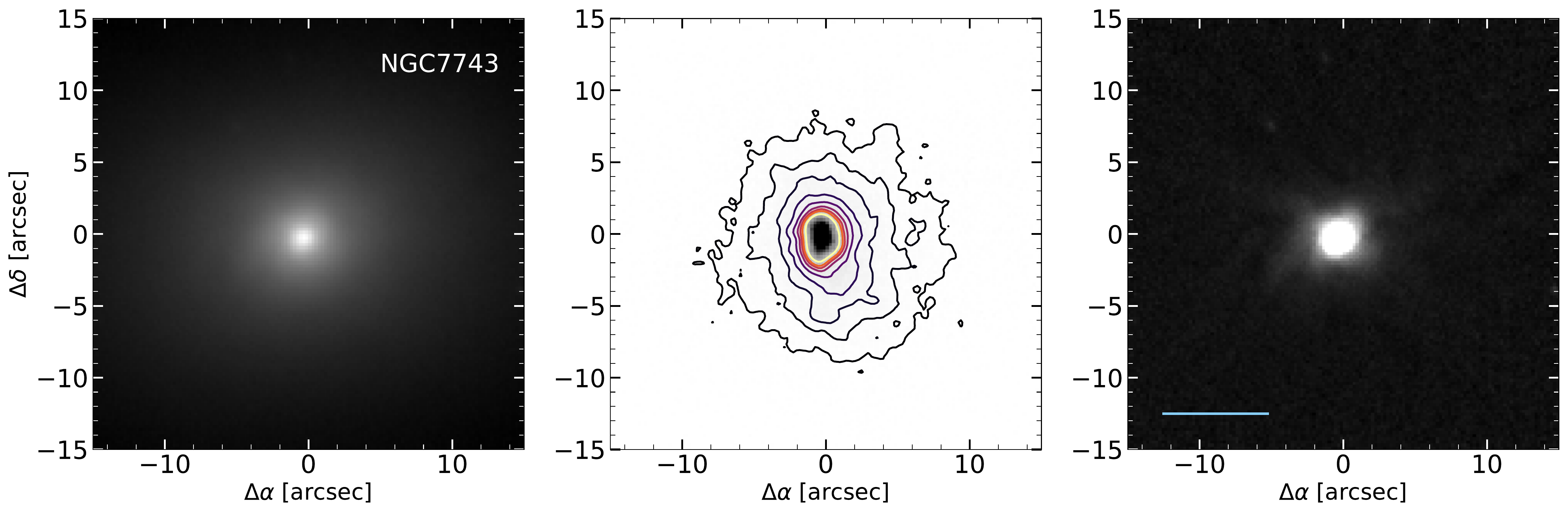}
   \caption{NGC\,7331 and NGC\,7743 H$\alpha$ emission. The complete description is in Fig.~\ref{Figure_AB1}.}  
   \label{Figure_AB10}
   \end{figure*}

\noindent \textit{NGC\,7331} (see Fig.~\ref{Figure_AB10}): This is a highly inclined spiral galaxy (\textit{i}\,=\,72\degr, see Table~\ref{Table_galaxieslisted}) located in a group along other 4 galaxies (NGC\,7335, NG\,7336, NGC\,7337, and NGC\,7340). The H$\alpha$ gas is asymmetric, resembling an outflow-like morphology. \cite{Mediavilla1997} studied the ionised gas emission, in particular the [O\,III] lines, that were modelled with 3 different components associated to: systemic velocity, a blueshifted component and a redshifted component. They interpret the components as coming from a disc and a shell of gas, although they do not discard the presence of an outflow/inflow. \cite{Battaner2003} suggest the existence of a massive stellar formation ring, through which there is infalling matter into the inner regions of the galaxy. Inflowing and outflowing material in this galaxy were also found using H\,I data from the THINGS survey \citep{Schmidt2016}. \\

\noindent \textit{NGC7743} (see Fig.~\ref{Figure_AB10}): We have classified the ionised gas morphology as \lq Core-halo\rq. The analysis of both long-slit and IFS spectroscopic data made by \cite{Katkov2011} conclude that the ionised gas present several components in the inner 1-2\arcsec, one of them probably produced by the interaction of the jet from the AGN with the ambient interstellar medium. This could be related to what we see in our image, that the H$\alpha$ contours in that regions shows a slightly different orientation than the outer contours (north-south vs northeast-southwest). The H$_{2}$ velocity field derived from K-band IFS with SINFONI allowed \cite{Davies2014} to conclude that the molecular gas is outflowing from the AGN. \\

\section{H\ensuremath{\alpha} and sharp-divided images}
\label{Appendix_B}
   
\end{appendix}


\end{document}